\documentclass[epj]{svjour}
\usepackage{graphicx}
\usepackage{color}

\def\mb#1{\setbox0=\hbox{$#1$}\kern-.025em\copy0\kern-\wd0
\kern-0.05em\copy0\kern-\wd0\kern-.025em\raise.0233em\box0}

\begin{document}
   \title{Kinetic theory of point vortices in two dimensions: analytical results and numerical simulations}

 \author{P.H. Chavanis\inst{1} and M. Lemou\inst{2}}

\institute{Laboratoire de Physique Th\'eorique (CNRS UMR 5152), 
Universit\'e Paul Sabatier, 118 route de Narbonne, \\
31062 Toulouse, France.
\email{chavanis@irsamc.ups-tlse.fr}\and
Math\'ematiques pour l'Industrie et la Physique (CNRS UMR 5640), 
 Universit\'e
Paul Sabatier, \\ 118 route de Narbonne,
31062 Toulouse, France.
\email{lemou@mip.ups-tlse.fr}}

\titlerunning{Kinetic theory of point vortices in two dimensions}

   \date{To be included later }

   \abstract{We develop the kinetic theory of point vortices in two-dimensional
hydrodynamics  and illustrate the main results of the
theory with numerical simulations. We first consider the evolution of
the system ``as a whole'' and show that the evolution of the
vorticity profile is due to resonances between different orbits of the
point vortices. The evolution stops when the profile of angular
velocity becomes monotonic even if the system has not reached the
statistical equilibrium state (Boltzmann distribution). In that case,
the system remains blocked in a quasi stationary state with a non
standard distribution. We also study the relaxation of a test vortex
in a steady bath of field vortices. The relaxation of the test vortex
is described by a Fokker-Planck equation involving a diffusion term
and a drift term. The diffusion coefficient, which is proportional to
the density of field vortices and inversely proportional to the shear,
usually decreases rapidly with the distance. The drift is proportional
to the gradient of the density profile of the field vortices and is
connected to the diffusion coefficient by a generalized Einstein
relation.  We study the evolution of the tail of the distribution
function of the test vortex and show that it has a front structure. We
also study how the temporal auto-correlation function of the position
of the test vortex decreases with time and find that it usually
exhibits an algebraic behavior with an exponent that we compute
analytically. We mention analogies with other systems with long-range
interactions.  
\PACS{
{05.20.-y}{Classical statistical mechanics}\and
 {05.45.-a}{Nonlinear dynamics and nonlinear dynamical systems}\and
{05.20.Dd}{Kinetic theory}\and
 {47.10.-g}{General theory in fluid dynamics}\and
{47.32.C-}{Vortex dynamics}} }

   \maketitle
%

\section{Introduction}
\label{sec_introduction}

Systems with long-range interactions have been the object of
considerable interest in recent years (see different contributions in
the book \cite{dauxois}). Their dynamics is
very rich and presents many interesting features
\cite{houches,pa1}. Therefore, the construction of a kinetic theory adapted to
such systems is a challenging problem with a lot of potential
applications. Several kinetic theories have been developed in the
past. The first kinetic theory was constructed by Boltzmann
\cite{boltzmann} for ideal gases. In that case, the particles do not
interact except during strong collisions. His results were later
extended by Landau
\cite{landau} in the case of Coulombian plasmas and by Chandrasekhar
\cite{chandra} (see a review by Kandrup \cite{kandrupREV}) in the
case of stellar systems. Developements and improvements of the
kinetic theory of Coulombian plasmas were made by Lenard
\cite{lenard} and Balescu \cite{balescu} using more formal
approaches allowing to take into account collective effects. They
showed in particular how collective effects can regularize the
logarithmic divergence of the diffusion coefficient at the Debye
length.  More recently, Bouchet \& Dauxois \cite{bouchet,bd}, Chavanis
{\it et al.} \cite{cvb,curious} and Chavanis \& Lemou \cite{cl} have
developed the kinetic theory of the Hamiltonian Mean Field (HMF)
model, a toy model of systems with long-range interactions which
possesses a lot of interesting properties and which can be studied in
great detail.  On the other hand, Chavanis \cite{pa1,landaud} and
Benedetti {\it et al.} \cite{benedetti} have worked out the kinetic
theory of a 2D Coulombian plasma and Valageas \cite{valageas} has
built up a kinetic theory for a 1D gravitational system in a
cosmological context. Finally, Chavanis \cite{action} has obtained a
general kinetic equation, written in angle-action variables, which is expected
to describes the dynamical evolution of a one-dimensional
inhomogeneous system of particles coupled by a weak long-range binary
potential of interaction.

In this paper, we shall consider the kinetic theory of point vortices
in two-dimensional hydrodynamics \cite{houches}. In that case, the
particles interact via a logarithmic potential in two dimensions. The
dynamical evolution of the system is due to long-range collisions
which involve interactions between vortices that can be at large
distances from each others (this will be referred to as ``distant
collisions''). The point vortex gas is probably the first physical
system for which long-range interactions and spatial delocalization
play a prevalent role in the kinetic theory. Indeed, it is not
possible to assume that the system is spatially homogeneous as done in
the kinetic theory of Coulombian plasmas (invoking Debye shielding
\cite{ichimaru}), in the kinetic theory of stellar systems
\footnote{Of course, stellar systems are spatially inhomogeneous but
the collision term is calculated as if the system were spatially
homogeneous (this results in a logarithmic divergence of the diffusion
coefficient at large scales). Then, spatial inhomogeneity is taken
into account in the inertial (Vlasov) term through the mean field
gravitational potential.}  (invoking a local approximation \cite{bt}),
and for the HMF model (above the critical energy $E_{c}$ where a
stable homogeneous phase exists [10-12,14]). Furthermore, the point
vortex gas is peculiar because point vortices have no inertia so that
the collision term in the kinetic equation directly acts in position
space. Therefore, the role of position plays the role of velocity in
usual kinetic theories.  This makes the case of point vortices
intermediate between kinetic theories of spatially homogeneous systems
for which the evolution occurs only in velocity space [4-12,14-16] and
kinetic theories of spatially inhomogeneous systems for which the
evolution occurs in velocity {\it and} position space \cite{curious},
or in energy space if we average over the orbits
\cite{valageas,action}.  Because of its connections with other systems
with long-range interactions, the kinetic theory of point vortices has
a broader interest than simply fluid mechanics.

A kinetic theory of point vortices in a shear flow 
has been first developed by Nazarenko \& Zakharov \cite{zac}. They
considered a multi-components system and assumed that the interaction
between vortices is shielded due to geophysical effects like
rotation. They studied in detail the {\it close} collisions between
two-screened particles (Stuart vortices) which are moved to each other
by the collective shear flow and developed a kinetic theory {\it
\`a la Boltzmann}.  They showed that the mean vorticity profile does
not change in time and that, due to collisions, the most intensive
vortices are concentrated in the regions of large total vorticity
while less intensive vortices are in small vorticity regions. For a
single species system, the collision integral cancels out
identically.

A kinetic theory of point vortices with equal
circulation has been developed more recently by Chavanis
[22-24,2] (see also related works in 
Chavanis \& Sire \cite{cs,cs2}) by analogy with the Brownian theory of
Chandrasekhar \cite{chandraB} in stellar dynamics.  In a first paper
\cite{preR}, he considered the relaxation of a test vortex in a
``sea'' of field vortices and derived a Fokker-Planck equation where
the evolution of the distribution function of the test vortex is due
to the competition between a diffusion term and a drift term. The
diffusion arises from the fluctuations of the velocity created by the
field vortices and the drift term is due to the response of the field
vortices to the perturbation caused by the test vortex, as in a
polarization process. Chavanis \cite{preR} calculated the expression
of the diffusion coefficient from the Kubo formula and the expression
of the drift term from a linear response theory. This is similar to
the calculations of Kandrup \cite{kandrup2} in stellar dynamics to
determine the diffusion coefficient and the friction force experienced
by a test star in a cluster of field stars. It is found that the
diffusion coefficient of point vortices is proportional to the local
vorticity created by the field vortices and inversely proportional to
the shear. On the other hand, the drift velocity is proportional to
the local vorticity gradient and inversely proportional to the
shear. Assuming that the vorticity profile of the field vortices is
positive, axisymmetric and decreasing with the distance, the
expression of the drift velocity shows that a test vortex with
positive circulation climbs the vorticity gradient and that a test
vortex with negative circulation descends the vorticity gradient. When
the field vortices have the Boltzmann distribution of statistical
equilibrium (thermal bath), the diffusion coefficient and the drift
coefficient (mobility) are related to each other by a generalization
of the Einstein relation involving a negative temperature. In a second
paper, Chavanis \cite{kin} developed a more complete kinetic theory of
point vortices by using the projection operator formalism of Willis \&
Picard \cite{wp}. From this general formalism, he obtained a kinetic
equation describing the evolution of the vortex system ``as a whole''.
This is the counterpart of the generalized Landau equation obtained by
Kandrup \cite{kandrup1} in stellar dynamics. This equation conserves
the energy and monotonically increases the Boltzmann entropy
($H$-theorem).  The computed collision integral is of
order $O(1/N)$ in a proper thermodynamic limit $N\rightarrow +\infty$
where the domain area $V$ is fixed and the circulation of the point
vortices scales like $\gamma\sim 1/N$ (so that the total circulation
$\Gamma=N\gamma$ is of order unity). This collision integral takes
into account the influence of two-body correlations. For $N\rightarrow
+\infty$, the correlations are negligible and we recover the 2D Euler
equation which describes a ``collisionless'' evolution. This is the
counterpart of the Vlasov equation in plasma physics. At order
$O(1/N)$, the kinetic theory shows that the ``collisional'' evolution
is due to a condition of resonance between the trajectories of point
vortices that can be far away. Therefore, this approach takes into
account {\it distant} collisions between vortices while the approach
developed by Nazarenko
\& Zakharov \cite{zac} focuses on {\it close} collisions. This is why
the collision integral derived in \cite{kin} can be non-zero for a
single species system of vortices while it cancels out identically in
\cite{zac}. The
above-mentioned kinetic theory, developed at the order $O(1/N)$,
describes the evolution of the system on a timescale of order
$Nt_{D}$, were $t_{D}$ is the dynamical time. Furthermore, it assumes
that the point vortices are transported by the collective shear flow
(meanfield velocity) rather than, say, triple collisions. It is valid
therefore when the shear is sufficiently strong in the system. If we
implement a thermal bath approximation to describe the relaxation of a
test vortex in a {\it fixed} distribution of field vortices, we
recover the Fokker-Planck equation involving the terms of diffusion
and drift obtained in \cite{preR}. In that case, the relaxation time
scales like $(N/\ln N)t_{D}$.

Similar problems have been studied independently by Dubin and
collaborators [31-34] in the context of
non-neutral plasmas under a strong magnetic field, a system isomorphic
to the point vortex system. Dubin \& O'Neil \cite{dn} developed a
kinetic theory of these systems, starting from the Klimontovich
equation \cite{pitaevskii} and using methods of plasma physics. Their
kinetic theory is able to take into account collective effects that
are ignored in the approach of Chavanis \cite{kin}. However, the
kinetic equation obtained by Chavanis \cite{kin} captures the main
properties of the dynamics and can be solved numerically more easily,
an advantage that will be used in the present paper. In recent works,
Dubin and collaborators
[32-34] studied the dynamics of a test vortex in a
background shear and provided very nice numerical simulations and
laboratory experiments of this process. On a theoretical point of
view, they derived the expressions of the diffusion coefficient and of
the drift velocity of the test vortex. These expressions are
consistent with those obtained previously by Chavanis
\cite{preR,kin}. They also 
addressed the form of the cut-off to the logarithmic divergence that
occurs in these quantities. However, they introduced these terms of
diffusion and drift as independent effects \footnote{The reason is
that Schecter \& Dubin
\cite{schecter} considered the drift of a test vortex in a shear
created by a {\it smooth} vorticity field (without fluctuation) that
is solution of the 2D Euler equation while Chavanis \cite{preR,kin}
considered the drift of a test vortex in a shear produced by a {\it
discret} collection of $N$ point vortices. The fluctuations, due to
finite $N$ effects, give rise to the diffusion process and the
inhomogeneity of the averaged vorticity profile gives rise to the
drift. Since there are no discrete effects in the situation considered
by Schecter \& Dubin \cite{schecter}, the test vortex experiences only
a drift due to the inhomogeneity of the vorticity background. Although
the systems are different, the expressions of the drift obtained by
Chavanis \cite{preR,kin} from the Liouville equation and by Schecter
\& Dubin \cite{schecter} from the 2D Euler equation are the same. } and did
not derive the Einstein relation connecting the diffusion to the drift
nor the Fokker-Planck equation governing the evolution of a test
vortex in a fixed distribution of field vortices.

The object of the paper is to further develop the kinetic theory of
point vortices and numerically solve the corresponding kinetic
equations in order to illustrate the basic results of the theory.  The
kinetic theory is important to describe the relaxation of the system
towards the Boltzmann distribution predicted by equilibrium
statistical mechanics for $t\rightarrow +\infty$
[36-41]. It is also necessary to determine the
timescale of the collisional relaxation and to prove whether or not
the system will truly relax towards Boltzmann statistical equilibrium. Indeed,
the convergence towards Boltzmann statistical equilibrium,
which  is based on an assumption of ergodicity, is not
firmly established for complex systems with long-range interactions
such as point vortices.  A first reason is that such systems exhibit
non-markovian effects and spatial delocalization so that
the monotonic increase of the Boltzmann entropy is
difficult to prove and could even be wrong in a strict sense
\cite{kin}. It is only when additional approximations are included
(markovian approximation, neglect of three-body correlations,..) that
the Boltzmann $H$-theorem is recovered
\footnote{These approximations are expected to be valid in a proper
thermodynamic limit $N\rightarrow +\infty$ with fixed $\eta=\beta
N\gamma^{2}$ and $\epsilon=E/N^{2}\gamma^{2}$. Therefore, the
$H$-theorem holds in that limit $N\rightarrow +\infty$. }. On the other
hand, the evolution of the system is due to resonances between
different orbits and it may happen that the evolution stops before the
system has reached statistical equilibrium because there is no
resonance anymore. In that case, the system is blocked in a quasi
stationary state (QSS) which can persist for a very long time. One
object of the paper is to discuss these issues and perform numerical
simulations in order to illustrate the particularity of the point
vortex dynamics.

The paper is organized as follows. In Sec. \ref{sec_pvg}, we briefly
discuss the statistical mechanics of the point vortex gas at
equilibrium and extend the theory to the case where the point vortices
have different circulations. The predictions of
equilibrium thermodynamics will be compared with the results of the
kinetic theory throughout the paper. In Sec. \ref{sec_evo}, we
develop the kinetic theory of point vortices and describe the
evolution of the system ``as a whole''.  In
Sec. \ref{sec_k}, we generalize the kinetic theory of \cite{kin},
based on the projection operator formalism, to a multi-species point
vortex gas. Considering an axisymmetric evolution, we obtain an
explicit kinetic equation for the evolution of the system at the order
$O(1/N)$. We provide a simpler derivation of this kinetic equation
than the one given in
\cite{kin}. We also stress some technical difficulties associated 
with the kinetic theory and propose some solutions to circumvent these
problems. In Sec. \ref{sec_c}, we show that the derived kinetic
equation conserves the vortex number of each species, the total energy
and the total angular momentum. Furthermore, it increases the
Boltzmann entropy ($H$-theorem). For a single species system, the
evolution of the point vortex gas on a timescale
$Nt_{D}$ is due to a condition of resonance
$\Omega(r,t)=\Omega(r',t)$ between different orbits $r\neq r'$ that
can be satisfied only when the profile of angular velocity is
non-monotonic. As a result, the Boltzmann distribution corresponding
to statistical equilibrium is not the only stationary solution of the
kinetic equation. For example, any distribution with a monotonic
profile of angular velocity is a steady state solution.  Therefore,
the dynamical evolution stops when the profile of angular velocity
becomes monotonic even if the system has not reached the Boltzmann
distribution (further evolution may occur on longer
timescale due to terms of order $1/N^{2}$,  $1/N^{3}$..., in the kinetic
theory corresponding to the influence of three-body, or higher,
correlations). This kinetic blocking, described in
Sec. \ref{sec_block}, is illustrated numerically in Sec.
\ref{sec_num}. In Sec. \ref{sec_rel}, we consider the relaxation of
a test vortex in a bath of field vortices. In
Sec. \ref{sec_f}, we show that the stochastic process is described by
a Fokker-Planck equation involving a term of diffusion and a term of
drift. This Fokker-Planck equation can be derived from the projection
operator formalism. We extend the approach of \cite{kin} by
considering a test particle with a circulation $\Gamma_{0}$ that can
be different from the circulation $\gamma$ of the field particles. In
Sec. \ref{sec_d}, we discuss specifically the terms of diffusion and
drift. For a thermal bath of field vortices, we show that they are
related to each other by a generalized Einstein relation
$\xi=D\beta\Gamma_{0}$. Furthermore, the diffusion coefficient $D(r)$
depends on the position and decreases rapidly with the distance,
leading to anomalous dynamical behaviors. In Sec. \ref{sec_intr}, we
derive the first and second moments of the radial position increment
from the Fokker-Planck equation. In Appendix
\ref{sec_fs}, we directly compute these moments from the Hamiltonian
equations of motion. We can therefore justify the Fokker-Planck
equation by this alternative approach instead of advocating the
projection operator formalism. In Sec. \ref{sec_ss}, we precisely
discuss the scaling of the relaxation time with $N$. We distinguish
the collisional timescale $\sim Nt_{D}$ associated with the evolution
of the system ``as a whole'' from the collisional timescale $(N/\ln
N)t_{D}$ associated with the evolution of a test vortex in a bath. In
Sec. \ref{sec_gen}, we consider different features of the
Fokker-Planck equation. In Sec. \ref{sec_t}, we study the evolution of
the tail of the distribution function of the test vortex and show that
it has a front structure. We characterize the displacement of the
front and its shape. In Sec. \ref{sec_co}, we consider the temporal
auto-correlation function of the position of the test vortex and show
that, in cases of physical interest, it decreases algebraically with
time. The exponent of algebraic decay is computed analytically and
compared with direct numerical simulations. In Secs. \ref{sec_asy} and
\ref{sec_examples}, we consider explicit examples of 
Fokker-Planck equations corresponding to typical bath distributions of
the field vortices. In order to be complete, we
describe various examples of flow profiles. However, we present
numerical simulations only for the gaussian vortex.

\section{The point vortex gas at equilibrium}
\label{sec_pvg}

\subsection{The statistical equilibrium state}
\label{sec_s}

Basically, the dynamical evolution of a system of point vortices in
two dimensions is described by the Hamiltonian equations of motion
\cite{newton}:
\begin{eqnarray}
\gamma_{i}{dx_{i}\over dt}={\partial H\over\partial y_{i}}, \qquad \gamma_{i}{dy_{i}\over dt}=-{\partial H\over\partial x_{i}},
\label{s1}
\end{eqnarray}
\begin{eqnarray}
H=-{1\over 2\pi}\sum_{i<j}\gamma_{i}\gamma_{j}\ln |{\bf r}_{i}-{\bf r}_{j}|,
\label{s2}
\end{eqnarray}
where $\gamma_{i}$ is the circulation of point vortex $i$.  A
particularity of this Hamiltonian system, first noticed by Kirchhoff
\cite{kirchhoff}, is that the coordinates $(x,y)$ of the point
vortices are canonically conjugate. The point vortex system conserves
several quantities. The number $N_{a}$ of vortices of species $a$, or
equivalently the total circulation of each species
$\Gamma_{a}=N_{a}\gamma_{a}$, is conserved. The system also conserves
the energy $E=H$. There are additional conserved quantities depending
on the geometry of the domain. The angular momentum
$L=\sum_{i}\gamma_{i} r_{i}^{2}$ is conserved in an infinite domain or
in a disk and the linear impulse ${\bf P}=\sum_{i}\gamma_{i}{\bf
r}_{i}$ is conserved in an infinite domain or in a channel. In an
unbounded domain, we must consider vortices of the same circulation
otherwise they would form pairs (dipoles) and escape to infinity, so
there is no equilibrium. In a bounded domain, the Hamiltonian has to
be modified so as to take into account the contribution of vortex
images. Note that the Hamiltonian (\ref{s2}) does not
involve a kinetic term $\sum_{i}m_{i}v_{i}^{2}/2$ usually present for
material particles.  This is because point vortices have no
inertia. Therefore, a point vortex produces a velocity, not an
acceleration (or a force), contrary to other systems of particles in
interaction like electric charges in a plasma \cite{landau} or stars
in a galaxy
\cite{chandra}. In a sense, this interaction is related to the
conception of motion according to Descartes, in contrast to Newton.

Assuming ergodicity, the point vortex gas is expected to achieve a
statistical equilibrium state for $t\rightarrow +\infty$. The
statistical mechanics of point vortices is very peculiar and was first
discussed by Onsager \cite{onsager}. He considered
box-confined configurations and  showed that for sufficiently large
energies, point vortices of the same sign tend to
``attract'' each other and group themselves to form ``clusters'' or
``supervortices'' similar to the large-scale vortices observed in the
atmosphere of giant planets. When all the vortices
have the same sign, this leads to monopoles (cyclones or anticyclones)
and when vortices have positive and negative signs, this leads
to dipoles (a pair of cyclone/anticlone) or even tripoles.  These
organized states are characterized by {\it negative}
temperatures. This is due to the fact that the structure function
$\Phi(E)=\int_{H\le E}dx_{1}dy_{1}...dx_{N}dy_{N}$ is finite for
$E\rightarrow +\infty$ since the phase space coincides with the
configuration space: $\Phi(+\infty)=\int
dx_{1}dy_{1}...dx_{N}dy_{N}=V^{N}$ where $V$ is the area of the
system. The existence of negative temperatures for point vortices
should not cause surprise. For material particles, the temperature is
a measure of the velocity dispersion and it must be positive. Indeed,
it appears in the Maxwell distribution $e^{-\beta v^2/2}$ (or in the
partition function) which can be normalized only for
$\beta>0$. However, since point vortices have no inertia, there is no
such term in the equilibrium distribution of point vortices (see
below) and the temperature can be negative.

To obtain more quantitative results, Joyce \& Montgomery \cite{jm} and
Lundgren \& Pointin \cite{lp} considered a mean field
approximation. This is valid in a proper thermodynamic limit 
$N\rightarrow +\infty$ in such a way that the normalized temperature
$\eta=\beta N\gamma^{2}$ and the normalized energy
$\epsilon=E/N^{2}\gamma^{2}$ are fixed (for a single species
system). On physical grounds, it is reasonable to
consider that the area of the domain and the total circulation of the
vortices are of order unity. Then, by rescaling the parameters appropriately,
the proper thermodynamic limit corresponds to $N\rightarrow +\infty$
with $\gamma\sim 1/N$, $V\sim 1$, $E\sim 1$ and $\beta\sim N$
\cite{houches}. Since the coupling constant $\beta\gamma^{2}=\eta/N\sim 1/N$  goes to zero for $N\rightarrow +\infty$, we are considering a {\it weak} long-range potential of interaction. In that limit, we can neglect the correlations
between vortices and the statistical equilibrium distribution of
point vortices is given by the Boltzmann distribution
\begin{eqnarray}
P_{eq}({\bf r})=A e^{-\beta\gamma\psi({\bf r})}, \label{s3}
\end{eqnarray}
where $\psi$ is the stream function. Using the Poisson equation
$\Delta\psi=-\omega$ with the smooth vorticity $\omega({\bf
r})=N\gamma P({\bf r})$, the statistical equilibrium state is
obtained by solving the Boltzmann-Poisson equation
\begin{eqnarray}
-\Delta\psi=N\gamma A e^{-\beta\gamma\psi},
\label{s4}
\end{eqnarray}
and substituting the resulting stream function in Eq. (\ref{s3}).
Joyce \& Montgomery
\cite{jm} obtained the equilibrium distribution (\ref{s3}) by
maximizing the entropy of the point vortex gas at fixed circulation
and energy.  Lundgren \& Pointin \cite{lp} derived Eq. (\ref{s4}) from
the equilibrium BBGKY hierarchy of equations in the limit
$N\rightarrow +\infty$.  A rigorous derivation of the mean field
equations was provided by [39-41].

On the other hand, at sufficiently large negative energies, the
temperature is positive. In that case, like-sign vortices tend to
``repell'' each other. When all the vortices of the system have the
same sign, they accumulate on the boundary of the domain. This regime
can again be described by the mean field theory of
[37-41]. Alternatively, when the system is neutral, opposite-sign
vortices tend to ``attract'' each other resulting in a spatially
homogeneous distribution with strong correlations between vortices. In
that case, the point vortex gas is similar to a Coulombian
plasma. This is the situation considered by Fr\"ohlich \& Ruelle
\cite{fr}. As discussed by Ruelle \cite{ruelle}, we may expect a phase
transition (related to the Kosterlitz-Thouless transition) as a
function of the temperature.  At large (positive) temperatures, the
system is in a ``conducting phase'' (in the plasma analogy) with free
vortices that can screen external ``charges''. In that case, there can
be Debye shielding like for a Coulombian plasma. At low (positive)
temperatures, opposite-sign vortices tend to form pairs $(+,-)$ and
the gas is in a ``dielectric phase'' where all charges are bound
forming dipolar pairs. These ``dipoles'' are similar to ``atoms''
$(+e,-e)$ in plasma physics. In that case, there is no screening.

Summarizing, there are two very different regimes in the point
vortex gas that correspond to different thermodynamic limits. In the
meanfield theory of [37-41], the thermodynamic limit
corresponds to $N\rightarrow +\infty$ in such a way that the size $V$
of the system is fixed and the interaction between vortices is weak
since the ``charge'' of the vortices tends to zero as $\gamma\sim
1/N$. In that case, the system is spatially inhomogeneous, the
correlations between vortices are negligible for $N\rightarrow
+\infty$, there is no shielding and the temperature can be
negative. The physics of the problem is controlled by the one-body
distribution function $P({\bf r})$. Alternatively, Fr\"ohlich \&
Ruelle \cite{fr} consider a neutral system of $2N$ positive and
negative point vortices with circulation $\pm\gamma$ and take the
usual thermodynamic limit $N,V\rightarrow +\infty$ with $N/V$ and
$E/N$ fixed (with $\gamma\sim 1$). In that case, the system is
spatially homogeneous in average, the temperature is positive and the
physics of the problem is controlled by the two-body correlation
function. In the screening phase (at high positive temperatures),
there is Debye shielding and the correlation function tends to zero
exponentially rapidly.

\subsection{Generalization to a collection of circulations}
\label{sec_gg}

In this paper, we shall restrict ourselves to the situation 
described in [37-41]. In this section, we
generalize the maximum entropy method of Joyce
\& Montgomery \cite{jm} in order to determine the statistical
equilibrium state of a system of point vortices with different
circulations. This will be useful to interpret the results of
Sec. \ref{sec_block}. Following the Boltzmann procedure, we divide the
domain into a very large number of microcells with size $h^{2}$
(ultimately $h\rightarrow 0$).  A microcell can be occupied by an
arbitrary number of point vortices.  We shall now group these
microcells into macrocells each of which contains many microcells but
remains nevertheless small compared to the extension of the whole
system. We call $\nu$ the number of microcells in a macrocell. A
macrostate is specified by the number $\lbrace n_{ia}
\rbrace$ of point vortices of circulation $\gamma_{a}$ in the
macrocell $i$ (irrespective of their position in the cell) while a
microstate is specified by their precise position in the cell. Using a
combinatorial analysis, the number of microstates
corresponding to the macrostate 
$\lbrace n_{ia} \rbrace$ is
\begin{eqnarray}
W(\lbrace n_{ia} \rbrace )=\prod_{a}N_{a}!\prod_{i}\frac{\nu^{n_{ia}}}{n_{ia}!}.
\label{gg1}
\end{eqnarray}
We introduce $n_{a}({\bf r})$ the smooth density of point vortices of
species $a$ in ${\bf r}$. The vorticity of species $a$ is
then $\omega_{a}({\bf r})=\gamma_{a} n_{a}({\bf r})$. If we define the
Boltzmann entropy by $S=\ln W$, use the Stirling formula for
$n_{ia}\gg 1$ and take the continuum limit, we get
\begin{eqnarray}
S=-\sum_{a}\int {\omega_{a}\over \gamma_{a}}\ln {\omega_{a}\over \gamma_{a}} \ d{\bf r}.
\label{gg2}
\end{eqnarray}
The Boltzmann entropy (\ref{gg2}) measures the number of microstates
corresponding to the macrostate specified by $\lbrace \omega_{a}({\bf
r})\rbrace$. At statistical equilibrium, the system is expected to be
in the most probable macrostate, i.e. the one that is the most
represented at the microscopic level. Assuming that all the
microstates are equiprobable, the equilibrium distribution is obtained
by maximizing the Boltzmann entropy (\ref{gg2}) while conserving the
circulation of each species $\Gamma_{a}=\int \omega_{a} d{\bf r}$ and
the mean field energy $E={1\over 2}\int \omega\psi d{\bf r}$.  The
vorticity and the streamfunction are related to each other by the
Poisson equation
\begin{eqnarray}
-\Delta\psi=\omega=\sum_{a}\omega_{a}.
\label{gg5}
\end{eqnarray}
Finally, in an infinite domain or in a disk, we must also conserve the
angular momentum $L=\int \omega r^{2} d{\bf r}$.
Introducing Lagrange multipliers and writing the first order variations as
\begin{eqnarray}
\delta S-\sum_{a}\alpha_{a}\delta\Gamma_{a}-\beta\delta E-{1\over 2}\beta\Omega_{L}\delta L=0,
\label{gg7}
\end{eqnarray}
we find that the most probable state is
\begin{eqnarray}
\omega_{a}({\bf r})=A_{a}e^{-\beta\gamma_{a}\psi'({\bf r})},
\label{gg8}
\end{eqnarray}
where
\begin{eqnarray}
\psi'=\psi+{1\over 2}\Omega_{L}r^{2},
\label{gg9}
\end{eqnarray}
is the relative stream function. This describes a flow that is steady
in a frame rotating with angular velocity $\Omega_{L}$ (see Appendix
\ref{sec_eul}). Therefore, the equilibrium state is obtained by
solving the multi-species Boltzmann-Poisson equation
\begin{eqnarray}
-\Delta\psi=\sum_{a}A_{a}e^{-\beta\gamma_{a}\psi'({\bf r})},
\label{gg10}
\end{eqnarray}
and substituting the resulting stream function back into Eq.
(\ref{gg8}). The Lagrange multipliers can then  be related to the
integral constraints.  We note that the vorticity profiles of
different species are related to each other by
\begin{eqnarray}
\left ({\omega_{a}\over A_{a}}\right )^{1/\gamma_{a}}=\left ({\omega_{b}\over A_{b}}\right )^{1/\gamma_{b}},
\label{gg11}
\end{eqnarray}
hence
\begin{eqnarray}
\omega_{a}({\bf r})=C_{ab}|\omega_{b}({\bf r})|^{\gamma_{a}/\gamma_{b}},
\label{gg12}
\end{eqnarray}
where $C_{ab}$ is independent on the position. Assuming that
$\gamma_{a}>0$, $\gamma_{b}>0$ and that $\omega_{b}({\bf r})$
decreases with the distance (which corresponds to equilibrium states
with $\beta<0$), this relation indicates that intense vortices
($\gamma_{a}>\gamma_{b}$) are more concentrated at the center, on
average, than weaker vortices. On the other hand, Eq. (\ref{gg12})
shows that opposite sign vortices tend to separate (at negative
temperatures where the mean field theory applies). More generally,
Eq. (\ref{gg12}) characterizes the seggregation between vortices with
different circulations.

\section{Evolution of the system as a whole}
\label{sec_evo}

The equilibrium statistical mechanics tells nothing concerning the
timescale of the relaxation towards statistical equilibrium. It is
furthermore not obvious whether the system of point vortices will
truly relax towards statistical equilibrium or not.  Indeed, the
Boltzmann distribution (\ref{s3}) is based on a hypothesis of
ergodicity and on the assumption that all the accessible microstates
are equiprobable. This is essentially a {\it postulate}. In order to
precisely answer these questions we need to develop a kinetic theory
of point vortices.

\subsection{Kinetic theory of point vortices}
\label{sec_k}

There are different methods to obtain a kinetic equation for a
system of point vortices. One approach is to start from the
Klimontovich equation and develop a quasi-linear theory as in plasma
physics \cite{dn}. Another possibility is to start from the
Fokker-Planck equation and calculate the first $\langle \Delta
r\rangle $ and second moments $\langle (\Delta r)^2\rangle$ of the
increments of position of a point vortex due to the interaction with
the other vortices. This approach is developed in Appendix
\ref{sec_fs}. A third possibility is to start from the Liouville
equation and use the projection operator formalism \cite{kin}. An
interest of this formalism is to yield a general non-Markovian
equation that is valid for flows that are not necessarily
axisymmetric. The other formalisms assume from the begining that the
distribution of point vortices is axisymmetric and work in Fourier
space (for the angular variables). By contrast, the projection
operator formalism remains in physical space which enlightens the
basic physics. Indeed, even if the formalism is abstract and
complicated \cite{wp}, the final kinetic equation takes a rather nice
form which bears a clear physical meaning
\footnote{Recently, it has been found that this
kinetic equation could also be obtained from a BBGKY-like hierarchy
\cite{bbgky}. This considerably simplifies the formalism. The
collision integral corresponds to the term of order $O(1/N)$ in a
systematic expansion of the equations of the hierarchy in powers of
$1/N$ when $N\rightarrow +\infty$ with fixed domain area $V$ and fixed total
circulation $\Gamma=N\gamma$.}. The drawback of that approach is
that it ignores collective effects. For axisymmetric flows, these
collective effects have been taken into account in the approach of
Dubin \& O'Neil
\cite{dn}.

We shall here extend the kinetic theory of Chavanis \cite{kin},
based on the projection operator formalism, to the case of a
multi-components point vortex gas. We shall not repeat the intermediate
steps of the calculations that can be found in
\cite{kin}.  Generalizing these calculations in order to include a
distribution of circulations among the point vortices, we obtain a
kinetic equation of the form
\begin{eqnarray}
{\partial\omega_{a}\over\partial t}+\langle {\bf V}\rangle\cdot \nabla\omega_{a}={\partial\over\partial r^{\mu}}\sum_{b}\int_{0}^{t}d\tau \int d{\bf r}_{1} {\cal V}^{\mu}(1\rightarrow 0)\nonumber\\
\times {\cal G}(t,t-\tau) 
\biggl \lbrack \gamma_{b}{\cal V}^{\nu}(1\rightarrow 0){\partial\over\partial r^{\nu}}\nonumber\\
+\gamma_{a} {\cal V}^{\nu}(0\rightarrow 1){\partial\over\partial r_{1}^{\nu}}\biggr \rbrack \omega_{a}({\bf r},t-\tau)\omega_{b}({\bf r}_{1},t-\tau),\nonumber\\
\label{k1}
\end{eqnarray}
where ${\cal G}(t,t-\tau)$ is the Green function associated with the
averaged Liouville operator constructed with the mean field velocity
$\langle {\bf V}\rangle({\bf r},t)=-{\bf z}\times\nabla\psi$. In
words, this means that we must perform the time integration by moving
the point vortices between $t$ and $t-\tau$ with the
mean field velocity. On the other hand, ${\cal V}^{\mu}(1\rightarrow
0,t)={V}^{\mu}(1\rightarrow 0,t)-\langle V^{\mu}\rangle ({\bf r},t)$
is the fluctuating velocity by unit of circulation created by particle
$1$ located in ${\bf r}_{1}$ on particle $0$ located in ${\bf r}$ (see
Appendix \ref{sec_m}). To obtain this closed kinetic equation, we have
implicitly assumed that three-body correlations are negligible. This
neglect is valid at order $1/N$ in the proper thermodynamic limit
$N\rightarrow +\infty$ defined above. In the general kinetic equation
(\ref{k1}), we clearly see the terms of diffusion and drift (first and
second terms in the r.h.s.) and their connection to a generalized form
of Kubo formula (the integral over time of the velocity
auto-correlation function). These points will be further developed in
the sequel. The ratio of the ``collision'' term (right hand side) on
the advective term (left hand side) is of order $1/N$ in the
thermodynamic limit (see Sec. \ref{sec_ss}).  Therefore, this kinetic
equation is valid at order $O(1/N)$ in an expansion of the collision
term in powers of $1/N$. Thus, it describes the evolution of the
system on a timescale of the order $N t_D$ where $t_D$ is the
dynamical time.

We must
distinguish two regimes in the evolution of the point vortex system:

{\it (i) Collisionless regime}: For fixed $t$ and $N\rightarrow
+\infty$, the collision term on the r.h.s. of Eq. (\ref{k1}) vanishes
and this equation reduces to the 2D Euler equation
\cite{kin}:
\begin{eqnarray}
{\partial\omega_{a}\over\partial t}+\langle {\bf V}\rangle\cdot \nabla\omega_{a}=0.
\label{k2}
\end{eqnarray}
This is the counterpart of the Vlasov equation in plasma physics
\cite{ichimaru} or stellar dynamics \cite{bt} (it is valid for each
species of particles). It can be directly derived from the Liouville
equation by neglecting correlations between vortices, i.e. by assuming
that the $N$-body distribution function is a product of $N$ one-body
distribution functions \cite{kin}. The Euler equation (\ref{k2})
describes the collisionless evolution of the point vortex system on a
timescale $t\ll Nt_{D}$ where $t_{D}$ is the dynamical time. For $N\gg
1$, the validity of the Vlasov regime can be very long in
practice. Starting from an unstable initial condition,
the 2D Euler-Poisson system can undergo a ``violent relaxation''
driven by purely mean field effects \cite{houches}. This form of
collisionless relaxation leads to the rapid emergence of a quasi
stationary state (QSS) on the coarse-grained scale
[47-50].

{\it (ii) Collisional regime}: The ``collision'' term (right hand side
in Eq. (\ref{k1})) takes into account finite $N$ effects.  It can be
viewed as the first order correction of the Euler/Vlasov equation in
an $N^{-1}$ expansion (see, e.g., \cite{houches} p.  260). Therefore,
the collision term in the kinetic equation (\ref{k1}) manifests itself
on a timescale $\sim Nt_{D}$. In this paper, we shall
be essentially interested by this ``slow collisional evolution''.

In the following, we shall assume that the vorticity profile is a
stable stationary solution of the 2D Euler equation, so that it
evolves under the sole effect of collisions (finite $N$ effects) on a
timescale $Nt_D$. Therefore, we shall not describe the process of
violent relaxation [47-50] due to mean field effects
that takes place during the collisionless regime on a timescale
$t_{D}$ (see the Conclusion for a discussion of the different regimes
of the dynamics). We thus start from a stable stationary solution of the  2D
Euler equation (possibly resulting from a phase of violent relaxation)
and discuss its collisional evolution.  If we restrict ourselves to
axisymmetric flows, the kinetic equation (\ref{k1}) becomes
\begin{eqnarray}
{\partial\omega_{a}\over\partial t}={1\over r}{\partial\over\partial
r} r\sum_{b}\int_{0}^{t}d\tau\int d{\bf r}_{1} {V}_{r}(1\rightarrow
0,t)\nonumber\\
\times \biggl \lbrack \gamma_{b}{V}_{r}(1\rightarrow 0,t-\tau){\partial\over\partial r}
+\gamma_{a}{V}_{r_{1}}(0\rightarrow 1,t-\tau){\partial\over\partial r_{1}}\biggr \rbrack\nonumber\\
\times  \omega_{a}({r},t-\tau)\omega_{b}({r}_{1},t-\tau),\qquad\qquad
\label{k3}
\end{eqnarray}
where $V_{r}={\bf V}\cdot \hat{\bf r}$ is the radial component of the
velocity in the direction $\hat{\bf r}={\bf r}/r$ of particle $0$ (we
adopt the same convention for particle $1$).  Since the collision term
is of order $1/N$, the smooth distribution $\omega_{a}(r,t)$ evolves
on a timescale of order $Nt_{D}$ at least, which is much larger than
the timescale on which the fluctuations have essential
correlations. Thus, we can make a Markov approximation
$\omega({r},t-\tau)\simeq
\omega({r},t)$ and extend the time integral to $+\infty$
(see, however, the end of this section). In the same order of
approximations, we can consider that, to leading order in $1/N$, the
point vortices follow circular trajectories given by
\begin{eqnarray}
r(t-\tau)=r, \qquad \theta(t-\tau)=\theta-\Omega(r,t)\tau,
\label{k4}
\end{eqnarray}
where $(r,\theta)$ specify the position of a point vortex at time $t$
and $\Omega(r,t)=\langle V\rangle_{\theta}(r,t)/r$ is the angular velocity
corresponding to the mean field velocity $\langle V\rangle_{\theta}(r,t)$.  Within
these approximations, the kinetic equation (\ref{k3}) becomes
\begin{eqnarray}
{\partial\omega_{a}\over\partial t}={1\over r}{\partial\over\partial r} r\sum_{b}\int_{0}^{+\infty}d\tau\int d{\bf r}_{1} {V}_{r}(1\rightarrow 0,t)\nonumber\\
\times\biggl \lbrack \gamma_{b}{V}_{r}(1\rightarrow 0,t-\tau){\partial\over\partial r}+\gamma_{a}{V}_{r_{1}}(0\rightarrow 1,t-\tau){\partial\over\partial r_{1}}\biggr \rbrack \nonumber\\
\times \omega_{a}({r},t)\omega_{b}({r}_{1},t),\qquad\qquad
\label{k5}
\end{eqnarray}
where (see Appendix \ref{sec_m})
\begin{eqnarray}
V_{r}(1\rightarrow 0,t)=-{1\over 2\pi}{r_{1}\sin(\theta-\theta_{1})\over r_{1}^{2}+r^{2}-2r r_{1}\cos(\theta-\theta_{1})},
\label{k6}
\end{eqnarray}
\begin{eqnarray}
V_{r}(1\rightarrow 0,t-\tau)=-{1\over 2\pi}{r_{1}\sin(\theta-\theta_{1}-\Delta\Omega\tau)\over r_{1}^{2}+r^{2}-2r r_{1}\cos(\theta-\theta_{1}-\Delta\Omega\tau)},\nonumber\\
\label{k7}
\end{eqnarray}
with $\Delta\Omega=\Omega(r,t)-\Omega(r_{1},t)$. Since
${V}_{r_{1}}(0\rightarrow 1)=-(r/r_{1})
{V}_{r}(1\rightarrow 0)$, we can rewrite the foregoing
equation in the form
\begin{eqnarray}
{\partial\omega_{a}\over\partial t}={1\over r}{\partial\over\partial r} r\sum_{b}\int_{0}^{+\infty}d\tau\int_{0}^{2\pi}d\theta_{1} \int_{0}^{+\infty} r r_{1}d{r}_{1} {V}_{r}(1\rightarrow 0,t)\nonumber\\
\times{V}_{r}(1\rightarrow 0,t-\tau)\biggl ( \gamma_{b}{1\over r}{\partial\over\partial r}-\gamma_{a}{1\over r_{1}}{\partial\over\partial r_{1}}\biggr )  \omega_{a}({r},t)\omega_{b}({r}_{1},t).\nonumber\\
\label{k8}
\end{eqnarray}
The kinetic equation involves the function
\begin{eqnarray}
M=\int_{0}^{+\infty}d\tau\int_{0}^{2\pi}d\theta_{1} {V}_{r}(1\rightarrow 0,t){V}_{r}(1\rightarrow 0,t-\tau).\qquad 
\label{k9}
\end{eqnarray}
Using Eqs. (\ref{k6})-(\ref{k7}), it is shown in \cite{kin,houches} (see also Appendix \ref{sec_m}) that
\begin{eqnarray}
M=-{1\over 4 r^{2}}\delta(\Delta\Omega)\ln\biggl\lbrack 1-\biggl ({r_{<}\over r_{>}}\biggr )^{2}\biggr\rbrack,
\label{k10}
\end{eqnarray}
where $r_{<}$ (resp. $r_{>}$) denotes the smallest (resp. largest) of
$r$ and $r_{1}$. Therefore, the final form of the kinetic equation is
\begin{eqnarray}
{\partial\omega_{a}\over\partial t}=-{1\over 4r}{\partial\over\partial r} \sum_{b} \int_{0}^{+\infty}  r' d{r}' \delta (\Omega-\Omega')\ln\biggl\lbrack 1-\biggl ({r_{<}\over r_{>}}\biggr )^{2}\biggr\rbrack\nonumber\\
\times
\biggl ( \gamma_{b}\omega_{b}'{1\over r}{\partial\omega_{a}\over\partial r}-\gamma_{a}\omega_{a}{1\over r'}{\partial\omega'_{b}\over\partial r'}\biggr ),\qquad\qquad
\label{k11}
\end{eqnarray}
where $\omega'$ stands for $\omega(r',t)$ and $\Omega'$ stands for
$\Omega(r',t)$. This generalizes the kinetic equation obtained by
Chavanis \cite{kin} for a single species of point vortices. 
The angular velocity and the vorticity are expressed in terms of the
orthoradial velocity by
\begin{eqnarray}
\langle V\rangle_{\theta}=-{\partial\psi\over \partial r}=\Omega r, \qquad \omega={1\over r}{\partial\over \partial r}(r\langle V\rangle_{\theta}),
\label{k12}
\end{eqnarray}
so that
\begin{eqnarray}
\omega={1\over r}{\partial \over \partial r}(r^{2}\Omega).
\label{k13}
\end{eqnarray}
For future convenience, it will be useful to write the logarithm as
\begin{eqnarray}
\ln\biggl\lbrack 1-\biggl ({r_{<}\over r_{>}}\biggr
)^{2}\biggr\rbrack = -\sum_{n=1}^{+\infty}{1\over n}\biggl
({r_{<}\over r_{>}}\biggr )^{2n}. \label{f5}
\end{eqnarray}
We note that, for a multi-species system, a logarithmic
divergence $\sum_{n=1}^{+\infty}1/n$ appears in the kinetic equation
(\ref{k11}) when $r'=r$ and $b\neq a$. This problem will be discussed
specifically in Sec. \ref{sec_f}. We shall heuristically regularize
the divergence by introducing an upper cut-off $\Lambda$ in the series
(\ref{f5}) when necessary, writing  $\sum_{n=1}^{\Lambda}{1\over n}
({r_{<}/ r_{>}})^{2n}$. 

We also emphasize that the Dirac function arising in Eq. (\ref{k11})
does not make sense for values of  $r$ and $r'$ such that
$\Omega(r)=\Omega(r')$ and $(\partial\Omega/\partial r)(r')=0$ (in
that case, the identity (\ref{num1}) clearly breaks down).
Mathematically speaking, the ``function'' $x\rightarrow \delta(x^2)$
has no sense even in the space of distributions. To overcome this
problem, we can notice that this Dirac distribution comes out
formally from Eq. (\ref{k9}) where the time integration takes over
all the interval $[0,+\infty)$. Therefore, one  could expect that if
the time averaging is taken  on a finite interval $[0,t)$ only, as
in the initial integral (\ref{k3}), then the resulting kernel $M(t)$
will be well defined as a smoothing approximation to Eq.
(\ref{k10}). The corresponding computation is given in  Appendix B
and leads to
\begin{eqnarray}
M(t)={1\over 4\pi r^{2}}\frac{1}{\Delta\Omega}\arctan\biggl\lbrack
\frac{R^2\sin(t\Delta\Omega )}{1-R^2\cos(t\Delta\Omega )}
\biggr\rbrack, \label{k14}
\end{eqnarray}
where $R={r_{<}/r_{>}}$. For $t\rightarrow +\infty$, we recover
formula (\ref{k10}). However, the regularized expression (\ref{k14})
does not suffer the problem discussed above (we may also wonder
whether collective effects \cite{dn} that we have neglected can regularize
or not the integral in the situation mentioned above).

\subsection{Condition of resonance and kinetic blocking}
\label{sec_block}

For a single species of point vortices, the kinetic equation
(\ref{k11}) becomes \cite{kin}:
\begin{eqnarray}
{\partial\omega\over\partial t}=-{\gamma\over 4r}{\partial\over\partial r} \int_{0}^{+\infty}  r' d{r}' \delta (\Omega-\Omega')\ln\biggl\lbrack 1-\biggl ({r_{<}\over r_{>}}\biggr )^{2}\biggr\rbrack\nonumber\\
\times
\biggl (\omega'{1\over r}{\partial\omega\over\partial r}-\omega {1\over r'}{\partial\omega'\over\partial r'}\biggr ).\qquad\qquad
\label{block1}
\end{eqnarray}
This kinetic equation is the counterpart of the Landau equation
describing the dynamical evolution of a 3D plasma or a stellar system
as a whole \cite{kandrup1}. As is well-known, the
Landau equation yields a logarithmic divergence
$\int_{0}^{+\infty}dk/k$ at small and large scales. The small scale
divergence is regularized at the Landau length, corresponding to a
deflection at $90^{o}$ of the particles' orbits so that the linear
trajectory approximation made by Landau is not valid anymore. The
large scale divergence is regularized, in plasma physics, by the Debye
shielding (as shown by the Lenard-Balescu treatment of collective
effects) and, in stellar dynamics, by the finite size of the system
(Jeans length). For the single species point vortex gas, we stress
that, contrary to the Landau equation, there is {\it no} logarithmic
divergence in the kinetic equation (\ref{block1}).  Therefore,
although there is no Debye shielding in the single species point
vortex gas, the kinetic equation (\ref{block1}) is well-posed
mathematically
\footnote{The situation is different for a 
unidirectional flow where a logarithmic divergence occurs at large
scales for the single species point vortex system (see Appendix
E.2. of \cite{kin}). In that case, it must be regularized by invoking
the finite extent of the system (like for self-gravitating systems) or geophysical effects like the  finite Rossby
radius of deformation  that plays a role
similar to the Debye length in plasma physics.}.  As discussed in
\cite{dn,kin,houches}, the ``collisional'' evolution of point vortices
described by Eq. (\ref{block1}) is due to a condition of
resonance which can be satisfied only if the profile of angular
velocity is non-monotonic
\footnote{We recall that, within our assumptions, this non-monotonic
angular velocity profile must be stable with respect to the 2D Euler
equation so as to avoid a ``violent relaxation'' process driven by
mean field effects.  It is indicated in \cite{dn} that such profiles
can be realized experimentally \cite{dmf}.}. The current of vorticity
in $r$ is due to long-range collisions with point vortices in $r'\neq
r$ whose orbits satisfy the condition $\Omega(r,t)=\Omega(r',t)$. The
self-interaction at $r=r'$ does not produce transport since the term
in parenthesis vanishes identically. When the profile of angular
velocity becomes monotonic, the evolution stops and the system becomes
``frozen'' in a QSS satisfying $\partial\omega/\partial t=0$. This QSS
usually differs from the statistical equilibrium state as will
be shown numerically in Sec. \ref{sec_num}. In that case, the
relaxation towards statistical equilibrium (if it really happens)
takes place on a timescale larger than $Nt_{D}$ which is not described
by the present approach. If we want to describe this regime, we need
to take into account terms of order $1/N^{2}$ or smaller in the
kinetic theory.  They are associated with three-body (or higher)
correlation functions
\cite{kin}.

The situation is different if the system consists in a collection of
point vortices with different circulations.  Assuming that the profile
of angular velocity is monotonic and using
$\delta(\Omega-\Omega')=\delta(r-r')/|\frac{\partial
\Omega}{\partial r}(r)|$, the kinetic equation (\ref{k11}) becomes
\begin{eqnarray}
{\partial\omega_{a}\over\partial t}={1\over
4r}\ln\Lambda{\partial\over\partial r}\sum_{b} {1\over |\frac{\partial
\Omega}{\partial r}(r,t)|} \biggl
(\gamma_{b}\omega_{b}{\partial\omega_{a}\over\partial
r}-\gamma_{a}\omega_{a}{\partial\omega_{b}\over\partial r}\biggr ),\nonumber\\
\label{block2}
\end{eqnarray}
where a cut-off $\Lambda$ has been introduced in Eq. (\ref{f5}) so
that $\ln\Lambda\sim\sum_{n=1}^{\Lambda}{1}/{n}$.  We see that the
diffusion current {\it per species} does not vanish anymore when there
are at least two species of particles in the system \footnote{As
shown by Dubin \cite{dubin}, the kinetic theory is only valid for
retrograde vortices such that $(\gamma_{a}+\gamma_{b})\Sigma<0$. The sum
in Eq. (\ref{block2}) must take into account this
constraint.}. In that case, the kinetic equation
(\ref{block2}) becomes very similar to the one obtained by Nazarenko
\& Zakharov \cite{zac} \footnote{The prefactor in front of the
parenthesis is, however, different because we are not modelling the
collisions exactly in the same way. We are considering collisions with
large impact parameters while Nazarenko \& Zakharov \cite{zac} consider close 
binary collisions; see the discussion in the Introduction.}. The transport of point vortices of
species $a$ is caused by collisions with point vortices of other
species at the same radial distance $r=r'$. However, using the
anti-symmetry of the collision term, we can easily see that the global
vorticity distribution $\omega=\sum_{a}\omega_{a}$ does not change with time,
i.e.
\begin{eqnarray}
{\partial\omega\over\partial t}=0.
\label{block3}
\end{eqnarray}
Finally, using the $H$-theorem of Sec. \ref{sec_hth}, it is simple
to show that the stationary solution of Eq. (\ref{block2}) (for all
species)  corresponds to
\begin{eqnarray}
\gamma_{b}\omega_{b}{\partial\omega_{a}\over\partial
r}=\gamma_{a}\omega_{a}{\partial\omega_{b}\over\partial r},
\label{block4}
\end{eqnarray}
for any $a$ and $b$. This is equivalent to
\begin{eqnarray}
\omega_{a}({r})=C_{ab}|\omega_{b}({r})|^{\gamma_{a}/\gamma_{b}},
\label{block5}
\end{eqnarray}
where $C_{ab}$ is independent on ${r}$. This relation was previously
obtained in \cite{zac,dubin}. It is similar to the relation (\ref{gg12})
derived for point vortices at statistical equilibrium. From Eq. (\ref{block4}) or (\ref{block5}),
we find that the vorticity of each species can be written
\begin{eqnarray}
\omega_{a}({r})=A_{a}e^{-\beta\gamma_{a}\chi(r)},
\label{block6}
\end{eqnarray}
where $A_{a}$ and $\beta$ are some constants and $\chi(r)$ is
determined by the initial conditions. However, contrary to the case
of statistical equilibrium (\ref{gg8}) considered in Sec.
\ref{sec_gg}, $\chi(r)$ does not represent in general the stream
function \cite{dubin}.

\subsection{Conservation laws and H-theorem}
\label{sec_c}

In this section, we show that the kinetic equation (\ref{k11}) respects the
conservation laws of the point vortex dynamics and increases the
Boltzmann entropy ($H$-theorem). We write the kinetic equation as
\begin{eqnarray}
{\partial\omega_{a}\over\partial t}=-{1\over r}{\partial J_{a}\over\partial r},
\label{c1}
\end{eqnarray}
where
\begin{eqnarray}
J_{a}={1\over 4} \sum_{b} \int_{0}^{+\infty}  r' d{r}' \delta (\Omega-\Omega')\ln\biggl\lbrack 1-\biggl ({r_{<}\over r_{>}}\biggr )^{2}\biggr\rbrack\nonumber\\
\times
\biggl ( \gamma_{b}\omega_{b}'{1\over r}{\partial\omega_{a}\over\partial r}-\gamma_{a}\omega_{a}{1\over r'}{\partial\omega'_{b}\over\partial r'}\biggr ),
\label{c2}
\end{eqnarray}
denotes the diffusion current.  Due to the conservative form of
Eq. (\ref{c1}), it is clear that the total circulation
$\Gamma_{a}=\int \omega_{a}d{\bf r}$ of each species of point vortices
is conserved provided that $J_{a}$ vanishes at the frontiere of the
domain.

\subsubsection{Boltzmann distribution}
\label{sec_bol}

The Boltzmann distribution
\begin{eqnarray}
\omega_{a}=A_{a}e^{-\beta\gamma_{a}\psi'},
\label{c3}
\end{eqnarray}
is a stationary solution of the kinetic equation (\ref{k11}). Indeed, using
\begin{eqnarray}
{\partial\omega_{a}\over\partial r}=-\beta\gamma_{a}\omega_{a}{\partial\psi'\over\partial r}=-\beta\gamma_{a}\omega_{a}\left ({\partial\psi\over\partial r}+\Omega_{L} r\right )\nonumber\\
=\beta\gamma_{a}\omega_{a} r(\Omega-\Omega_{L}),
\label{c4}
\end{eqnarray}
we find that the term in parenthesis in Eq. (\ref{c2}) is equal to  
\begin{eqnarray}
\biggl ( \gamma_{b}\omega_{b}'{1\over r}{\partial\omega_{a}\over\partial r}-\gamma_{a}\omega_{a}{1\over r'}{\partial\omega'_{b}\over\partial r'}\biggr )=\beta \gamma_{a}\gamma_{b}\omega_{a}\omega_{b}'(\Omega-\Omega').\nonumber\\
\label{c2bis}
\end{eqnarray}
Therefore, the integrand in Eq. (\ref{c2}) is proportional to
\begin{eqnarray}
   \delta (\Omega-\Omega')(\Omega-\Omega')=0.
\label{c5}
\end{eqnarray}
Therefore, the current vanishes and the Boltzmann distribution is a
stationary solution of the kinetic equation (\ref{k11}). However, this
is not the only stationary solution. For a single species system, any
vorticity distribution with a monotonic profile of angular velocity is
a stationary solution of the kinetic equation (\ref{block1}) since
$\Omega(r)\neq
\Omega(r')$ for any $r\neq r'$ (and for $r=r'$ the term in parenthesis
vanishes). For a multi-species system, the steady distributions of the
kinetic equation (\ref{k11}) with a monotonic profile of angular
velocity are given by Eq. (\ref{block6}). They are in general
different from the Boltzmann distribution.

\subsubsection{Conservation of energy}
\label{sec_ene}

The time variation of energy can be written
\begin{eqnarray}
\dot E=\int \psi {\partial\omega\over\partial t}d{\bf r}=\sum_{a}\int_{0}^{+\infty} \psi {\partial\omega_{a}\over\partial t}2\pi r dr\nonumber\\
=2\pi \sum_{a}\int_{0}^{+\infty} J_{a}{\partial\psi\over\partial r} dr=-2\pi \sum_{a}\int_{0}^{+\infty} J_{a}\Omega r dr.
\label{c6}
\end{eqnarray}
To get the first equality, we have used the Poisson equation
(\ref{gg5}) and integrated by parts twice. To get the third
equality, we have used Eq. (\ref{c1}) and integrated by parts.
Inserting the current (\ref{c2}) in Eq. (\ref{c6}), we get
\begin{eqnarray}
\dot E=-{\pi\over 2} \sum_{a,b}\int_{0}^{+\infty} r r' dr dr'   \delta (\Omega-\Omega')\Omega\ln\biggl\lbrack 1-\biggl ({r_{<}\over r_{>}}\biggr )^{2}\biggr\rbrack\nonumber\\
\times \biggl ( \gamma_{b}\omega_{b}'{1\over r}{\partial\omega_{a}\over\partial r}-\gamma_{a}\omega_{a}{1\over r'}{\partial\omega'_{b}\over\partial r'}\biggr ).\qquad\qquad
\label{c7}
\end{eqnarray}
Interchanging the dummy variables $a$, $b$ and $r$, $r'$, we obtain
\begin{eqnarray}
\dot E={\pi\over 2} \sum_{a,b}\int_{0}^{+\infty} r r' dr dr'   \delta (\Omega-\Omega')\Omega'\ln\biggl\lbrack 1-\biggl ({r_{<}\over r_{>}}\biggr )^{2}\biggr\rbrack\nonumber\\
\times\biggl ( \gamma_{b}\omega_{b}'{1\over r}{\partial\omega_{a}\over\partial r}-\gamma_{a}\omega_{a}{1\over r'}{\partial\omega'_{b}\over\partial r'}\biggr ).\qquad\qquad
\label{c8}
\end{eqnarray}
Taking the half-sum of these two expressions, we find that
\begin{eqnarray}
\dot E=-{\pi\over 4} \sum_{a,b}\int_{0}^{+\infty} r r' dr dr'   \delta (\Omega-\Omega')(\Omega-\Omega')\nonumber\\
\times\ln\biggl\lbrack 1-\biggl ({r_{<}\over r_{>}}\biggr )^{2}\biggr\rbrack
\biggl ( \gamma_{b}\omega_{b}'{1\over r}{\partial\omega_{a}\over\partial r}-\gamma_{a}\omega_{a}{1\over r'}{\partial\omega'_{b}\over\partial r'}\biggr ).
\label{c9}
\end{eqnarray}
Using Eq. (\ref{c5}),  we conclude that $\dot E=0$.

\subsubsection{Conservation of angular momentum}
\label{sec_am}

The time variation of angular momentum can be written
\begin{eqnarray}
\dot L=\int  {\partial\omega\over\partial t}r^{2} d{\bf r}=\sum_{a}\int_{0}^{+\infty}  {\partial\omega_{a}\over\partial t}r^{2} 2\pi r dr\nonumber\\
=4\pi \sum_{a}\int_{0}^{+\infty} J_{a}r dr.
\label{c10}
\end{eqnarray}
Inserting the current (\ref{c2}) in Eq. (\ref{c10}), we get
\begin{eqnarray}
\dot L=-\pi \sum_{a,b}\int_{0}^{+\infty} r r' dr dr'   \delta (\Omega-\Omega')\ln\biggl\lbrack 1-\biggl ({r_{<}\over r_{>}}\biggr )^{2}\biggr\rbrack\nonumber\\
\times\biggl ( \gamma_{b}\omega_{b}'{1\over r}{\partial\omega_{a}\over\partial r}-\gamma_{a}\omega_{a}{1\over r'}{\partial\omega'_{b}\over\partial r'}\biggr ).\qquad\qquad
\label{c11}
\end{eqnarray}
Interchanging the dummy variables $a$, $b$ and $r$, $r'$, we obtain
\begin{eqnarray}
\dot L={\pi} \sum_{a,b}\int_{0}^{+\infty} r r' dr dr'   \delta (\Omega-\Omega')\ln\biggl\lbrack 1-\biggl ({r_{<}\over r_{>}}\biggr )^{2}\biggr\rbrack\nonumber\\
\times
\biggl ( \gamma_{b}\omega_{b}'{1\over r}{\partial\omega_{a}\over\partial r}-\gamma_{a}\omega_{a}{1\over r'}{\partial\omega'_{b}\over\partial r'}\biggr ).\qquad\qquad
\label{c12}
\end{eqnarray}
Taking the sum of the last two expressions, we conclude that  $\dot L=0$.

\subsubsection{H-theorem}
\label{sec_hth}

The time variation of the entropy (\ref{gg2}) can be written
\begin{eqnarray}
\dot S=-2\pi \sum_{a}\int_{0}^{+\infty} {1\over \gamma_{a}\omega_{a}}{\partial\omega_{a}\over\partial r}J_{a} dr.
\label{c13}
\end{eqnarray}
To get this expression, we have used Eq. (\ref{c1}) and integrated
by parts. Inserting the current (\ref{c2}) in Eq. (\ref{c13}), we
get
\begin{eqnarray}
\dot S=-{\pi\over 2} \sum_{a,b}\int_{0}^{+\infty} dr dr'{1\over \gamma_{a}\omega_{a}}r'{\partial\omega_{a}\over\partial r}  \delta (\Omega-\Omega')\nonumber\\
\times\ln\biggl\lbrack 1-\biggl ({r_{<}\over r_{>}}\biggr )^{2}\biggr\rbrack
\biggl ( \gamma_{b}\omega_{b}'{1\over r}{\partial\omega_{a}\over\partial r}-\gamma_{a}\omega_{a}{1\over r'}{\partial\omega'_{b}\over\partial r'}\biggr ).
\label{c14}
\end{eqnarray}
Interchanging the  dummy variables $a$, $b$ and $r$, $r'$, we obtain
\begin{eqnarray}
\dot S={\pi\over 2} \sum_{a,b}\int_{0}^{+\infty} dr dr'{1\over \gamma_{b}\omega'_{b}}r{\partial\omega'_{b}\over\partial r'}  \delta (\Omega-\Omega')\nonumber\\
\times\ln\biggl\lbrack 1-\biggl ({r_{<}\over r_{>}}\biggr )^{2}\biggr\rbrack
\biggl ( \gamma_{b}\omega_{b}'{1\over r}{\partial\omega_{a}\over\partial r}-\gamma_{a}\omega_{a}{1\over r'}{\partial\omega'_{b}\over\partial r'}\biggr ).
\label{c15}
\end{eqnarray}
Taking the half sum of these two expressions, we find that
\begin{eqnarray}
\dot S=-{\pi\over 4} \sum_{a,b}\int_{0}^{+\infty} r r'dr dr'{1\over \gamma_{a}\omega_{a}\gamma'_{a}\omega'_{a}}    \delta (\Omega-\Omega')\nonumber\\
\times\ln\biggl\lbrack 1-\biggl ({r_{<}\over r_{>}}\biggr )^{2}\biggr\rbrack
\biggl ( \gamma_{b}\omega_{b}'{1\over r}{\partial\omega_{a}\over\partial r}-\gamma_{a}\omega_{a}{1\over r'}{\partial\omega'_{b}\over\partial r'}\biggr )^{2}.\nonumber\\
\label{c16}
\end{eqnarray}
Since this quantity is positive, the entropy cannot decrease. An
$H$-theorem results: $\dot S\ge 0$. However, the
condition $\dot S=0$ does not only select the Boltzmann distribution.
It is satisfied for any steady solution of Eq. (\ref{k11}). 

\subsection{Numerical simulations}
\label{sec_num}

We have performed numerical simulations of the kinetic equation
(\ref{block1}) for a single species system of point vortices. To solve this equation, we have used the identity
\begin{eqnarray}
\delta(\Omega(r)-\Omega(r'))=\sum_{k}\frac{\delta(r'-r_{k})}{|\Omega'(r_{k})|},
\label{num1}
\end{eqnarray}
where $\lbrace r_{k}\rbrace$ is the set of resonant points which
satisfy the condition $\Omega(r_{k})=\Omega(r)$. Of course, this
relation is valid only for resonant points $r_k$ such that $$
\frac{\partial \Omega}{\partial r}(r_k) \ne 0.$$ If a resonant point
with $\Omega'(r_k)= 0$ is encountered, then the kernel of the
collision kinetic operator has to be modified and one of the many
different ways to regularize its expression is presented at the end of
section 3.1 (it may also be necessary  to reconsider the approximation
(\ref{k4}) when $\Omega'(r_{k})=0$). Assuming here that this situation
never occurs, the kinetic equation (\ref{block1}) becomes
\begin{eqnarray}
{\partial\omega\over\partial t}=-{\gamma\over 4r}{\partial\over\partial r} \sum_{k} \frac{1}{|\frac{\partial \Omega}{\partial r}(r_{k},t)|}\ln\biggl\lbrack 1-\biggl ({r_{<}\over r_{>}}\biggr )^{2}\biggr\rbrack\nonumber\\
\times
\biggl \lbrack \omega(r_{k},t){r_{k}\over r}{\partial\omega\over\partial r}-\omega {\partial\omega\over\partial r}(r_{k},t)\biggr \rbrack.
\label{num2}
\end{eqnarray}
where $r_{<}$ (resp. $r_{>}$) denotes the smallest (resp. largest)
of $r$ and $r_{k}$. This can be written in the more compact form
\begin{eqnarray}
{\partial \omega\over\partial t}={1\over r}{\partial\over\partial r}\biggl\lbrack r \biggl (D(r,t){\partial \omega\over\partial r}-\omega\eta(r,t)\biggr )\biggr\rbrack,
\label{num3}
\end{eqnarray}
with
\begin{eqnarray}
D(r,t)=-{\gamma\over 4r^{2}} \sum_{k} {r_{k}\over |\frac{\partial \Omega}{\partial r}(r_{k},t)|}
\ln\biggl\lbrack 1-\biggl ({r_{<}\over r_{>}}\biggr )^{2}
\biggr\rbrack \omega(r_{k},t),\nonumber\\
\label{num4}
\end{eqnarray}
\begin{eqnarray}
\eta(r,t)=-{\gamma\over 4r} \sum_{k} {1\over |\frac{\partial \Omega}{\partial r}(r_{k},t)|} \ln\biggl\lbrack 1-\biggl ({r_{<}\over r_{>}}\biggr )^{2}\biggr\rbrack
  {\partial\omega\over \partial r}(r_{k},t).\nonumber\\
\label{num5}
\end{eqnarray}

We have considered an initial vorticity field $\omega_{0}(r)$
associated with a non-monotonic profile of angular velocity
$\Omega_{0}(r)$. Specifically, we have taken $\omega_{0}(r)={\rm
exp}\lbrack -(r-0.2)^{2}/0.05\rbrack$ in a disk of radius $R=1$ (see
Figs. \ref{f} and \ref{Omega}). We numerically find that the system
evolves until the profile of angular velocity becomes monotonic (see
Fig. \ref{Omega}). In the present situation, the vorticity profile
also becomes monotonic (see Fig. \ref{f}). When there is no resonance,
the evolution stops and the system remains blocked in a stationary
state generically different from the Boltzmann distribution \footnote{This has been checked numerically by computing the term in
parenthesis in Eq. (\ref{block1}) for different couples of points $r$
and $r'$. This term is not proportional to
$\omega(r)\omega(r')\lbrack\Omega(r)-\Omega(r')\rbrack$ as would be
the case for a Boltzmann distribution according to
Eq. (\ref{c2bis}). Note that the tail of the distribution does not
evolve with time (since vortices in the tail are never in resonance
with vortices in the core) so that the tail is clearly
non-Boltzmannian. The previous check shows that the final distribution
of the core (that has evolved through resonances) is not Boltzmannian
neither. }. {\it Therefore, the main effect of long-range collisions
between point vortices is to make the profile of angular velocity
monotonic.}  To our knowledge, the kinetic theory of point vortices
\cite{dn,kin} is the first kinetic theory to exhibit such a behavior.
In usual kinetic theories developed for ordinary gas
\cite{boltzmann}, plasmas
\cite{landau} and stellar systems \cite{chandra}, the system described
by the Boltzmann equation (or by the Landau or Lenard-Balescu
equation) always relaxes towards the Boltzmann distribution (the case
of stellar systems is peculiar because of the phenomena of evaporation
and gravothermal catastrophe \cite{ijmpb} but the
convergence to the Boltzmann distribution is, however, the general
{\it tendency}). Alternatively, for the homogeneous phase of the HMF
model (or other one-dimensional systems), the collision term vanishes
identically \cite{bd,cvb,pa1} at the order $1/N$ so that there is no
evolution at all on a timescale $\sim N t_{D}$ (numerical simulations
\cite{yamaguchi} show that the HMF model relaxes towards statistical
equilibrium on a timescale $N^{1.7}t_{D}$). For point vortices, the
situation is intermediate between these two extremes. The system
described by the kinetic equation (\ref{block1}) evolves until the
profile of angular velocity becomes monotonic, then stops.  For the
inhomogeneous phase of the HMF model (or other one-dimensional
systems), we expect a behavior similar to that observed for point
vortices when we use angle-action variables, as discussed in
\cite{action}. However, the problem is more difficult to investigate
numerically because there is a richer variety of resonances. 

Let us conclude this section by some remarks that should be given
further consideration in future works:

\begin{figure}
\centering
\includegraphics[width=8cm]{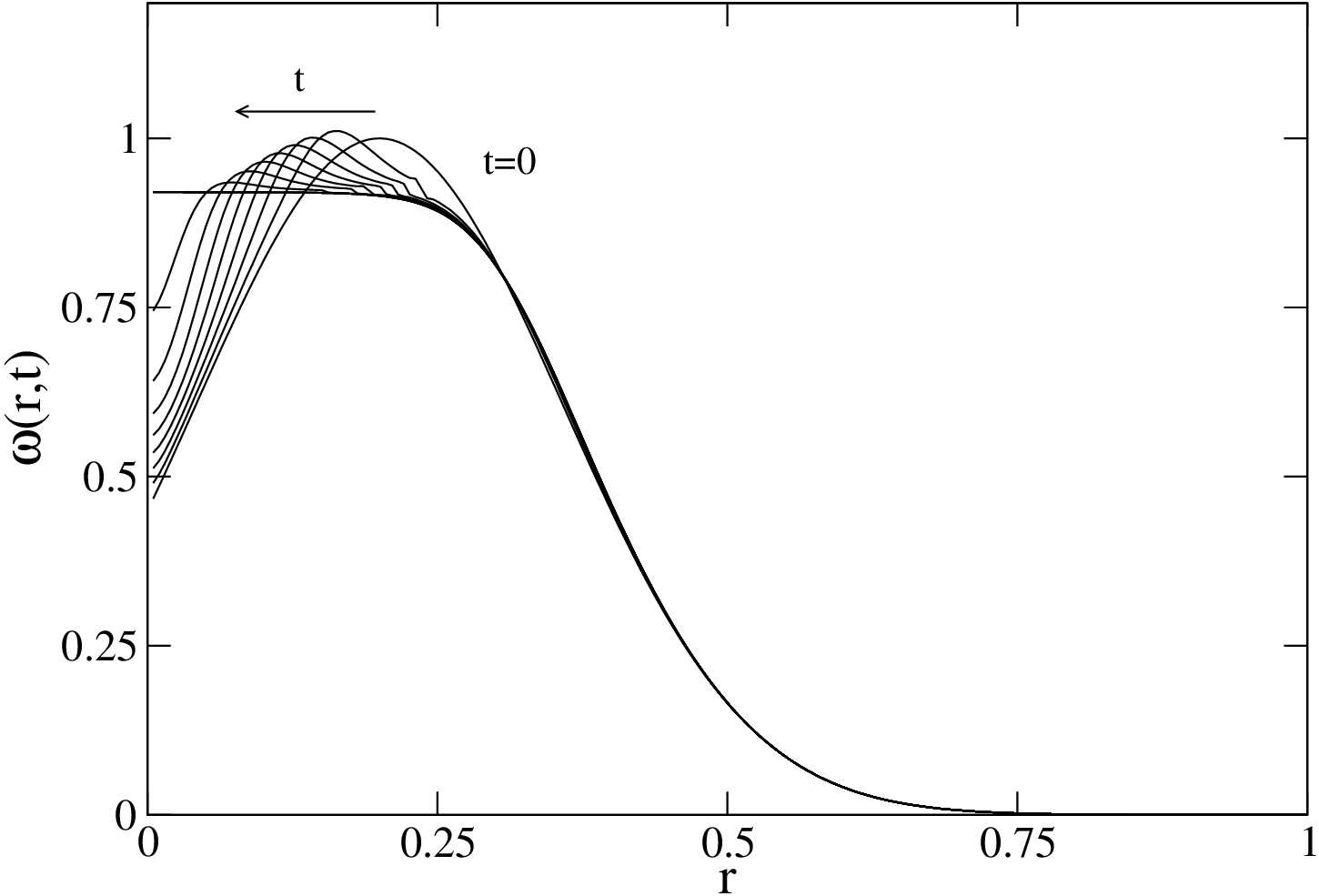}
\caption{Evolution of the vorticity profile obtained by solving numerically the kinetic equation (\ref{block1}). This kinetic equation  is  valid on a timescale $Nt_{D}$. On this timescale, the vortex gas does not reach statistical equilibrium but remains blocked in a QSS with a monotonic profile of angular velocity (see Fig. \ref{Omega}). }
\label{f}
\end{figure}

\begin{figure}
\centering
\includegraphics[width=8cm]{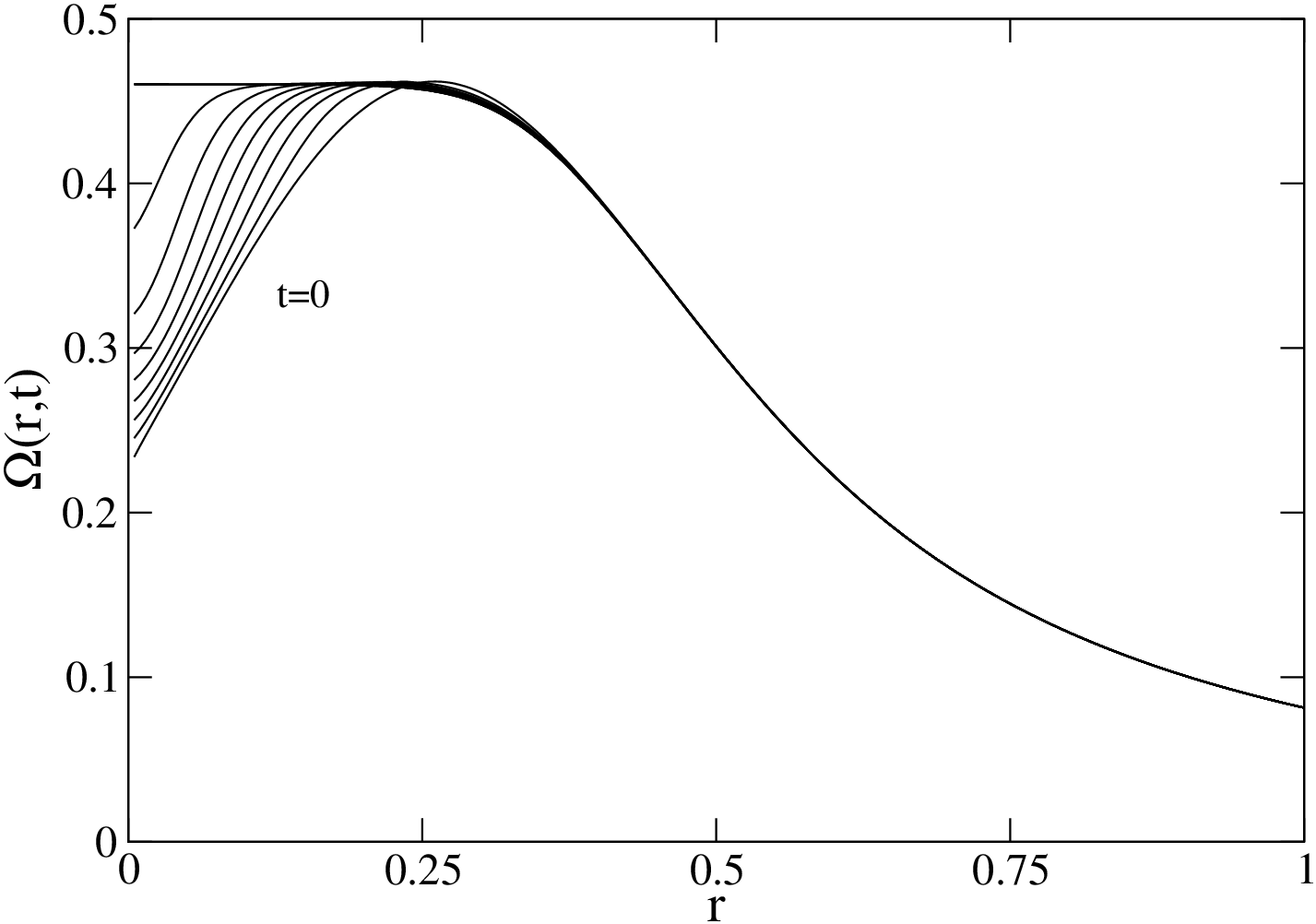}
\caption{Evolution of the profile of angular velocity corresponding
to the vorticity field represented in Fig. \ref{f}. The evolution
stops when the profile of angular velocity becomes monotonic so that
there is no resonance anymore. Note, parenthetically, that the
terminal point of the curve does not move. Indeed, according to
Eq. (\ref{dd3}), the angular velocity at $r=R$ is $\Omega=\Gamma/(2\pi
R^{2})$ which is constant.} \label{Omega}
\end{figure}

(i) The diffusion current vanishes at $r$ either because there is no
point $r'\neq r$ such that $\Omega(r')=\Omega(r)$ (no resonance) or
because the term in parenthesis vanishes for any $r'$ such that
$\Omega(r')=\Omega(r)$. Therefore, we have three possibilities: (i)
the profile of angular velocity is monotonic everywhere so that the
current vanishes because of the Dirac function (ii) the vorticity
profile is the Boltzmann distribution so that the current vanishes
because the term in parenthesis is zero at each point of resonance
(if any) (iii) we are in a mixed situation where there is no
resonance in certain regions and the term in parenthesis is zero at
each point of resonance in other regions.

(ii) As mentionned above, formula (\ref{num1}) is wrong when
$\Omega'(r_{k})=0$, i.e. when $r$ is in resonance with a point located
at an extremum of angular velocity \footnote{In the numerical
simulation reported in Figs. \ref{f} and \ref{Omega}, the diffusive
nature of the kinetic equation implies that
$\omega'(0,t)=\Omega'(0,t)=0$ for $t>0$ (even if this is not the case
initially). Therefore, at each time $t>0$, there exists a point $r>0$
which is in resonance with the point $r_{k}=0$ at which
$\Omega'(r_{k},t)=0$.  However, the point $r_{k}=0$ is
special. Indeed, the numerator of $D(r,t)$ is proportional to
$r_{k}\ln(1-(r_{k}/r)^2)\sim r_{k}^{3}$ and the numerator of
$\eta(r,t)$ is proportional to $\omega'(r_{k},t)\ln(1-(r_{k}/r)^2)\sim
r_{k}^{3}$. These terms compensate the term $\Omega'(r_{k},t)\sim
r_{k}\rightarrow 0$ in the denominator of
Eqs. (\ref{num4})-(\ref{num5}), so that there is no singularity:
$D(r,t)\sim\eta(r,t)\sim r_{k}^{3}/r_{k}\rightarrow 0$.}. This means
that initial conditions that have at least two extrema (in addition to
$r=0$) are not allowed by this model. Furthermore, it is not proven
that initial condition having at most one extremum will not generate
profiles with two extrema in finite time (although this seems
relatively unlikely). In other words, it is not clear whether the
model will break down or not at finite time for some initial
conditions. To overcome these problems, a smoothing strategy has been
proposed at the end of section 3.1, and developed in Appendix B.

(iii) If the profile of angular velocity presents zones of
resonances and zones of non-resonance, then the diffusion
coefficient (\ref{num4}) is discontinuous at the separation and this
can lead to numerical and physical instabilities.

Therefore, the complete study of the kinetic equation (\ref{block1})
is complicated. We have presented here just one example of
evolution. A more thorough (numerical and theoretical) study of this
equation would be certainly valuable. We also recall that the results
of this section assume that the vorticity profile $\omega_{0}(r)$ is
stable with respect to the 2D Euler equation. Now, a system with a
non-monotonic profile of angular velocity can be unstable to diocotron
modes \cite{davidson}. This diocotron (or Kelvin-Helmholtz)
instability, which is a non-axisymmetric instability, can develop in
the nonlinear regime and lead to a violent collisionless relaxation
[55-57]. The evolution is therefore very different
from the one reported in Figs. \ref{f} and \ref{Omega}. However, there
also exists systems with non-monotonic profile of angular velocity
that are stable with respect to the 2D Euler equation \cite{dmf,dn} and
that evolve under the sole effect of ``collisions'' like in Figs.
\ref{f} and \ref{Omega}. A better characterization of these regimes
(and their selection) would be valuable.

\section{Relaxation  of a test vortex in a sea of field vortices}
\label{sec_rel}

Another classical problem in kinetic theory concerns the description
of the relaxation of a test particle in a bath of field particles. If
we inject a test particle in a steady bath of field particles, it will
undergo a stochastic process that we wish to describe.  This problem
is simpler than the previous one (evolution of the $N$-body system as
a whole) because it assumes that the distribution of the field
particles is {\it fixed}, while in the preceding case it was evolving
self-consistently. This fixed distribution either represents the
statistical equilibrium state (thermal bath) which does not change at
all or a quasi-stationary distribution that evolves on a timescale
that is much larger than the typical relaxation time of the test
particle. We shall now consider this classical problem in the context
of two-dimensional point vortices.

\subsection{The Fokker-Planck equation}
\label{sec_f}

We consider the relaxation of a test vortex with circulation
$\Gamma_{0}$ in a ``sea'' of field vortices with circulation $\gamma$
(the field vortices play the role of the bath). In the $N\rightarrow
+\infty$ limit, the $N$-body distribution of the field vortices can be
written
\begin{eqnarray}
\mu_{bath}({\bf r}_{1},...,{\bf r}_{N})=\prod_{i} P_{bath}({\bf r}_{i}),
\label{f1}
\end{eqnarray}
where $P_{bath}({\bf r})$ is the steady probability distribution of a
single vortex of the bath. The vorticity profile of the bath is
$\omega({\bf r})=N\gamma P_{bath}({\bf r})$.  Using the projection
operator formalism, we find that the evolution of the density
probability $P({\bf r},t)$ of finding the test vortex in ${\bf r}$ at
time $t$ is governed by an equation of the form \cite{kin}:
\begin{eqnarray}
{\partial P\over\partial t}+\langle {\bf V}\rangle\cdot \nabla P={\partial\over\partial r^{\mu}}\int_{0}^{t}d\tau \int d{\bf r}_{1} {\cal V}^{\mu}(1\rightarrow 0,t)\nonumber\\
\times{\cal G}(t,t-\tau)\biggl \lbrack \gamma {\cal V}^{\nu}(1\rightarrow 0,t-\tau){\partial\over\partial r^{\nu}}\nonumber\\
+\Gamma_{0} {\cal V}^{\nu}(0\rightarrow 1,t-\tau){\partial\over \partial r_{1}^{\nu}}\biggr \rbrack P({\bf r},t-\tau)\omega({\bf r}_{1}).
\label{f2}
\end{eqnarray}
For an axisymmetric flow, we can repeat the same steps as in Sec.
\ref{sec_k} and we successively obtain
\begin{eqnarray}
{\partial P\over\partial t}={1\over r}{\partial\over\partial r} r\int_{0}^{+\infty}d\tau\int_{0}^{2\pi}d\theta \int_{0}^{+\infty} r r_{1}d{r}_{1} {V}_{r}(1\rightarrow 0,t)\nonumber\\
\times{V}_{r}(1\rightarrow 0,t-\tau)
\biggl ( \gamma {1\over r}{\partial\over\partial r}-\Gamma_{0}
{1\over r_{1}}{d\over dr_{1}}\biggr )  P({r},t)\omega({r}_{1}),\nonumber\\
\label{f3}
\end{eqnarray}
and
\begin{eqnarray}
{\partial P\over\partial t}=-{1\over 4r}{\partial\over\partial r} \int_{0}^{+\infty}  r' d{r}' \delta (\Omega-\Omega')\ln\biggl\lbrack 1-\biggl ({r_{<}\over r_{>}}\biggr )^{2}\biggr\rbrack\nonumber\\
\times
\biggl ( \gamma \omega'{1\over r}{\partial P\over\partial r}-\Gamma_{0} P{1\over r'}{d\omega'\over d r'}\biggr ).\qquad\qquad
\label{f4}
\end{eqnarray}
Equation (\ref{f4}) can be viewed as a Fokker-Planck equation (see
below) describing the relaxation of the test vortex. It can be
obtained directly from Eq. (\ref{k11}) by replacing $\omega_{a}$ by
the distribution of the test particle $P(r,t)$ and $\omega_{b}$ by
the static distribution of the bath $\omega(r)$. This procedure
transforms the integrodifferential equation (\ref{k11}) into a
differential equation (\ref{f4}).

Let us first assume that the field vortices are at statistical
equilibrium so that $\omega(r)=A e^{-\beta\gamma\psi'}$ represents
the Boltzmann distribution (where $\psi'$ denotes the relative
stream function). Introducing
\begin{eqnarray}
\frac{1}{r'}\frac{d\omega}{dr}(r')=-\beta\gamma\frac{1}{r'}\omega(r')\frac{d\psi'}{dr}(r')\nonumber\\
=\beta\gamma\omega(r')\lbrack\Omega(r')-\Omega_{L}\rbrack,
\label{fn1}
\end{eqnarray}
in Eq. (\ref{f4}), using the $\delta$-function allowing to replace
$\Omega'$ by $\Omega$, and using Eq. (\ref{fn1}) again with $r$
instead of $r'$, we find that the Fokker-Planck equation (\ref{f4})
can be rewritten in the form
\begin{eqnarray}
{\partial P\over\partial t}={1\over r}{\partial\over\partial r}\biggl\lbrack r  D(r)\biggl ({\partial P\over\partial r}+\beta {\Gamma_{0}} P {d\psi'\over d r}\biggr )\biggr\rbrack,
\label{fn2}
\end{eqnarray}
with a diffusion coefficient
\begin{eqnarray}
D(r)=-{\gamma\over 4r^{2}} \int_{0}^{+\infty}  r' d{r}' \delta (\Omega-\Omega')\ln\biggl\lbrack 1-\biggl ({r_{<}\over r_{>}}\biggr )^{2}\biggr\rbrack\omega(r').\nonumber\\
\label{fn3}
\end{eqnarray}
These equations are valid even if the profile of angular velocity is
non monotonic since the Boltzmann distribution is always a steady
state of the kinetic equation (\ref{k11}). This statistical
equilibrium distribution does not evolve at all. We see that the
diffusion coefficient in $r$ is due to interactions with vortices
whose orbits satisfy $\Omega(r')=\Omega(r)$. This includes the local
interaction with vortices at $r'=r$ but also the interactions with far
away vortices with $r'\neq r$. To our knowledge, this is the first
kinetic theory where the diffusion coefficient exhibits such a spatial
delocalization.

We now consider a bath with a monotonic profile of angular velocity
that is not necessarily the Boltzmann distribution of statistical
equilibrium. Indeed, we have seen in the previous section that any
Euler stable distribution with a monotonic profile of angular velocity
is a stationary solution of the kinetic equation
(\ref{block1}). Therefore, this profile does not evolve on a timescale
of order $Nt_{D}$ on which the kinetic equation (\ref{block1}) is
valid (but it may evolve on a longer timescale). We shall see that $N
t_{D}$ is precisely the timescale controlling the relaxation of the
test vortex, i.e. the time needed by the test vortex to acquire the
distribution of the bath.  Therefore, we can consider that the
distribution of the field vortices is frozen on this
timescale. Thus, let us consider a bath that is
not necessarily a thermal bath but that is a slowly evolving
out-of-equilibrium distribution. If the profile of angular velocity of
the bath is monotonic, then
\begin{eqnarray}
\delta(\Omega(r)-\Omega(r'))={\delta(r-r')\over |\frac{\partial
\Omega}{\partial r}(r)|}. \label{f6}
\end{eqnarray}
In that case,  the Fokker-Planck equation  (\ref{f4}) can be written
\begin{eqnarray}
{\partial P\over\partial t}={1\over 4}\ln\Lambda{1\over r}{\partial\over\partial r}\biggl\lbrack  {r\over |\Sigma(r)|}\biggl (\gamma\omega{\partial P\over\partial r}-\Gamma_{0} P{d\omega\over d r}\biggr )\biggr\rbrack,
\label{f7}
\end{eqnarray}
where $\Sigma(r)=r\Omega'(r)$ is the local shear created by the
field vortices and we have introduced the Coulomb factor
\begin{eqnarray}
\ln\Lambda=\sum_{n=1}^{+\infty}{1\over n}. \label{f8}
\end{eqnarray}
We note that the series diverges when $n \rightarrow +\infty$. This
divergent sum occurs because nearby vortices following unperturbed
orbits take a long time to separate. However, our theory breaks down
at small separation because for separations less than $d$ (say) we
cannot assume that the motion of the particles is given by the
unperturbed trajectory (\ref{k4}). In that case, one must consider the
detail of the interaction between neighboring vortices (note that a
similar treatment is required in 3D plasma physics and stellar
dynamics to regularize the logarithmic divergence of the diffusion
coefficient at small scales, i.e. at the Landau length
\cite{kandrupREV}). Phenomenologically, the logarithmic divergence can
be regularized by adding cutoffs so that $\ln\Lambda=\ln ( r/d)$
\cite{preR,kin}. A precise estimate of the lower cut-off $d$ has
been given by Dubin and collaborators
[32-34]. They propose to take $d={\rm Max}(\delta,l)$ where $l$ is the trapping distance $l=(2\gamma^{2}/\lbrack 4\pi (\gamma+\Gamma_{0})|\Sigma|\rbrack)^{1/2}$ and $\delta$ is the diffusion-limited minimum separation $\delta=(4D/|\Sigma|)^{1/2}$ (where $D$ is the diffusion coefficient 
given by Eq. (\ref{f10})). Orders of magnitude indicate that
$\delta/l\sim (D/\gamma)^{1/2}\sim (\ln\Lambda)^{1/2}$ and $r/l\sim
R\lbrack (\gamma+\Gamma_{0})|\Sigma|/\gamma^{2}\rbrack^{1/2}\sim
R\lbrack (\gamma+\Gamma_{0})N\gamma/\gamma^{2}R^{2}\rbrack^{1/2}\sim
\lbrack (\gamma+\Gamma_{0})/\gamma\rbrack^{1/2}N^{1/2}$. Therefore,
the logarithmic factor scales with $N$ as
\begin{eqnarray}
\ln\Lambda\sim{1\over 2}\ln N,
\label{lnl}
\end{eqnarray}
in agreement with the rough estimate given in \cite{kin}.

The Fokker-Planck equation (\ref{f7}) can be rewritten in the form
\begin{eqnarray}
{\partial P\over\partial t}={1\over r}{\partial\over\partial r}\biggl\lbrack r  D(r)\biggl ({\partial P\over\partial r}-{\Gamma_{0}\over\gamma} P{d\ln |\omega|\over d r}\biggr )\biggr\rbrack,
\label{f9}
\end{eqnarray}
with a diffusion coefficient
\begin{eqnarray}
D(r)={\gamma\over 4}\ln\Lambda {1\over |\Sigma(r)|}\omega(r).
\label{f10}
\end{eqnarray}
This is a drift-diffusion equation describing the evolution of the
test vortex in an ``effective potential'' 
\begin{eqnarray}
U_{\rm
eff}(r)=-(\Gamma_{0}/\gamma)\ln|\omega(r)|,
\end{eqnarray}
produced by the field
vortices \cite{kin}. The diffusion coefficient is proportional to the
density of field vortices $\omega(r)$ and inversely proportional to
the local shear $|\Sigma(r)|$ created by the background vorticity
distribution. The equilibrium distribution of the test vortex is
\begin{eqnarray}
P_{e}(r)\propto |\omega(r)|^{\Gamma_{0}/\gamma}, \label{f11}
\end{eqnarray}
up to a normalization factor. When the test vortex has the same
circulation  as the field vortices ($\Gamma_0=\gamma$), the test
vortex ultimately acquires the distribution of the bath: $P_{e}(r)\propto |\omega(r)|$. However, when the test vortex has a
circulation different from that of the bath, the distribution of the
test vortex differs from the distribution of the bath by a power
$\Gamma_{0}/\gamma$. If the field vortices are at statistical
equilibrium (thermal bath), their distribution is determined by the Boltzmann-Poisson equation
\begin{eqnarray}
\omega(r)=-\Delta\psi=Ae^{-\beta\gamma\psi'(r)}.
\label{f12}
\end{eqnarray}
Assuming that the   profile of angular velocity is monotonic,
we can use the Fokker-Planck equation (\ref{f9}) with
$\ln|\omega|=-\beta\gamma\psi'+\ln |A|$ to obtain \footnote{We note the
analogy of Eq. (\ref{f13}) with the ordinary Smoluchowski equation
describing the sedimentation of colloidal suspensions in a
gravitational field. Usually, the Smoluchowski equation is obtained
from the Fokker-Planck equation (Kramers equation) in a strong
friction limit where the inertia of the particles can be neglected.
In the present context, since the point vortices have no inertia,
the Fokker-Planck equation describing the relaxation of a test
particle directly has the form of a Smoluchowski equation in
physical space.}
\begin{eqnarray}
{\partial P\over\partial t}={1\over r}{\partial\over\partial r}\biggl\lbrack r D(r)\biggl ({\partial P\over\partial r}+\beta{\Gamma_{0}} P{d\psi'\over d r}\biggr )\biggr\rbrack,
\label{f13}
\end{eqnarray}
with a diffusion coefficient given by Eq. (\ref{f10}).  The Fokker-Planck equation (\ref{f13}) describing
the relaxation of a point vortex in a sea of field vortices is the
counterpart of the Kramers equation derived by Chandrasekhar
\cite{chandraB} to describe the relaxation of a test star in a
cluster. We emphasize that the diffusion coefficient in Eq.
(\ref{f13}) depends on the position $r$ of the test vortex
\cite{preR}. Similarly, in the Kramers-Chandrasekhar equation, the
diffusion coefficient of the test star depends on its velocity $v$
\cite{chandraB}.

\subsection{Diffusion coefficient and drift term}
\label{sec_d}

We can write the Fokker-Planck equation (\ref{f3}) in a form that
explicitly isolates the terms of diffusion and drift:
\begin{eqnarray}
{\partial P\over\partial t}={1\over r}{\partial\over\partial r}\biggl\lbrack r \biggl (D{\partial P\over\partial r}-P\eta\biggr )\biggr\rbrack.
\label{d1}
\end{eqnarray}
The diffusion coefficient is given by
\begin{eqnarray}
D=\gamma\int_{0}^{+\infty}d\tau\int d{\bf r}_{1} {V}_{r}(1\rightarrow 0,t){V}_{r}(1\rightarrow 0,t-\tau)\omega({r}_{1}).\nonumber\\
\label{d2}
\end{eqnarray}
It can be written in the form of a Kubo formula
\begin{eqnarray}
D=\int_{0}^{+\infty}\langle {\cal V}_{r}(t){\cal V}_{r}(t-\tau)\rangle \, d\tau,
\label{d3}
\end{eqnarray}
representing the time integral of the velocity auto-correlation
function \cite{kin}. The drift term is given by
\begin{eqnarray}
\eta=\Gamma_{0}\int_{0}^{+\infty}d\tau\int d{\bf r}_{1} {V}_{r}(1\rightarrow 0,t){V}_{r}(1\rightarrow 0,t-\tau) {r\over r_{1}}{d\omega_{1}\over dr_{1}}.\nonumber\\
\label{d4}
\end{eqnarray}
It can be seen  as a sort of generalized Kubo formula involving the {\it
gradient} of the density of field vortices instead of their density
itself. The physical origin of the drift can be understood by
developing a linear response theory
\cite{preR}. It arises as the response of the field vortices to the
perturbation caused by the test vortex, as in a polarization
process.

According to Eq. (\ref{f4}), the diffusion coefficient and the drift
term can be written more explicitly as
\begin{eqnarray}
D=-{\gamma\over 4r^{2}} \int_{0}^{+\infty}  r' d{r}' \delta (\Omega-\Omega')\ln\biggl\lbrack 1-\biggl ({r_{<}\over r_{>}}\biggr )^{2}\biggr\rbrack\omega(r'),\nonumber\\
\label{d5}
\end{eqnarray}
\begin{eqnarray}
\eta=-{\Gamma_{0}\over 4r} \int_{0}^{+\infty}   d{r}' \delta (\Omega-\Omega')\ln\biggl\lbrack 1-\biggl ({r_{<}\over r_{>}}\biggr )^{2}\biggr\rbrack
  {d\omega'\over d r'}.\nonumber\\
\label{d6}
\end{eqnarray}
When the profile of angular velocity is monotonic, using
Eq. (\ref{f6}), we find that the diffusion coefficient is given by
\begin{eqnarray}
D(r)={\gamma\over 4}\ln\Lambda {1\over |\Sigma(r)|}\omega(r).
\label{d6b}
\end{eqnarray}
This expression of the diffusion coefficient, with the shear reduction, was first derived in Chavanis \cite{preR,kin}. An equivalent expression has been obtained by Dubin \& Jin \cite{jin}. On the other hand, using Eq. (\ref{f6}), we find that the drift term is given by
\begin{eqnarray}
\eta={\Gamma_{0}\over 4}\ln\Lambda {1\over |\Sigma(r)|}{d\omega\over dr}(r).
\label{d6c}
\end{eqnarray}
The drift is proportional to the local vorticity gradient and
inversely proportional to the shear. Comparing with Eq. (\ref{d6b}),
we find that the drift velocity is connected to the diffusion coefficient
by the relation
\begin{eqnarray}
\eta={\Gamma_{0}\over\gamma}D(r){d\ln|\omega|\over dr}.
\label{d7}
\end{eqnarray}
This relation is valid for an arbitrary distribution of field vortices
provided that the profile of angular momentum is monotonic. This
expression was given in Chavanis \cite{kin} [see Eq. (123)].  We see
that the drift is directed along the vorticity gradient. Assuming that
the background vorticity $\omega(r)$ is positive and decreases
monotonically with the radius, we find that the test vortex ascends
the gradient if $\Gamma_{0}>0$ and descends the gradient if
$\Gamma_{0}<0$. The drift velocity $\eta=\langle V\rangle_{drift}$ can
also be calculated from a linear response theory \cite{preR}. In the
case of a unidirectional shear flow, combining Eqs. (21) and (23) of
\cite{preR}, it is found that
\begin{eqnarray}
\langle V\rangle_{drift}=\frac{\Gamma_{0}}{2\pi}\frac{1}{|\Sigma(y)|}\ln\Lambda \arctan\left (\frac{|{\Sigma}|}{2}t\right )\frac{d\omega}{dy},
\label{d7b}
\end{eqnarray}
where $\Sigma(y)=\langle V\rangle'(y)$ is the local shear. For
$t\rightarrow +\infty$, we obtain an expression equivalent to Eq.
(\ref{d6c}) for a cylindrical shear flow. These expressions for the
drift velocity are equivalent to those obtained by Schecter \& Dubin
\cite{schecter}  using the Euler equation. These authors showed that
expressions (\ref{d6c}) and (\ref{d7b}) are only valid for
retrograde vortices ($\Gamma_{0}>0$ if $\omega$ is positive and
decreasing). The linear response theory is not correct for prograde
vortices (on the basis of numerical simulations \cite{schecter}) so
that  nonlinear effects must be considered.

For a thermal bath (statistical equilibrium state), introducing  Eq.
(\ref{fn1}) in Eq. (\ref{d6}), using the property of the
$\delta$-function to replace $\Omega'$ by $\Omega$, and using Eq.
(\ref{fn1}) again with $r$ instead of $r'$, we find that the drift
term (\ref{d6}) takes the form
\begin{eqnarray}
\eta=-D\beta\Gamma_{0} {d\psi'\over dr}, \label{d8}
\end{eqnarray}
where $D$ is given by Eq. (\ref{d5}) in the general case and by Eq.
(\ref{d6b}) when the profile of angular velocity is monotonic.  In
vectorial form, the drift can be written
${\mb\eta}=-D\beta\Gamma_{0}\nabla\psi'$. It is perpendicular to the
relative mean field velocity $\langle {\bf V}'\rangle =-{\bf z}\times
\nabla\psi'$. In the analogy with the Brownian
motion \cite{preR,kin}, the drift coefficient can be seen as a sort of
``mobility''. It is related to the diffusion coefficient and to the
temperature (which takes negative values in cases of interest) by an
analogue of the Einstein relation \cite{preR}:
\begin{eqnarray}
\xi=D\beta\Gamma_{0}.
\label{d9}
\end{eqnarray}
We note that the {\it drift velocity}  \cite{preR} of a test vortex
is the counterpart of the Chandrasekhar {\it dynamical friction}
\cite{chandraB} experienced by a star in a cluster.

\subsection{Connection with the usual form of the Fokker-Planck equation}
\label{sec_intr}

For an axisymmetric system, the Fokker-Planck equation is usually
written in the form
\begin{eqnarray}
{\partial P\over\partial t}={1\over 2r}{\partial\over\partial r}\left\lbrack r{\partial\over\partial r}\left ( {\langle (\Delta r)^{2}\rangle\over \Delta t}P\right )\right\rbrack-{1\over r}{\partial\over\partial r}\left (rP {\langle \Delta r\rangle\over \Delta t}\right ),\nonumber\\
\label{d10}
\end{eqnarray}
where $\Delta r$ denotes the increment of position of the test
vortex in the radial direction. The second moment ${\langle (\Delta
r)^{2}\rangle/ 2\Delta t}$ represents the diffusion coefficient and
the first moment ${\langle \Delta r\rangle/\Delta t}$ represents the
drift velocity (the corresponding term in the Fokker-Planck equation
can be viewed as an advection term). Comparing Eq. (\ref{d10}) with
Eq. (\ref{d1}), we find that
\begin{eqnarray}
D={1\over 2}{\langle (\Delta r)^{2}\rangle\over \Delta t},
\label{d11}
\end{eqnarray}
and
\begin{eqnarray}
\eta={\langle \Delta r\rangle\over \Delta t}-{\partial D\over\partial r}.
\label{d12}
\end{eqnarray}
The second expression shows that the drift velocity $\langle \Delta
r\rangle /\Delta t$ is not exactly given by Eq. (\ref{d6c}) but that
there is an additional contribution $\partial D/\partial r$ (see
\cite{bbgky}).  Integrating Eq. (\ref{d6}) by parts, we have
\begin{eqnarray}
\eta={\Gamma_{0}\over 4r} \int_{0}^{+\infty}   d{r}'{\omega(r')} {\partial \over \partial r'}\delta (\Omega-\Omega')\ln\biggl\lbrack 1-\biggl ({r_{<}\over r_{>}}\biggr )^{2}\biggr\rbrack.\nonumber\\
\label{d13}
\end{eqnarray}
Inserting the expressions (\ref{d5}) and (\ref{d13}) in Eqs. (\ref{d11}) and (\ref{d12}), we find that the  second (diffusion) and first (drift) moments of the position increment of the test vortex can be put in the form
\begin{eqnarray}
{\langle (\Delta r)^{2}\rangle\over \Delta t}=-{\gamma\over 2r^{2}} \int_{0}^{+\infty}  r' d{r}' \delta (\Omega-\Omega')\nonumber\\
\times\ln\biggl\lbrack 1-\biggl ({r_{<}\over r_{>}}\biggr )^{2}\biggr\rbrack\omega(r'),\nonumber\\
\label{d14}
\end{eqnarray}
\begin{eqnarray}
{\langle \Delta r\rangle\over \Delta t}=-{1\over 4} \int_{0}^{+\infty}  rr' d{r}' \omega(r')\left ({\gamma\over r}{\partial\over\partial r}-{\Gamma_{0}\over r'}{\partial\over\partial r'}\right )\nonumber\\
\times\delta (\Omega-\Omega')\ln\biggl\lbrack 1-\biggl ({r_{<}\over r_{>}}\biggr )^{2}\biggr\rbrack {1\over r^{2}}.
\label{d15}
\end{eqnarray}
In Appendix \ref{sec_fs}, it is shown that these expressions can be
obtained directly from the equations of motion of the point vortices
in the large $N$ limit.  Therefore, starting from these expressions
(\ref{d14}) and (\ref{d15}), inserting them in Eq. (\ref{d10}), and
repeating the above calculations in the other way round, we can derive
the Fokker-Planck equation (\ref{f4}) for the relaxation of a test
vortex in a bath. If we now account for the fact that the distribution
of the bath in Eq. (\ref{f4}) is not stationary but slowly evolves
under the effect of collisions in a self-consistent way, we obtain the
integro-differential equation (\ref{k11}) for the evolution of the
system as a whole. This approach provides an alternative derivation of
the kinetic equations directly from the equations of motion (see
Appendix \ref{sec_fs}).

\subsection{Scaling of the collision term with $N$}
\label{sec_ss}

Let us consider the kinetic equation (\ref{block1}) for a single
species of point vortices.  We introduce the total circulation
$\Gamma=N\gamma$, the typical radius of the system $R$ (fixed by the
domain size or by the angular momentum) and the dynamical
time $t_{D}=R^{2}/\Gamma$. If we measure time in units of $t_{D}$
and distances in units of $R$, the kinetic equation describing the
evolution of the $N$-body system as a whole becomes
\begin{eqnarray}
{\partial\omega\over\partial t}=-{1\over 4N}{1\over r}{\partial\over\partial r} \int_{0}^{+\infty}  r' d{r}' \delta (\Omega-\Omega')\nonumber\\
\times\ln\biggl\lbrack 1-\biggl ({r_{<}\over r_{>}}\biggr )^{2}\biggr\rbrack
\biggl (\omega'{1\over r}{\partial\omega\over\partial r}-\omega{1\over r'}{\partial\omega'\over\partial r'}\biggr ).
\label{ss1}
\end{eqnarray}
This expression shows that the collision term scales like $1/N$.  This
implies that the kinetic equation (\ref{ss1}) correctly describes the
dynamics of the point vortex gas on a timescale $\sim Nt_{D}$ (without
$\ln N$ correction). We emphasize that this timescale does not
generically represent the collisional relaxation time $t_{relax}$
towards statistical equilibrium because the kinetic equation
(\ref{ss1}) does not relax towards the Boltzmann distribution in
general.  However, on this timescale, the system reaches a steady
state with a monotonic profile of angular velocity (in generic
situations) due to distant collisions (resonances).  Therefore, it is
a sort of collisional relaxation time that we shall denote
\begin{eqnarray}
t_{relax}^{*}\sim Nt_{D}, \label{ss1b}
\end{eqnarray}
where the asterix indicates that this is not, in general, the
relaxation time $t_{relax}$ towards statistical
equilibrium\footnote{There may be situations where the steady state
reached by the system has an approximate Boltzmann
distribution. Indeed, resonances have the tendency to ``push'' the
system towards the Boltzmann distribution (since the Boltzmann entropy
monotonically increases). Thus, if there are sufficient resonances
during the evolution, the system can converge to a distribution
{close} to the Boltzmann distribution on a timescale $\sim
Nt_{D}$. This is however not always the case. For example, in the
numerical simulation reported in Sec. \ref{sec_num}, we have checked
that the final distribution is relatively {far} from the Boltzmann
distibution, even in the core. }. The proper determination of the
relaxation time towards the Boltzmann distribution (exactly), and the
scaling of $t_{relax}$ with $N$, remains unknown.

Using the same normalization as before, the Fokker-Planck equation
(\ref{f9}) describing the relaxation of a test vortex in a {\it
fixed} distribution of $N$ field vortices can be written
\begin{eqnarray}
{\partial P\over\partial t}={1\over 8}{\ln N\over N}{1\over r}{\partial\over\partial r}\biggl\lbrack r  D(r)\biggl ({\partial P\over\partial r}-P{d\ln |\omega|\over d r}\biggr )\biggr\rbrack,
\label{ss3}
\end{eqnarray}
with a diffusion coefficient $D(r)= {\omega(r)/ |r\Omega'(r)|}$.
This expression shows that the typical timescale of the relaxation
of the test vortex towards the bath distribution is
\begin{eqnarray}
t_{relax}\sim {N\over\ln N}t_{D}.
\label{ss5}
\end{eqnarray}
Said differently, the relaxation time corresponds to the typical
time needed by the vortex to diffuse on a length $\sim R$, the domain
size. Thus $t_{relax}\sim R^{2}/D$. Using Eq. (\ref{d6b}), we get
$t_{relax}\sim R^{2}|\Sigma|/(\gamma\ln \Lambda\omega)\sim
R^{2}/(\gamma\ln N)\sim NR^{2}/(\Gamma\ln N)\sim (N/\ln N)t_{D}$.  We
recall that the $\ln N$ factor arises from the logarithmic divergence
of the diffusion coefficient when we make a bath approximation. There
is no such term in Eqs. (\ref{ss1}) and (\ref{ss1b}) when we study the
evolution of the system as a whole (this is different from the case of
stellar systems where the $\ln N$ term arises both in the
gravitational Landau equation describing the evolution of the system
as a whole and in the Kramers-Chandrasekhar equation describing the
relaxation of a test particle
\cite{kandrupREV}).  In Sec. \ref{sec_co}, we shall see that the
spectrum of the eigenvalues of the Fokker-Planck equation (\ref{f9})
has no gap when the diffusion coefficient decreases rapidly with the
distance. This implies that the relaxation towards the steady
distribution (\ref{f11}) is not exponential but rather
algebraic. Therefore, Eq.  (\ref{ss5}) is not an exponential
relaxation time. However, it provides an estimate of the timescale on
which ``collisional'' effects take place to drive the distribution of
the test vortex towards the distribution of the bath.

\section{General study of the Fokker-Planck equation}
\label{sec_gen}

\subsection{Relaxation of the tail of the distribution}
\label{sec_t}

Equation (\ref{f9}) is a particular case of the general Fokker-Planck
equation
\begin{eqnarray}
{\partial P\over\partial t}={1\over r^{d-1}}{\partial\over\partial r}\biggl\lbrack r^{d-1}  D(r)\biggl ({\partial P\over\partial r}+ P U'(r)\biggr )\biggr\rbrack,
\label{t1}
\end{eqnarray}
with $d=2$, $U(r)=-\mu\ln|\omega|(r)$ where $\mu=\Gamma_{0}/\gamma$,
and $D(r)$ is given by Eq. (\ref{f10}). The stationary solution of
this equation is
\begin{eqnarray}
P_{e}(r)=A e^{-U(r)},
\label{t2}
\end{eqnarray}
where $A$ is a normalization constant.  In an appropriate system of
coordinates, the relaxation of the tail of the distribution function
has a front structure. This has been studied by Potapenko {\it et al.}
\cite{potapenko} for Coulombian plasmas and by Chavanis \& Lemou
\cite{cl} in the general case. Let us briefly recall the main results
of the theory before considering its application to the case of point
vortices. If we make the change of variables $dx/dr=1/\sqrt{D(r)}$ and
introduce the function $u(r,t)=P(r,t)/P_{e}(r)$, we can rewrite the
Fokker-Planck equation (\ref{t1}) 
in the form of an advection-diffusion equation
\begin{eqnarray}
{\partial u\over\partial t}+V\lbrack r(x)\rbrack {\partial
u\over\partial x}={\partial^{2}u\over\partial x^{2}}, \label{t3}
\end{eqnarray}
with a velocity field
\begin{eqnarray}
V(r)=-\sqrt{D(r)}{d \over d r}\biggl \lbrace \ln \biggl \lbrack r^{d-1}e^{-U(r)}D^{1/2}(r)\biggr \rbrack\biggr\rbrace.
\label{t4}
\end{eqnarray}
The evolution of the position of the front $r_{f}(t)$ is determined by
the equation
\begin{eqnarray}
{dr_{f}\over dt}=\sqrt{D(r_{f})}V(r_{f}).
\label{t5}
\end{eqnarray}
Introducing $z=x-x_{f}(t)$, $u(x,t)=\phi(z,t)$ and $\tau=2t$, the profile of the front for sufficiently large times is given by
\begin{eqnarray}
\phi(z,\tau)=\Phi\left \lbrack {z\over\chi(\tau)}\right \rbrack,
\label{t6}
\end{eqnarray}
where
\begin{eqnarray}
\Phi(x)={1\over\sqrt{\pi}}\int_{x}^{+\infty}e^{-y^{2}}dy,
\label{t7}
\end{eqnarray}
is connected to the error function by $\Phi(x)={1\over 2}\lbrack 1-{\rm
erf}(x)\rbrack$ and  $\chi(\tau)$ is the function
\begin{eqnarray}
\chi^{2}(\tau)=2\int_{1}^{\tau}e^{\lbrack H(\tau)-H(\tau')\rbrack}d\tau',
\label{t8}
\end{eqnarray}
where $H(\tau)$ is a primitive of $h(\tau)=g(\tau/2)$ with
\begin{eqnarray}
g(t)=V'(r_{f}(t))\sqrt{D(r_{f}(t))}.
\label{t9}
\end{eqnarray}
When $g(t)=-1/(2t)$, a case that  often occurs, we have
$\chi(\tau)=\tau^{1/2}$. The derivation of these formulae is given
in \cite{cl}. In addition, it is shown that the parameter
controlling the validity of the approximations made in the theory is
\begin{eqnarray}
\epsilon(t)\equiv \biggl | {\sqrt{\pi}\over 4}{\chi(2t)\sqrt{D(r_{f})}\over V'(r_{f})/V''(r_{f})}\biggr |.
\label{t10}
\end{eqnarray}
The theory becomes asymptotically exact if $\epsilon(t)\rightarrow 0$
for $t\rightarrow +\infty$. If this condition is not fulfilled, the
theory can nevertheless provide a good description of the front
structure provided that $\epsilon$ is sufficiently small (see explicit examples in \cite{cl}).

\subsection{Correlation functions}
\label{sec_co}

In this section, we determine the asymptotic behavior of the
temporal correlation functions $\langle A(0)A(t)\rangle$ of the
Fokker-Planck equation (\ref{t1}) for $t\rightarrow +\infty$.  When
the spectrum of the linearized Fokker-Planck equation (\ref{t1})
has a gap $\mu$ at the spectral value $0$, the correlation functions tend to zero
exponentially rapidly as $e^{-\mu t}$. However, when the
spectrum has no gap in $[0, +\infty[$, the decay can be
slower (algebraic). This is the case in particular when the diffusion
coefficient $D(r)$ decreases rapidly with the distance (see below). This
problem has been considered by different authors in various contexts
[59-62,11] and we shall mainly rely upon
their results.  In particular, the situation that we investigate here is
closely related to the one considered by Bouchet \& Dauxois \cite{bd}
for the HMF model.

In the following, we briefly recall the general theory developed in
the Appendix B of Marksteiner {\it et al.} \cite{mark} (see also
[60-62,11]) to compute the asymptotic behaviour of a
correlation function associated to a Fokker-Planck equation with
logarithmic potential and derive new results that will be useful in
the sequel.  If we make the change of variables $dx/dr=1/\sqrt{D(r)}$
and $f(x,t)=\sqrt{D(r)}P(r,t)S_{d}r^{d-1}$ (where $S_{d}$ denotes the
surface of a $d$-dimensional sphere of unit radius), we can rewrite
the Fokker-Planck equation (\ref{t1}) in the form of a 1D
Fokker-Planck equation with a constant diffusion coefficient
\begin{eqnarray}
{\partial {f}\over\partial t}={\partial\over\partial x}\biggl ({\partial {f}\over\partial x}+{f}{\partial\Phi\over\partial x}\biggr ),
\label{co1}
\end{eqnarray}
and a  potential
\begin{eqnarray}
\Phi(x)=- \ln \biggl \lbrack r^{d-1}e^{-U(r)}D^{1/2}(r)\biggr \rbrack.
\label{co2}
\end{eqnarray}
We note that $\Phi'(x)=V\lbrack r(x)\rbrack$ where $V(r)$ is the
velocity field introduced in Eq. (\ref{t4}).

We recall that, in our context, the Fokker-Planck equations are
written in radial coordinates $r\geq 0$ or $x\geq 0$ and that the
current must vanish at the origin. Therefore equations (\ref{t1}) or
(\ref{co1}) are subject to the following radial boundary condition at
$r=0$ or $x=0$: \begin{eqnarray}
\biggl ({\partial P\over\partial r}+ P U'(r)\biggr )|_{r=0}=0, \ \
\mbox{or} \ \ \ \biggl ({\partial {f}\over\partial
  x}+{f}{\partial\Phi\over\partial x}\biggr )|_{x=0}=0.\nonumber\\
\label{RBC}
\end{eqnarray}
This is equivalent to say that the resulting solution can be
extended to an {\em even}  solution in  $]-\infty,+\infty[$. In
particular the total mass on $r,x\ge 0$ is conserved and a
normalization condition can be assumed
\begin{eqnarray}
\int_{0}^{+\infty}f(x,t)dx=1.
\label{co3}
\end{eqnarray}
Instead, the analysis developed by  Marksteiner {\it et al.}
\cite{mark} concerns the case of {\em odd} functions.  This means
that they consider a similar Fokker-Planck equation in which the
boundary condition  at $r=0$ is:
\begin{eqnarray}
P(t,r=0)=0, \ \
\mbox{or} \ \ \ f(t,x=0)=0.
\label{DBC}
\end{eqnarray}
In this case, the mass on $r,x\ge 0$ is not conserved and the
distribution function $f(x,t)$ goes to $0$ for large time.

We now give the main steps of the derivation of the asymptotic
behavior of the temporal correlation functions in the natural context
of the present study (this means that we incorporate the boundary
condition (\ref{RBC}) with the normalization condition (\ref{co3}))
and compare the theoretical results with numerical simulations.  The
stationary solution of Eqs. (\ref{co1})-(\ref{RBC}) can be written
\begin{eqnarray}
f_{e}(x)={1\over Z}e^{-\Phi(x)},
\label{co4}
\end{eqnarray}
where $Z=\int_{0}^{+\infty} e^{-\Phi(x)}dx$ is the normalization
constant. We shall assume that $\Phi(x)\sim \alpha\ln x$ for
$x\rightarrow +\infty$. This is indeed a situation that occurs in
the kinetic theory of point vortices (see below). The stationary
solution then behaves like $f_{e}\sim x^{-\alpha}$ for $x\rightarrow
+\infty$ and it is normalizable provided that $\alpha>1$.  Let
$W(x,t;x_{0},0)$ be the solution of Eq. (\ref{co1}) with the initial
condition $f(x,0)=\delta(x-x_{0})$. For any function $A(x)$, the
temporal correlation function is defined by
\begin{eqnarray}
\langle A(0)A(t)\rangle=\int A(x_{0})A(x)W(x,t;x_{0},0)f_{e}(x_{0})dx_{0}dx.\nonumber\\
\label{co5}
\end{eqnarray}
This can be rewritten
\begin{eqnarray}
\langle A(0)A(t)\rangle=\int_{0}^{+\infty} A(x)f_{A}(x,t)dx, \label{co6}
\end{eqnarray}
where $f_{A}(x,t)$ is the solution of Eq. (\ref{co1}) with initial
condition $f_{A}(x,0)=A(x)f_{e}(x)$. Note that its total ``mass''
is $\int_{0}^{+\infty} f_A(x,0)dx=\langle A\rangle$
where $\langle A\rangle$ is the average value of $A(x)$ at
equilibrium. This implies that $f_{A}(x,+\infty)=\langle A\rangle
f_{e}(x)$ since the mass is conserved.

\begin{figure}
\vskip0.5cm
\centering
\includegraphics[width=8cm]{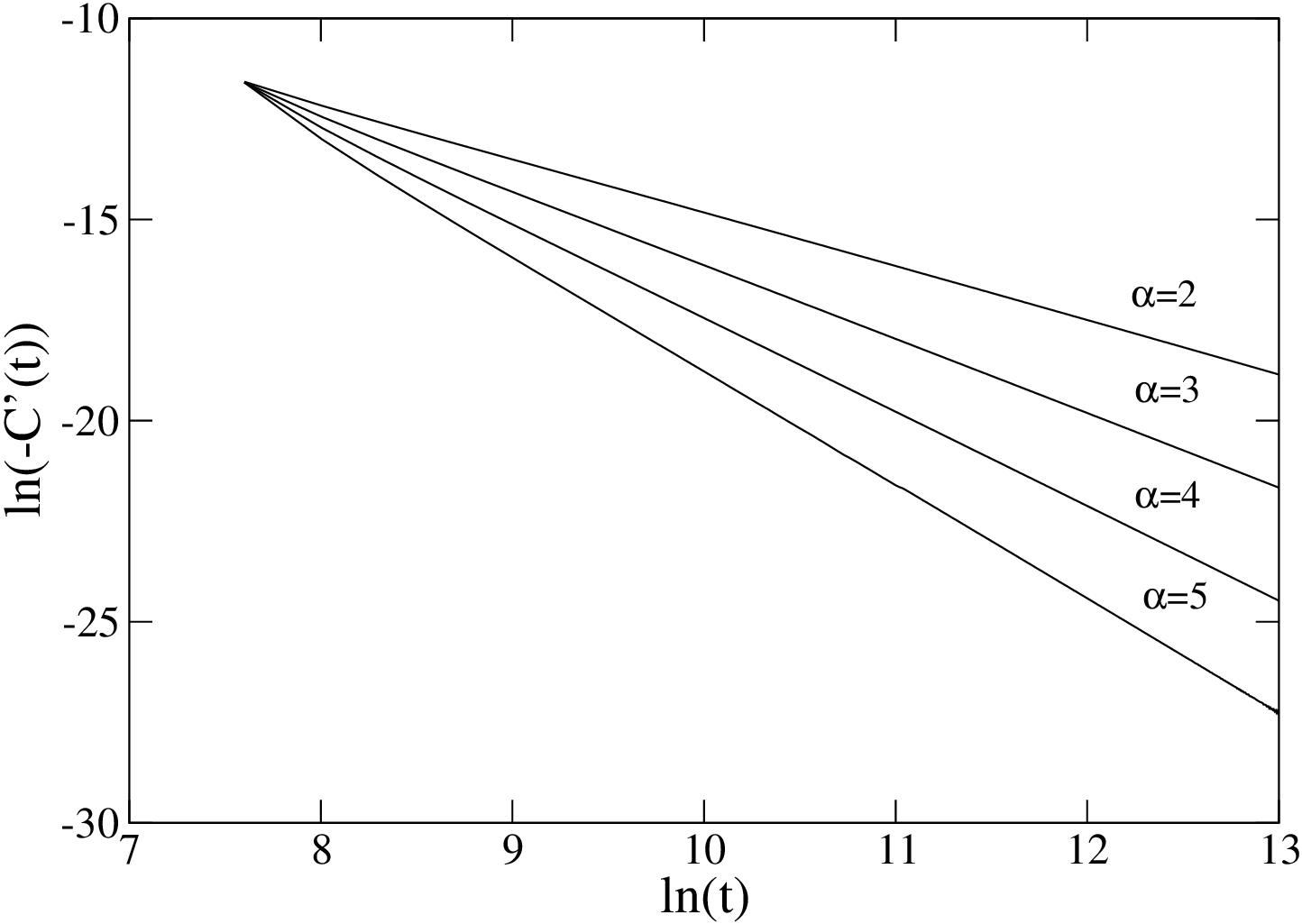}
\caption{Evolution of the correlation function (\ref{co16}) obtained
by solving the Fokker-Planck equation (\ref{co1}) with a logarithmic
potential $\Phi(x)=\alpha\ln x$ for $x\ge 1$ and
$\Phi(x)=\frac{1}{2}\alpha (x^2-1)$ for $x\le 1$. We have used the
boundary condition (\ref{RBC}) corresponding to even functions
$f(x,t)$. The correlation function is defined by Eqs. (\ref{co14}) and
(\ref{co6}) with $A(x)=\ln x$ ($\delta=1$) for $x\ge 1$ and $A(x)=0$
for $x\le 1$. We have considered different values of $\alpha=2,3,4,5$
(for clarity, the curves have been shifted to the same origin). To get
rid of the constant $\langle A\rangle$, we have computed the time
derivative $C'(t)$ of the correlation function (this is a good
strategy otherwise it is not clear whether we should use the exact or
the numerical value of $\langle A\rangle$ which differ due to finite
resolution). The slope obtained numerically clearly depends on
$\alpha$ and is found to be in good agreement with the theoretical
prediction (\ref{co16}), up to logarithmic corrections. }
\label{alpha23456pair}
\end{figure}

Setting $\psi=f e^{\Phi/ 2}$, we can transform Eq. (\ref{co1}) into a Schr\"odinger equation with imaginary time
\begin{eqnarray}
{\partial {\psi}\over\partial t}={\partial^{2}\psi\over\partial x^{2}}-V_{s}(x)\psi,
\label{co7}
\end{eqnarray}
with a potential $V_{s}(x)={1\over 4}(\Phi')^{2}-{1\over 2}\Phi''$. Looking for
solutions of the form $\psi\sim e^{-E_{k}t}\psi_{k}(x)$ with
$E_{k}=k^{2}$ ($k\ge 0$), we obtain the eigenvalue equation
\begin{eqnarray}
-{d^{2}\psi_{k}\over dx^{2}}+V_{s}(x)\psi_{k}=E_{k}\psi_{k},
\label{co8}
\end{eqnarray}
with $V_{s}(x)\sim \gamma/x^{2}$ for $x\rightarrow +\infty$ where
$\gamma=\alpha (\alpha+2)/4$. Then, the general solution of
Eq. (\ref{co1}) with initial
condition $f_{A}(x,0)=A(x)f_{e}(x)$ can be written
\begin{eqnarray}
f_A(x,t)=e^{-\Phi(x)/2}\left \lbrack
a_{0}\psi_{0}(x)+\int_{0}^{+\infty}
a(k)\psi_{k}(x)e^{-E_{k}t}dk\right\rbrack. \nonumber\\
\label{co9}
\end{eqnarray}
We assume that the functions $\psi_{k}(x)$ are properly orthonormalized, such that
$\int \psi_{k}(x)\psi_{k'}(x)dx=\delta(k-k')$ and $\int
\psi_{0}(x)\psi_{k}(x)dx=\delta_{k,0}$. Taking $t=0$ in Eq. (\ref{co9}), 
comparing with the initial condition, multiplying by $\psi_{k'}(x)$, integrating on $x$ and using the orthonormalization conditions, we obtain after simple calculations
\begin{eqnarray}
a(k)={1\over Z}\int_{0}^{+\infty} A(x)e^{-\Phi(x)/2}\psi_{k}(x)dx,
\label{co10}
\end{eqnarray}
\begin{eqnarray}
a_{0}={1\over Z}\int_{0}^{+\infty} A(x)e^{-\Phi(x)/2}\psi_{0}(x)dx.
\label{co11}
\end{eqnarray}
On the other
hand, considering the limit $t\rightarrow +\infty$, we find that
$\psi_{0}(x)\propto f_{e}(x)e^{\Phi(x)/2}\propto e^{-\Phi(x)/2}$.
Therefore, the normalized ground state is
\begin{eqnarray}
\psi_{0}(x)={1\over\sqrt{Z}}e^{-\Phi(x)/2}.
\label{co12}
\end{eqnarray}
Inserting this expression in Eq. (\ref{co11}), we find that
\begin{eqnarray}
a_{0}={1\over\sqrt{Z}}\int_{0}^{+\infty}
A(x)f_{e}(x)dx={1\over\sqrt{Z}}\langle A\rangle. \label{co13}
\end{eqnarray}
 Inserting Eq. (\ref{co9}) in Eq. (\ref{co6})
and using Eqs. (\ref{co10}) and (\ref{co13}), we finally obtain
\begin{eqnarray}
C(t)\equiv \langle A(0)A(t)\rangle-\langle A\rangle^{2}=Z\int_{0}^{+\infty}a(k)^{2}e^{-E_{k}t}dk.\nonumber\\
\label{co14}
\end{eqnarray}

To obtain the large time behavior of the correlation function, we
need to determine the equivalent of $a(k)$ for $k\rightarrow 0$. To
that purpose, we need the form of the eigenfunctions $\psi_{k}(x)$
for $k\rightarrow 0$. In the case where $V_{s}(x)\sim \gamma/x^{2}$
for $x\rightarrow +\infty$, corresponding to the logarithmic
Fokker-Planck equation, it is possible to solve the eigenvalue
equation (\ref{co8}) by match asymptotics \cite{mark,bd} and obtain
$\psi_{k}(x)$ for $k\rightarrow 0$. Then, we can determine how the
correlation function (\ref{co14}) decreases for $t\rightarrow
+\infty$ depending on the behavior of the function $A(x)$ for
$x\rightarrow +\infty$. Using the results of Marksteiner {\it et
al.} \cite{mark}, we find that when $A(x)\sim x^{n}$ the correlation
function decreases algebraically like
\begin{eqnarray}
C(t)\sim t^{-\xi}, \qquad \xi=-n+{\alpha-1\over 2}.
\label{co15}
\end{eqnarray}
This result has been given previously by Lutz \cite{lutz} and Bouchet
\& Dauxois \cite{bd}. When $A(x)\sim (\ln x)^{1/\delta}$, using the
results of Marksteiner {\it et al.} \cite{mark}, we find that the
correlation function decreases like
\begin{eqnarray}
C(t)\sim {(\ln t)^{2/\delta}\over t^{\alpha-1\over 2}}.
\label{co16}
\end{eqnarray}
This expression has been checked numerically for different values of
$\alpha$ (see Fig. \ref{alpha23456pair}). For $\alpha=3$, we recover
the result $C(t)\sim (\ln t)^{2/\delta}/t$ given by  Bouchet \& Dauxois
\cite{bd} which turns out to be valid only for this particular value 
of $\alpha$.

\begin{figure}
\centering
\includegraphics[width=8cm]{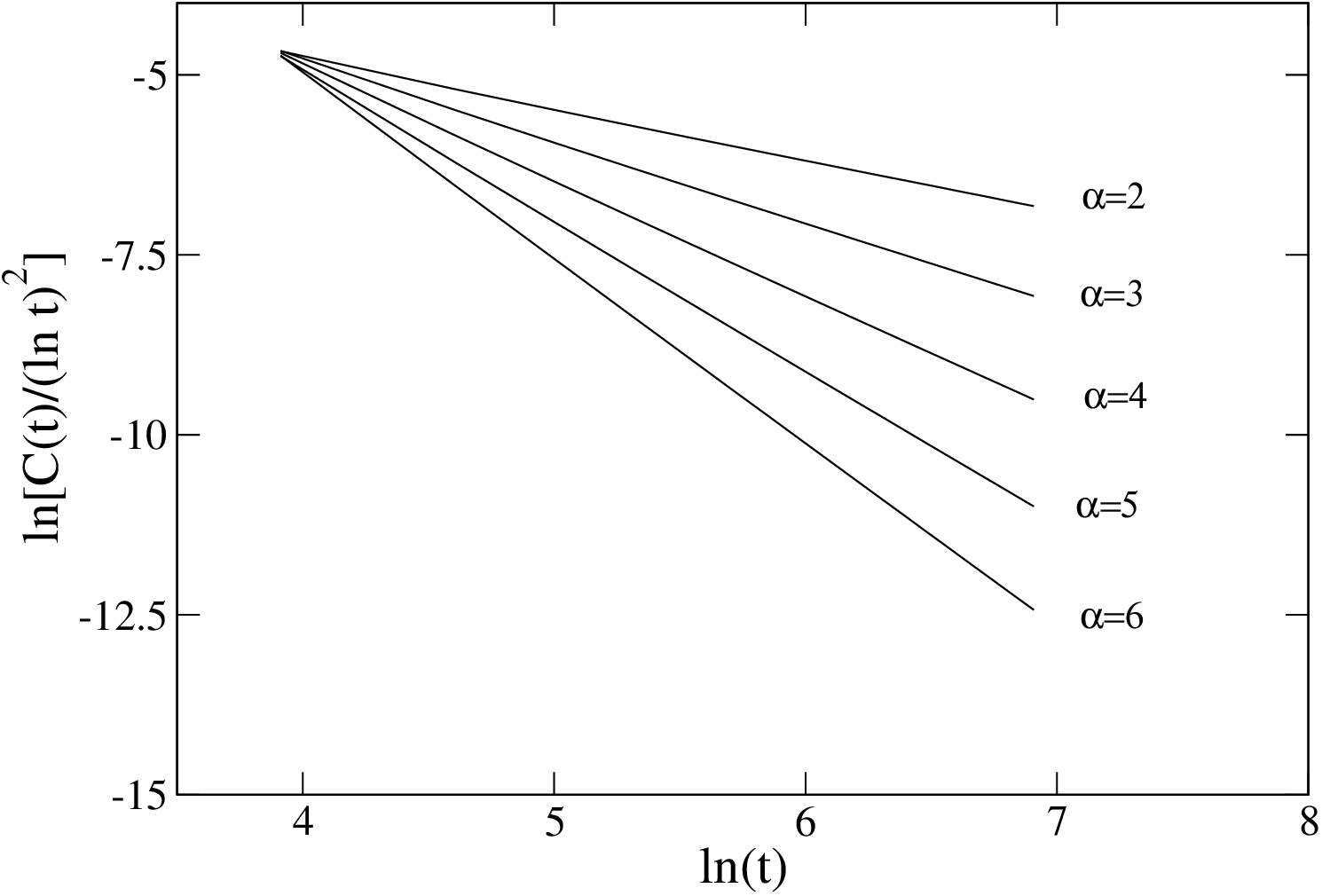}
\caption{Same as Fig. \ref{alpha23456pair}, except that we solve the
Fokker-Planck equation with the boundary condition (\ref{DBC})
corresponding to odd functions $f(x,t)$. We have considered
different values of $\alpha=2,3,4,5,6$. In the present case $\langle
A\rangle =0$ so that we can directly plot $C(t)$. The slope obtained
numerically clearly depends on $\alpha$ and is found to be in good
agreement with the theoretical prediction (\ref{co16}). This is the
same value as in Fig. \ref{alpha23456pair} but we note that the
scaling regime is reached much more rapidly. } \label{alpha23456}
\end{figure}

The above results can be easily adapted to the case where the
boundary condition is given by Eq. (\ref{DBC}) instead of Eq.
(\ref{RBC}). In that case, $f_e(x)$ is not a stationary solution of Eqs.
(\ref{co1})-(\ref{DBC}) because it does not satisfy the boundary
condition at $r=x=0$. The stationary solution of Eqs.
(\ref{co1})-(\ref{DBC}) is $f(x)=0$. This implies that
$\psi_0(x)=0$ and $a_0=0$. As a result, Eq. (\ref{co14}) remains
valid provided that we take $\langle A\rangle =0$. Furthermore, since the
asymptotic behaviors of Eqs. (\ref{co15}) and (\ref{co16}) obtained from Eq. (\ref{co14}) only depend on the
large $x$ behavior of the functions $\psi_{k}(x)$ and not on the
boundary condition at $x=0$, these results are valid both for the boundary conditions (\ref{DBC})
and (\ref{RBC}).  This has been checked numerically in  Figs.
\ref{alpha23456pair} and \ref{alpha23456}. Interestingly, these numerical simulations  show that the scaling
regime is reached more rapidly when the boundary condition is given by
Eq. (\ref{DBC}) instead of Eq. (\ref{RBC}).

\section{General asymptotic results}
\label{sec_asy}

We shall now apply the preceding results to the case of point vortices
described by the Fokker-Planck equation (\ref{f9}). We first give
general results obtained by considering the asymptotic behavior of the
diffusion coefficient.

\subsection{Asymptotic behavior of the diffusion coefficient}
\label{sec_dd}

The angular velocity is determined from the vorticity by solving the
differential equation
\begin{eqnarray}
\omega={1\over r}{d\over dr}(r^{2}\Omega).
\label{dd1}
\end{eqnarray}
If $\Omega_{1}$ is a particular solution of this equation, the general
solution is $\Omega(r)=\Omega_{1}(r)+{C/ r^{2}}$
where $C$ is an arbitrary constant. One could {\it a priori} consider
profiles of angular velocity that are divergent for $r\rightarrow 0$.
However, this situation is not physical and in general, the constant
$C$ is determined in order to avoid the singularity at $r=0$. In that
case, the profile of angular velocity is given by
\begin{eqnarray}
\Omega(r)={1\over r^{2}}\int_{0}^{r}\omega(s)s \; ds. \label{dd3}
\end{eqnarray}
Asymptotically, we have for $r\rightarrow +\infty$:
\begin{eqnarray}
\Omega(r)\sim {\Gamma\over 2\pi r^{2}},
\label{dd4}
\end{eqnarray}
where $\Gamma$ is the circulation.  Then, using Eq. (\ref{f10}), we
obtain the asymptotic behavior of the diffusion coefficient
\begin{eqnarray}
D\sim {\pi\gamma\ln\Lambda\over 4\Gamma} r^{2}\omega(r)\propto r^{2}\omega(r).
\label{dd5}
\end{eqnarray}
More generally, combining Eqs. (\ref{f10}) and (\ref{dd1}), we find that
the diffusion coefficient can be expressed in terms of the angular velocity
by
\begin{eqnarray}
D(r)={\gamma\over 4}\ln \Lambda {(r^{2}\Omega)'\over |r^{2}\Omega'|}.
\label{dd6}
\end{eqnarray}
Alternatively, for a given expression of the diffusion coefficient
$D(r)$, the above equation is a first order differential equation for
$\Omega(r)$. It can be easily integrated, leading to
\begin{eqnarray}
\Omega(r)={\rm exp}\biggl\lbrace -\int {2dr\over \lbrack 1-{4\epsilon D(r)\over \gamma\ln\Lambda}\rbrack r}\biggr\rbrace,
\label{dd7}
\end{eqnarray}
where $\epsilon=\pm 1$ depending on the sign of the shear
($\epsilon=+1$ if $\Omega'>0$ and $\epsilon=-1$ if $\Omega'<0$). We
note that if $D(r)\rightarrow 0$ for $r\rightarrow +\infty$, which is
the situation of physical interest, we get $\Omega(r)\sim {C/r^{2}}$ for $r\rightarrow +\infty$.

\subsection{Power law}
\label{sec_p}

We first assume that $\omega\propto r^{-\alpha}$ for $r\rightarrow
+\infty$. The circulation is finite if $\alpha>2$ and the angular
momentum is finite if $\alpha>4$. The diffusion coefficient and
the potential behave like
\begin{eqnarray}
D\sim D_{0} r^{-(\alpha-2)}, \qquad U(r)\sim \mu\alpha\ln r.
\label{p1}
\end{eqnarray}
In the following, we shall take $D_{0}=1$ which can always be achieved
by rescaling the time appropriately. The stationary solution
(\ref{t2}) of the Fokker-Planck equation (\ref{t1}) decreases like
$P_{e}\sim r^{-\alpha\mu}$. It is normalizable provided that
$\alpha\mu>2$ and the variance $\langle r^{2}\rangle$ is finite
provided that $\alpha\mu>4$.

Let us first study the properties of the front
structure, using the results of Sec. \ref{sec_t}. The change of
variables $dx/dr=1/\sqrt{D(r)}$ leads to $r\sim (\alpha
x/2)^{2/\alpha}$ for $r\rightarrow +\infty$. On the other hand, the
velocity field (\ref{t4}) behaves like $V(r)\sim \lbrack
(2\mu+1)\alpha/2-2\rbrack r^{-\alpha/2}$. According to Eq. (\ref{t5}),
the position of the front evolves like $r_{f}(t)\sim
\lbrack \alpha ((2\mu+1)\alpha/2-2)t\rbrack^{1/\alpha}$. On the other hand, the function defined by Eq. (\ref{t9}) behaves like $g(t)\sim -1/(2t)$ for $t\rightarrow +\infty$. Therefore, according to Eq. (\ref{t6}), we find that the profile of the front characterizing the tail of the distribution for large times is given by
\begin{eqnarray}
u(r,t)\sim \Phi\left\lbrack {2\over \alpha}\
{r^{\alpha/2}-\sqrt{\alpha\lbrack {(2\mu+1)\alpha\over 2}-2\rbrack
t}\over (2t)^{1/2}}\right\rbrack.
\label{p2}
\end{eqnarray}
For $t\rightarrow +\infty$, the parameter $\epsilon(t)$ controlling the validity of the theory tends to
\begin{eqnarray}
\epsilon={1\over 4}\left ({\alpha\over 2}+1\right )\sqrt{4\pi\over \alpha (({2\mu+1})\alpha-4)}.
\label{p2esp}
\end{eqnarray}
Since $\epsilon(t)$ does not vanish for $t\rightarrow +\infty$, the
theory is only marginally applicable. However, it can still provide a
good description of the front structure when $\epsilon$ is
sufficiently small \cite{cl}.

Let us now study the temporal auto-correlation function of the
position $\langle r(0)r(t)\rangle$, using the results of Sec.
\ref{sec_co}. The potential defined by Eq. (\ref{co2}) behaves like
$\Phi(x)\sim \lbrack ((2\mu+1)\alpha-4)/\alpha\rbrack \ln x$. On the
other hand, in terms of the variable $x$, we need to determine the
correlation of $A(x)=r(x)\sim x^{2/\alpha}$. According to Eq.
(\ref{co15}), we find that
\begin{eqnarray}
\langle r(0)r(t)\rangle-\langle r\rangle^{2} \sim t^{-{(\mu\alpha-4)\over\alpha}} .
\label{p3}
\end{eqnarray}

\subsection{Stretched exponential}
\label{sec_e}

We shall now assume that $\omega\sim Ae^{-\lambda r^{\delta}}$ for
$r\rightarrow +\infty$. The Gaussian case corresponds to $\delta=2$
and the exponential case to $\delta=1$. The diffusion coefficient
and the potential behave like
\begin{eqnarray}
D\sim D_{0} r^{2}e^{-\lambda r^{\delta}}, \qquad  U(r)\sim \mu\lambda r^{\delta}.
\label{e1}
\end{eqnarray}
As before, we shall take $D_{0}=1$ without loss of generality.  The
stationary solution (\ref{t2}) of the Fokker-Planck equation
(\ref{t1}) decreases like $P_{e}\sim e^{-\lambda\mu r^{\delta}}$ and
all the moments of $r$ are well-defined.

Let us first study the properties of the front
structure, using the results of Sec. \ref{sec_t}. The relation between $x$ and $r$, and the velocity field $V(r)$ are asymptotically given by
\begin{eqnarray}
x\sim {2\over \lambda\delta r^{\delta}}e^{{\lambda\over 2}r^{\delta}}, \qquad V(r)\sim {(2\mu+1)\lambda\over 2}\delta r^{\delta}e^{-{\lambda\over 2}r^{\delta}}.\nonumber\\
\label{e2}
\end{eqnarray}
According to Eq. (\ref{t5}), the position of the front evolves like
\begin{eqnarray}
{e^{\lambda r_{f}^{\delta}}\over r_{f}^{2\delta}}={(2\mu+1)\lambda^{2}\over 2}\delta^{2}t.
\label{e3}
\end{eqnarray}
To leading order, we have $r_{f}(t)\sim (\ln t/\lambda)^{1/\delta}$. On the other hand,  the function defined by Eq. (\ref{t9}) behaves like  $g(t)\sim -1/(2t)$ for $t\rightarrow +\infty$. Therefore, according to Eq. (\ref{t6}), we find that the profile of the front characterizing the tail of the distribution for large times is given by
\begin{eqnarray}
u(r,t)=\Phi \left \lbrack {2\over\lambda\delta}{{1\over r^{\delta}}e^{{\lambda\over 2}r^{\delta}}-\sqrt{{(2\mu+1)\lambda^{2}\over 2}\delta^{2}t}\over (2t)^{1/2}}\right \rbrack.
\label{e4}
\end{eqnarray}
For $t\rightarrow +\infty$, the parameter $\epsilon(t)$ controlling the validity of the theory tends to
\begin{eqnarray}
\epsilon={1\over 8}\sqrt{4\pi\over 2\mu+1}.
\label{e4esp}
\end{eqnarray}
Since $\epsilon(t)$ does not vanish for $t\rightarrow +\infty$, the
theory is only marginally applicable. However, we shall see in Sec.
\ref{sec_g} that the theory still provides a good description of the
front structure.

Let us now study the temporal auto-correlation function $\langle
r(0)r(t)\rangle$ of the position, using the results of Sec.
\ref{sec_co}. Since $r\sim ((2/\lambda)\ln x)^{1/\delta}$, the
potential defined by Eq. (\ref{co2}) behaves like $\Phi(x)\sim
(2\mu+1)\ln x$. In terms of the variable $x$, we need to determine
the correlation of $A(x)=r(x)\sim (\ln x)^{1/\delta}$. Using Eq.
(\ref{co16}), we find that
\begin{eqnarray}
\langle r(0)r(t)\rangle-\langle r\rangle^{2} \sim {(\ln t)^{2/\delta}\over t^{\mu}}.
\label{e5}
\end{eqnarray}

\subsection{Behaviour of position-correlations according to the considered 
vorticity profile}
\label{sec_news}

We here qualitatively describe how the position-correlation
function decreases depending on the considered vorticity profile.
According to Eq. (\ref{co1}), the relaxation towards equilibrium is
controlled by the strengh of the drift term $\Phi(x)\sim\alpha_{0}\ln
x$, i.e. by the exponent $\alpha_{0}$. Therefore, for large
$\alpha_{0}$, the decorrelation should be fast, i.e. $\xi$ should be
large, which is indeed the case according to Eqs. (\ref{co15}) and
(\ref{co16}). Now, if the vorticity profile decreases like
$\omega(r)\sim r^{-\alpha}$, we find that
$\alpha_{0}=2\mu+1-4/\alpha$. Therefore, $\alpha_{0}$ is a
monotonically increasing function of $\alpha$ tending to
$\alpha_{0}=2\mu+1$ for $\alpha\rightarrow +\infty$. On the other
hand, the case $\omega(r)\sim e^{-\lambda r^{\delta}}$ leads to
$\alpha_{0}=2\mu+1$ which is consistent with the case $\omega(r)\sim
r^{-\alpha}$ with $\alpha\rightarrow +\infty$.  We conclude therefore
that when the vorticity profile decreases rapidly ($\alpha$ large, or
stretched exponential), the system decorrelates rapidly ($\alpha_{0}$
large hence $\xi$ large). The physical reason is not simple because
the vorticity profile enters both in the diffusion coefficient and in
the drift velocity in Eq. (\ref{t1}) so that the decorrelation is a
combined effect of these two terms. As discussed above, the situation 
is more easily interpreted from Eq. (\ref{co1}). 

\section{Particular examples}
\label{sec_examples}

In this section, we treat explicit examples corresponding to typical distributions of the field vortices.

\subsection{Gaussian vortex}
\label{sec_g}

We first consider the case where the distribution of the field vortices
corresponds to the statistical equilibrium state (thermal bath). As discussed in Sec. \ref{sec_pvg}, it is obtained by maximizing the entropy of the point vortex gas
\begin{eqnarray}
S=-\int {\omega\over\gamma}\ln {\omega\over\gamma} d{\bf r},
\label{g1}
\end{eqnarray}
at fixed circulation, energy and angular momentum. This leads to the Boltzmann-Poisson equation
\begin{eqnarray}
\omega=-\Delta\psi=A e^{-\beta\gamma (\psi+{1\over 2}\Omega_{L} r^{2})}.
\label{g2}
\end{eqnarray}
This maximization problem provides a condition of thermodynamical
stability. According to Appendix \ref{sec_eul}, the functional
(\ref{g1}) can also be interpreted as a particular $H$-function (or a
Casimir functional). In this sense, a vorticity profile $\omega({\bf
r})$ that maximizes $S$ at fixed $\Gamma$, $E$ and $L$ is a
nonlinearly dynamically stable stationary solution of the 2D Euler
equation. We conclude that for the single species point vortex gas,
thermodynamical stability implies nonlinear dynamical stability.  

In
this section, we consider the case where $\beta\rightarrow 0$ and
$\Omega_{L}\rightarrow +\infty$ in such a way that the product
$\lambda\equiv
\gamma\beta\Omega_{L}/2$ remains finite. The vorticity profile is then
given by
\begin{eqnarray}
\omega=\omega_{0} e^{-\lambda r^{2}}.
\label{g3}
\end{eqnarray}
This will be called the {\it gaussian vortex} (note that this is a
particular case of Sec. \ref{sec_e} with $\delta=2$). The
Fokker-Planck equation describing the relaxation of a test vortex in a
gaussian bath is
\begin{eqnarray}
{\partial P\over\partial t}={1\over r}{\partial\over\partial r}\biggl\lbrack r  D(r)\biggl ({\partial P\over\partial r}+2\lambda \mu P r\biggr )\biggr\rbrack.
\label{g6}
\end{eqnarray}
The stationary solution is
\begin{eqnarray}
P_{e}(r)={\lambda\mu\over \pi}e^{-\lambda\mu r^{2}}.
\label{g6b}
\end{eqnarray}
It has been normalized such that $\int P_{e} d{\bf r}=1$. Using Eq. (\ref{dd3}), the profile of angular velocity of the bath is
\begin{eqnarray}
\Omega={\omega_{0}\over 2\lambda r^{2}}(1-e^{-\lambda r^{2}}).
\label{g4}
\end{eqnarray}
Therefore, according to Eq.  (\ref{f10}), the diffusion coefficient is
given by
\begin{eqnarray}
D(r)= {D_{0}\lambda r^{2}\over e^{\lambda r^{2}}-\lambda r^{2}-1},
\label{g5}
\end{eqnarray}
with $D_{0}=(1/4)\gamma\ln\Lambda$ (in the following, we shall take
$D_{0}\lambda=1$ by rescaling the time appropriately). We note that
the shear vanishes at $r=0$ leading to a divergence of the diffusion
coefficient like $D(r)\sim r^{-2}$. This indicates a failure
\footnote{The assumptions made in the kinetic theory assume a
relatively strong shear so that the equations of motion for the point
vortices are dominated by the mean field trajectories (\ref{k4}). In
the absence of shear, the expression of the diffusion coefficient is
different as discussed in \cite{cs,houches} and \cite{dubin}. Thus,
the expression (\ref{g5}) is only valid at large distances $r\gg 1$.} 
of the kinetic theory for $r\rightarrow 0$. For $r\rightarrow
+\infty$, on the other hand, we have $D(r)\propto r^{2}e^{-\lambda
r^{2}}$. Since we are particularly interested by the tail of the
distribution, we shall extend this expression of the diffusion
coefficient for all $r$, thereby circumventing the problems arising
for $r\rightarrow 0$ (this regularization shall not change the
asymptotic results for $r\rightarrow +\infty$).

\begin{figure}
\centering
\includegraphics[width=8cm]{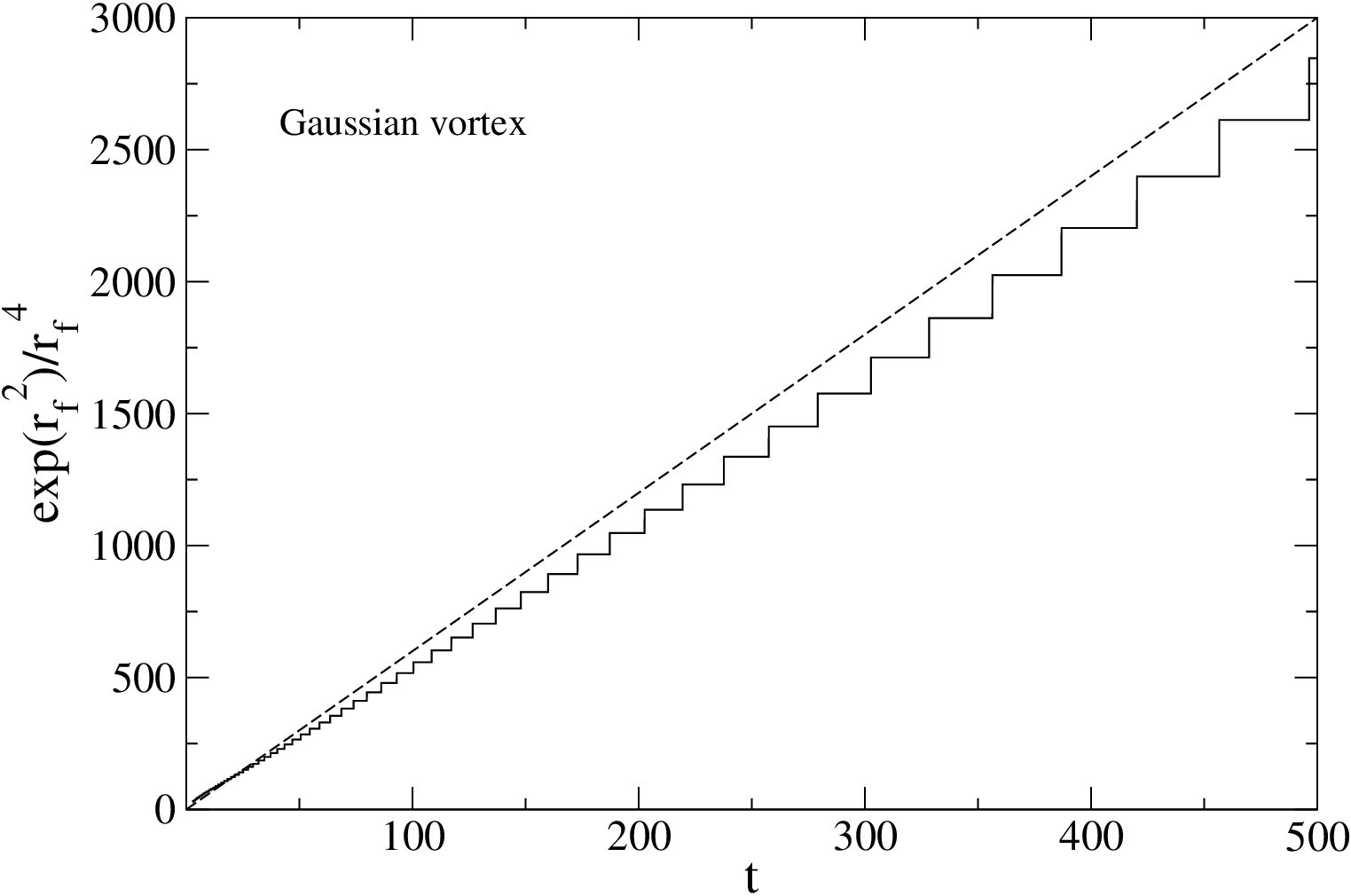}
\caption{Evolution of the front position $r_{f}(t)$ for the Fokker-Planck equation (\ref{g6}) with a diffusion coefficient $D(r)= r^{2}e^{-\lambda r^{2}}$ (we have taken $\mu=\lambda=1$). The details of the procedure are given in \cite{cl}. The result obtained numerically is in good agreement with the theoretical prediction (\ref{g7}). The agreement is, however, not perfect because the asymptotic value of $\epsilon\simeq 0.256$ is not exactly zero. As a result, the front position (solid line) evolves less rapidly than predicted theoretically (dashed line).  }
\label{front-gaussian}
\end{figure}

\begin{figure}
\centering
\includegraphics[width=8cm]{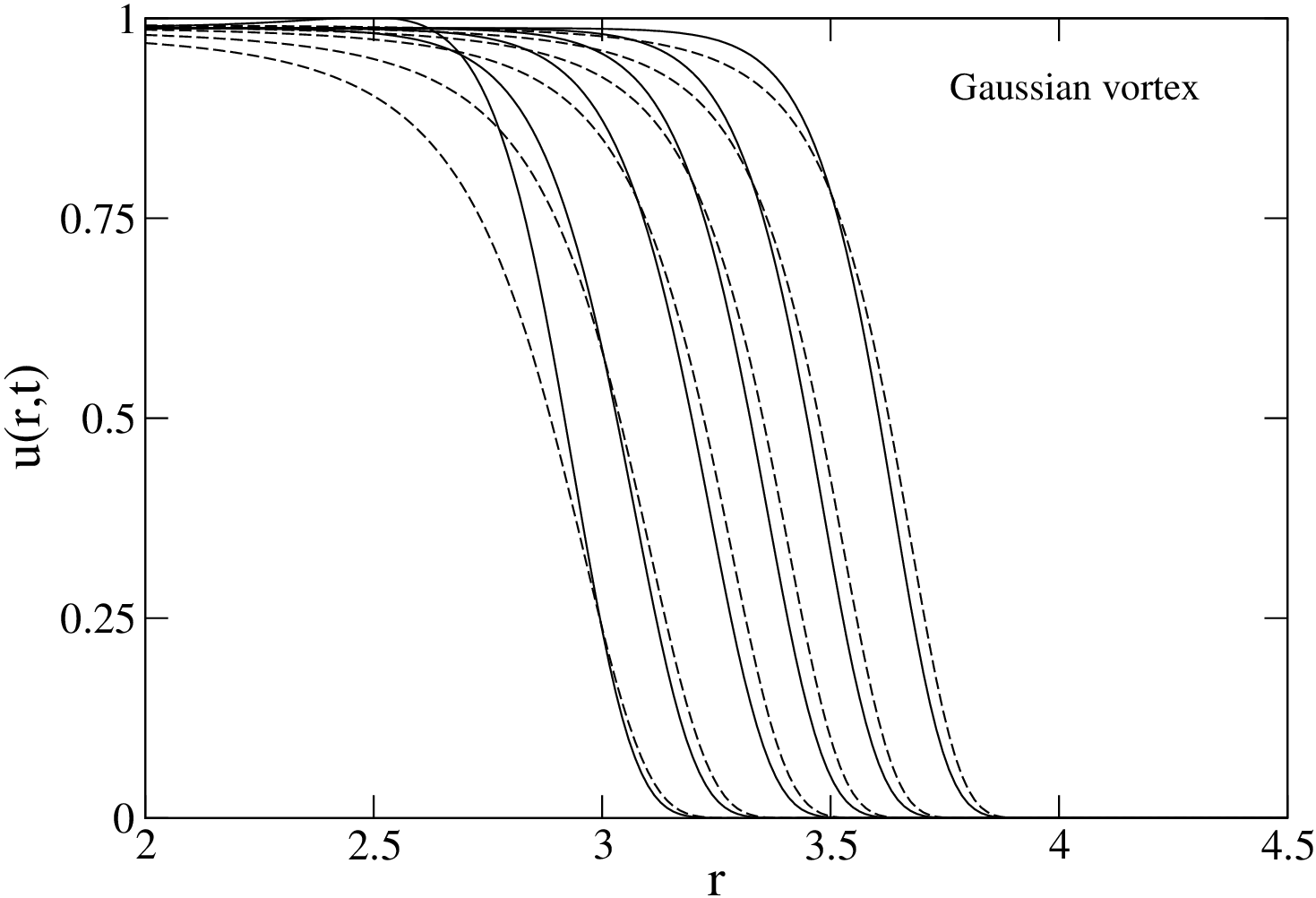}
\caption{Evolution of the front profile $u(r,t)$ for the Fokker-Planck equation (\ref{g6}) with a diffusion coefficient $D(r)= r^{2}e^{-\lambda r^{2}}$ (we have taken $\mu=\lambda=1$). The solid lines correspond to the numerical simulation and the dashed lines to the theoretical prediction.  The initial condition corresponds to the Heaviside function: $P(r,0)=1/(\pi a^{2})$ if $r\le a$ and $P(r,0)=0$ if $r\ge a$ with $a=2$.}
\label{prof-gauss}
\end{figure}

Let us first discuss the front structure of the distribution
that is solution of the Fokker-Planck equation (\ref{g6}). According
to Eq. (\ref{e3}), the evolution of the front position is given by
\begin{eqnarray}
{e^{\lambda r_{f}^{2}}\over r_{f}^{4}}=2(2\mu+1)\lambda^{2}t.
\label{g7}
\end{eqnarray}
For very large times, it can be approximated by $r_{f}(t)\sim (\ln
t/\lambda)^{1/2}$ so that the evolution is very slow (logarithmic). On
the other hand, according to Eq. (\ref{e4}), the front profile is
given by
\begin{eqnarray}
u(r,t)=\Phi \left \lbrack {1\over\lambda}{{1\over r^{2}}e^{{\lambda\over 2}r^{2}}-\sqrt{2(2\mu+1)\lambda^{2}t}\over (2t)^{1/2}}\right \rbrack.
\label{g8}
\end{eqnarray}
These theoretical results are compared in Figs. \ref{front-gaussian}
and \ref{prof-gauss} with direct numerical simulations of the
Fokker-Planck equation (\ref{g6}). For $\mu=1$, which is the value
considered in the numerical simulations, the parameter (\ref{e4esp})
controlling the validity of the theory is $\epsilon={1\over
8}(4\pi/3)^{1/2}\simeq 0.256...$. Since $\epsilon(t)$ does not tend to
zero, the agreement with theory is not perfect, but the relatively
small value of $\epsilon$ explains why the theory gives however a fair
description of the numerical results. In the present
case, we find that the front evolves less rapidly than predicted by
the theory (assuming $\epsilon\rightarrow 0$) but this tendency is not
general. In \cite{cl}, we have found other examples where the front
evolves more rapidly than predicted by the theory. Understanding this
difference requires to develop a theory at order $O(\epsilon)$ in an
expansion of the solutions of the front equation in powers of
$\epsilon\rightarrow 0$.

\begin{figure}
\centering
\includegraphics[width=8cm]{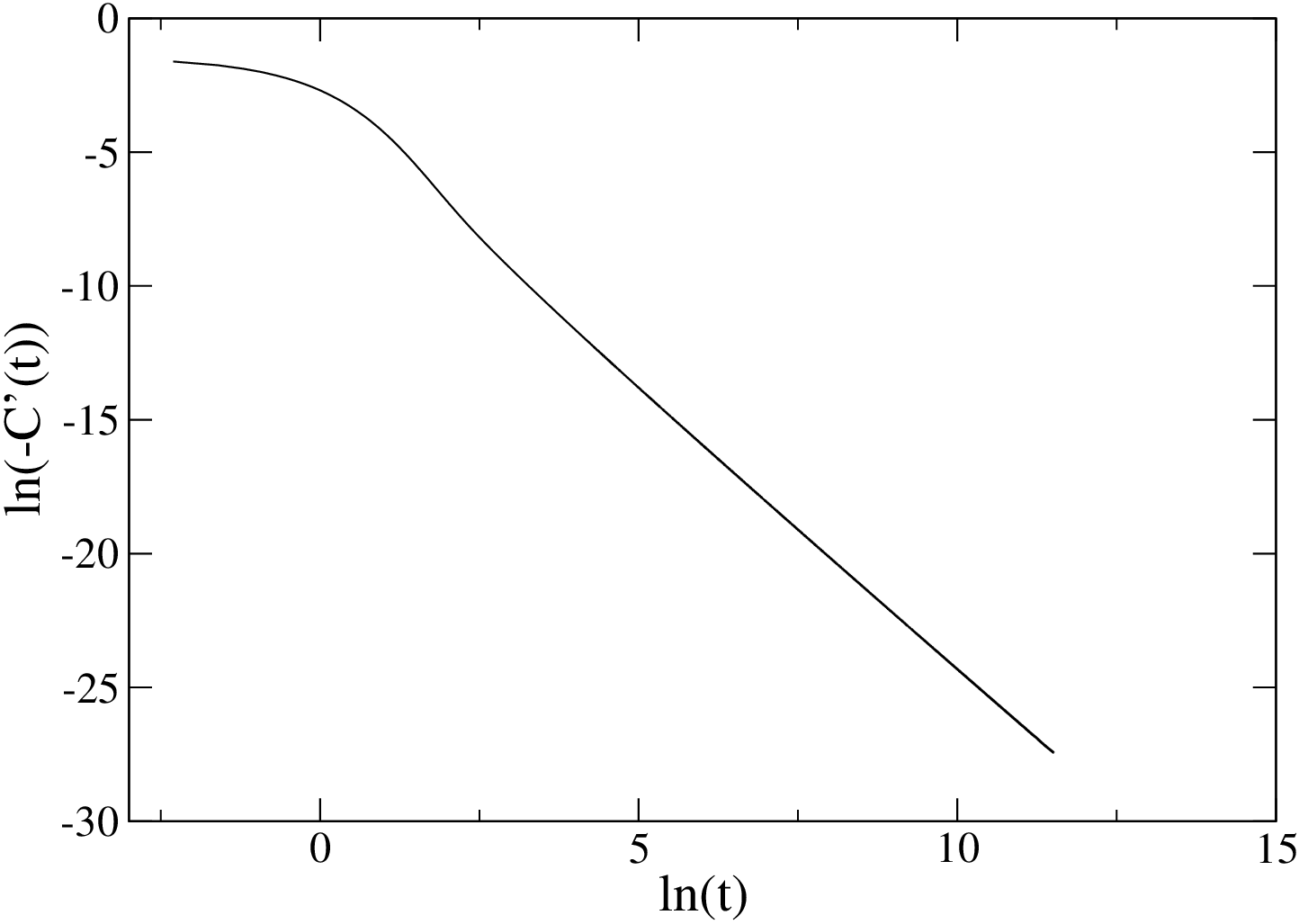}
\caption{Time evolution of the derivative of the correlation function $C(t)$. It is obtained from Eq. (\ref{g10}) by solving the Fokker-Planck equation (\ref{g6}) with the initial condition $P(r,0)=(1/\pi)r e^{-r^{2}}$ (we have taken $\lambda=\mu=1$). For large times, we find that the asymptotic slope of the curve is $2.0704...$, in good agreement with the theoretical value $2$ (up to logarithmic corrections).  }
\label{pentecorrelation}
\end{figure}

\begin{figure}
\centering
\includegraphics[width=8cm]{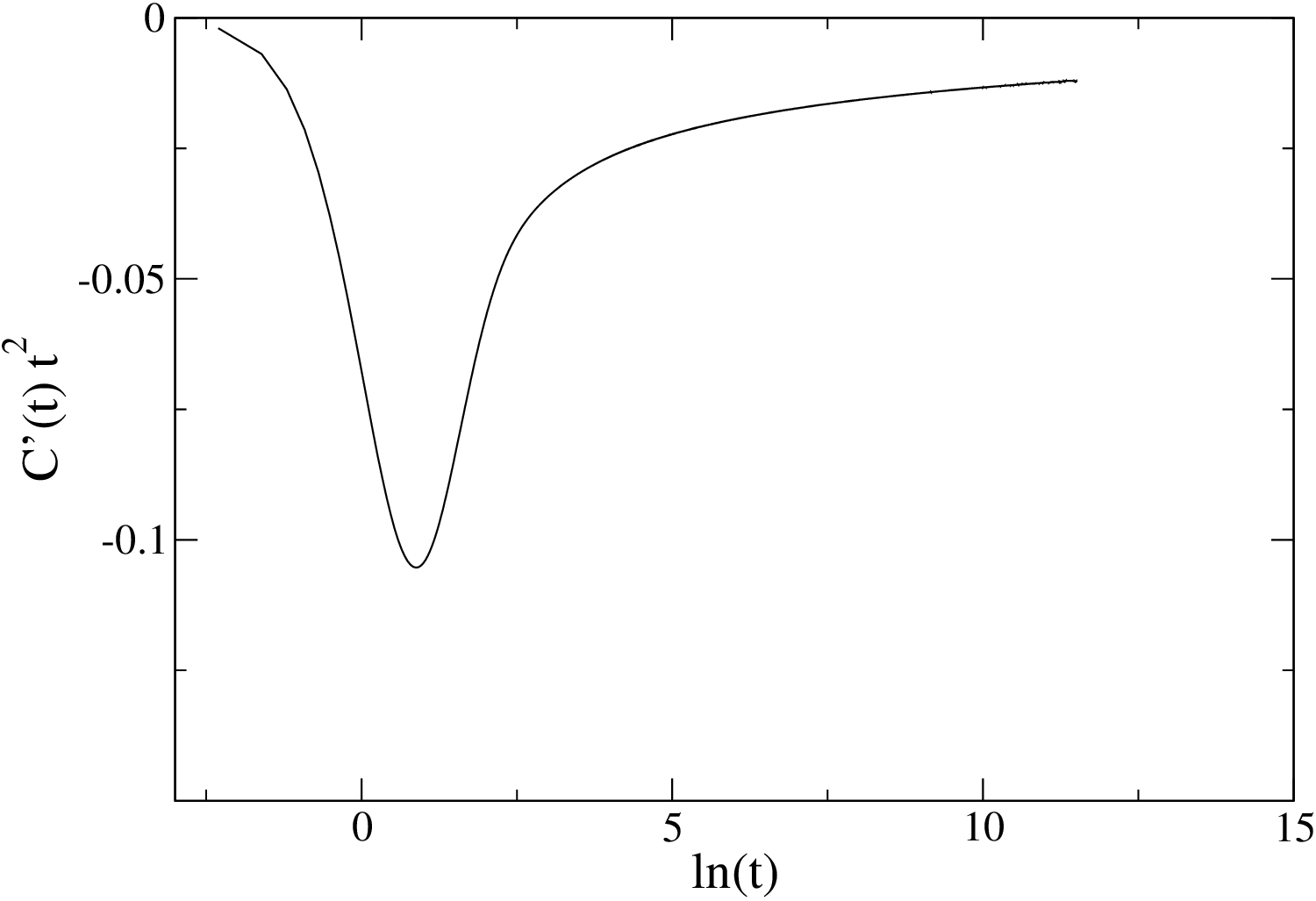}
\caption{This curve suggests that, for large times, the correction
to the algebraic decay of the correlation function is logarithmic as
expected from the theory. } \label{logcorrelation}
\end{figure}

Finally, according to Eq. (\ref{e5}), the temporal correlation
function of the position of the test vortex is predicted to decrease
like
\begin{eqnarray}
C(t)\equiv \langle r(0)r(t)\rangle-\langle r\rangle^{2} \sim {\ln t\over t^{\mu}}.
\label{g9}
\end{eqnarray}
To check this prediction, the function $\langle r(0)r(t)\rangle$ is computed numerically from the expression
\begin{eqnarray}
\langle r(0)r(t)\rangle =\int r_{0}r W({\bf r},t; {\bf r}_{0},0)P_{e}({\bf r}_{0})d{\bf r}_{0}d{\bf r}\nonumber\\
= \int r P_{R}({\bf r},t)d{\bf r},
\label{g10}
\end{eqnarray}
where $P_{R}({\bf r},t)=P_{R}(r,t)$ is the solution of the Fokker-Planck equation (\ref{g6}) with initial condition $P_{R}({\bf r},0)=r P_{e}({\bf r})$. On the other hand, $\langle r(t)\rangle$ is defined by
\begin{eqnarray}
\langle r(t)\rangle =\int r W({\bf r},t; {\bf r}_{0},0)P_{e}({\bf r}_{0})d{\bf r}_{0}d{\bf r}= \int r P_{E}({\bf r},t)d{\bf r},\nonumber\\
\label{g11}
\end{eqnarray}
where $P_{E}({\bf r},t)=P_{E}(r,t)$ is the solution of the
Fokker-Planck equation (\ref{g6}) with initial condition $P_{E}({\bf
r},0)=P_{e}({\bf r})$. Since $P_{E}({\bf r},t)=P_{e}({\bf r})$ for all
times (this is the stationary solution), we find that $\langle
r(t)\rangle=\langle r\rangle$ is the average of $r$ with the
equilibrium distribution (\ref{g6b}). Thus, $\langle
r\rangle={1\over 2}(\pi/\lambda\mu)^{1/2}$. We also note that $\int
P_{R}d{\bf r}=\langle r\rangle$ so that $P_{R}(r,+\infty)=\langle
r\rangle P_{e}(r)$. Therefore, according to Eq. (\ref{g10}), $\langle
r(0)r(t)\rangle\rightarrow
\langle r\rangle^{2}$ for $t\rightarrow +\infty$. This is why we
need to subtract the term $\langle r\rangle^{2}$ in the correlation
function (\ref{g9}). To avoid numerical errors
in the evaluation of $\langle r\rangle$ (due to finite resolution), we
have chosen to study the derivative of the correlation function which is
predicted to behave like
\begin{eqnarray}
C'(t) \sim -{\ln t\over t^{\mu+1}}.
\label{g12}
\end{eqnarray}
The results of the numerical simulations are represented in
Figs. \ref{pentecorrelation} and \ref{logcorrelation}. Figure
\ref{pentecorrelation} shows that the correlation function decays
algebraically with the correct exponent predicted by the theory (up to
logarithmic corrections considered in Fig. \ref{logcorrelation}).

In conclusion, we find that the relaxation of the distribution function
$P(r,t)$ to the equilibrium state (\ref{g6b}) is  peculiar. The evolution
of the front position is very slow (logarithmic) and the temporal
correlation function decreases algebraically (up to logarithmic
corrections). These results differ from the traditional exponential
relaxation in the usual Brownian theory. They are due to the rapid
(gaussian) decrease of the diffusion coefficient with $r$. The same
behaviours have been found for the HMF model \cite{bd,cl}.

\subsection{Isothermal vortex in a disk}
\label{sec_i}

We now consider the case of an {\it isothermal vortex} (\ref{g2}) in a disk
of radius $R$ and assume that $\Omega_{L}=0$. This corresponds to a
particular value of the angular momentum  $L_{0}(E,\Gamma)$ for a given
energy $E$ and circulation $\Gamma$. The equilibrium profile is then
determined by solving the Boltzmann-Poisson equation
\begin{eqnarray}
-{1\over r}{d\over dr}\left (r{d\psi\over dr}\right )=Ae^{-\beta\gamma\psi}=\omega_{0}e^{-\beta\gamma(\psi-\psi_{0})},
\label{i1}
\end{eqnarray}
with $\psi'(0)=0$ and $\psi(R)=0$. This equation can be solved analytically, see e.g. \cite{caglioti,houches}.  The vorticity profile is explicitly given by
\begin{eqnarray}
\omega={4\Gamma\over \pi R^{2}(\eta+4)}{1\over \bigl (1-{\eta\over\eta+4}{r^{2}\over R^{2}}\bigr )^{2}},
\label{i2}
\end{eqnarray}
where $\eta=\beta\gamma\Gamma/(2\pi)$ is a normalized temperature.
From this distribution, we can determine the mean field energy. The
caloric curve $\beta(E)$ is given in \cite{houches}. There exists
solutions for any value of the normalized energy $\epsilon=2\pi
E/(N^{2}\Gamma^{2})\ge 0$. The corresponding temperature satisfies
$\eta>\eta_{c}=-4$, i.e. $\beta>\beta_{c}=-8\pi/(\gamma\Gamma)$. For
$\beta>0$, the vorticity increases with the distance and for
$\beta<0$, the vorticity decreases with the distance.  For
$\beta=\beta_{c}$, we have $\omega({\bf r})=\Gamma\delta({\bf r})$ so
that $\epsilon\rightarrow +\infty$. Typical vorticity profiles are
shown in \cite{houches} depending on the value of the temperature.

The Fokker-Planck equation describing the motion of a test vortex in a thermal bath specified by the distribution (\ref{i2}) is
\begin{eqnarray}
{\partial P\over\partial t}={1\over r}{\partial\over\partial r}\biggl\lbrack r  D(r)\biggl ({\partial P\over\partial r}-{4\mu P\eta r\over (\eta+4)R^{2}-\eta r^{2}}\biggr )\biggr\rbrack.
\label{i5}
\end{eqnarray}
The profile of angular velocity corresponding to the
isothermal vortex (\ref{i2}) is
\begin{eqnarray}
\Omega=-{2\Gamma\over \pi R^{2} (\eta+4)}{1\over 1-{\eta\over\eta+4}{r^{2}\over R^{2}}}.
\label{i3}
\end{eqnarray}
Therefore, according to Eq. (\ref{f10}), the diffusion
coefficient is given by
\begin{eqnarray}
D(r)={\gamma\over 4}\ln\Lambda {\eta+4\over |\eta|}{R^{2}\over r^{2}}.
\label{i4}
\end{eqnarray}
Again, this expression is valid only for $r\gg 1$. Setting
$x=(1/2)(r/R)^{2}$, $C=-\eta/(\eta+4)$, $D_{0}={\gamma}\ln\Lambda
{(\eta+4)/ |\eta|}$ and $\tau=D_{0}t/4R^{2}$, we can rewrite the
Fokker-Planck equation as
\begin{eqnarray}
{\partial P\over\partial \tau}={\partial\over\partial x}\biggl ({\partial P\over\partial x}+{\Gamma_{0}\over\gamma}{4CP\over 1+2Cx}\biggr ),
\label{i6}
\end{eqnarray}
with $0\le x\le 1$. This is of the form of Eq. (\ref{co1}) with a logarithmic potential $\Phi(x)=2\mu\ln(1+2Cx)$. With the change of variable $\psi=Pe^{\Phi/2}=(1+2Cx)^{\mu}P$, it can be transformed into a Schr\"odinger equation
with imaginary time
\begin{eqnarray}
{\partial {\psi}\over\partial \tau}={\partial^{2}\psi\over\partial x^{2}}-V_{s}(x)\psi,
\label{i7}
\end{eqnarray}
with a potential
\begin{eqnarray}
V_{s}(x)={\lambda C^{2}\over (1+Cx)^{2}},
\label{i8}
\end{eqnarray}
where $\lambda=4\mu (1+\mu)$. Since the domain is bounded, we cannot
apply the results of Secs. \ref{sec_t} and \ref{sec_co} which assume
that the tail of the distribution extends to infinity. Therefore, in
the next section, we shall consider the isothermal vortex in an
unbounded domain.

\subsection{Isothermal vortex in an infinite domain}
\label{sec_v}

In an infinite domain, extending the calculations of \cite{houches},
we find that the Boltzmann-Poisson equation (\ref{i1}) has solution
only for the critical temperature
\begin{eqnarray}
\beta_{c}=-\frac{8\pi}{\gamma \Gamma}, \label{v1}
\end{eqnarray}
and that the vorticity profile is given by the family of solutions
\begin{eqnarray}
\omega(r)={\omega_{0}\over (1+{\pi\omega_{0}\over\Gamma}r^{2})^{2}},
\label{v2}
\end{eqnarray}
parameterized by the central vorticity $\omega_{0}$. Since $\omega\sim
r^{-4}$ at large distances, we note that the angular momentum diverges
logarithmically. From the Poisson equation (\ref{gg5}) with the vorticity field (\ref{v2}), we find that the streamfunction is given by
\begin{eqnarray}
\psi(r)=-{\Gamma\over 4\pi}\ln\left (\frac{\Gamma}{\pi\omega_{0}}+r^{2}\right ),
\label{v3}
\end{eqnarray}
where we have assumed that $\psi+{\Gamma\over 2\pi}\ln r\rightarrow 0$
for $r\rightarrow +\infty$ in order to determine the Gauge
constant. The energy of interaction $E={1\over 2}\int\omega\psi d{\bf
r}$ is finite and its value is
\begin{eqnarray}
E=-{\Gamma^{2}\over 8\pi}+{\Gamma^{2}\over 8\pi}\ln\left ({\pi\omega_{0}\over \Gamma}\right ).
\label{v4}
\end{eqnarray}
This relation determines the central density $\omega_{0}$ as a
function of $E$. The caloric curve is a straight line with constant
temperature $\beta=\beta_{c}$ for all energies. We note that the
kinetic energy $E'={1\over 2}\int {\bf u}^{2}d{\bf r}={1\over 2}\int
(\nabla\psi)^{2}d{\bf r}$ diverges logarithmically. It differs from
the energy of interaction $E$ because the boundary terms do not
vanish in the present case.

The Fokker-Planck equation describing the relaxation of a test vortex in a thermal bath specified by the distribution (\ref{v2}) is
\begin{eqnarray}
{\partial P\over\partial t}={1\over r}{\partial\over\partial r}\left \lbrack r D(r)\left ({\partial P\over\partial r}+{4\mu P r\over {\Gamma\over \pi\omega_{0}}+r^{2}}\right )\right \rbrack.
\label{v7}
\end{eqnarray}
The profile of angular velocity corresponding to the isothermal vortex (\ref{v2}) is
given by
\begin{eqnarray}
\Omega(r)={\omega_{0}\over 2\left (1+{\pi\omega_{0}\over \Gamma}r^{2}\right )}.
\label{v5}
\end{eqnarray}
Therefore, the diffusion coefficient is
\begin{eqnarray}
D(r)={\gamma\Gamma\over 4\pi \omega_{0}}\ln\Lambda {1\over r^{2}}.
\label{v6}
\end{eqnarray}
Setting $x=(\pi\omega_{0}/2\Gamma)r^{2}$, $D_{0}=\pi\gamma\ln\Lambda$ and $\tau=D_{0}\omega_{0}t/4\Gamma$, the Fokker-Planck equation can be rewritten
\begin{eqnarray}
{\partial P\over\partial \tau}={\partial\over\partial x}\left ({\partial P\over\partial x}+{4\mu P \over 1+2x}\right ),
\label{v8}
\end{eqnarray}
with $x\in [0,\infty[$. This is of the form of Eq. (\ref{co1}) with a
logarithmic potential $\Phi(x)=2\mu\ln(1+2x)$. With
the change of variable $\psi=Pe^{\Phi/2}=(1+2x)^{\mu}P$,
it can be transformed into a Schr\"odinger equation of the form (\ref{i7})
with a potential
\begin{eqnarray}
V_{s}(x)={\lambda \over (1+2x)^{2}},
\label{v9}
\end{eqnarray}
where $\lambda=4\mu(1+\mu)$. In the following, we set
$(\gamma\Gamma/4\pi\omega_{0})\ln\Lambda=1$ and
$\Gamma/\pi\omega_{0}=1$ which can always be achieved by rescaling the
time and the distances appropriately. Then, the results of
Sec. \ref{sec_gen} can be directly applied with $\alpha=4$. Concerning
the tail of the distribution, the position of the front is given by
\begin{eqnarray}
r_{f}(t)\sim (16\mu t)^{1/4},
\label{v10}
\end{eqnarray}
and the profile of the front is
\begin{eqnarray}
u(r,t)\sim \Phi\left\lbrack 2\ {r^{2}-\sqrt{16\mu t}\over (2t)^{1/2}}\right\rbrack.
\label{v11}
\end{eqnarray}
As discussed in Sec. \ref{sec_t}, the theory is only marginally applicable. However, it can give a fair description of the front if the parameter $\epsilon=(3/8)\sqrt{\pi/2\mu}$ is sufficiently small (for $\mu=1$, we have $\epsilon\simeq 0.47$). On the other hand, the temporal correlation function decreases like
\begin{eqnarray}
\langle r(0)r(t)\rangle-\langle r\rangle^{2} \sim t^{-(\mu-1)} .
\label{v12}
\end{eqnarray}
We note that we must impose $\mu>1$ otherwise the variance $\langle
r^{2}\rangle$ of the equilibrium state diverges.

\subsection{Polytropic vortex}
\label{sec_poly}

As discussed in Sec. \ref{sec_evo}, the distribution of the bath can
be any stable stationary solution of the 2D Euler equation with a
monotonic profile of angular velocity, not necessarily the statistical
equilibrium state. A general criterion of nonlinear dynamical
stability is given in Appendix \ref{sec_eul}. As an explicit
example, let us consider an $H$-function (or Casimir integral) of the
form
\begin{eqnarray}
S_{q}=-{1\over q-1}\int (\omega^{q}-\omega)d{\bf r}.
\label{poly1}
\end{eqnarray}
A vorticity distribution which maximizes this functional at fixed
circulation and energy (and angular momentum) is nonlinearly
dynamically stable with respect to the 2D Euler equation. The
variational problem (\ref{eul6}) leads to a relationship between the vorticity and the stream function of the form
\begin{eqnarray}
\omega=\left\lbrack\mu-{\beta (q-1)\over q}\psi'\right \rbrack^{1\over q-1}.
\label{poly2}
\end{eqnarray}
We shall call these vortices {\it polytropic vortices} \cite{gen}
because they are analogous to stellar polytropes in astrophysics
\cite{bt}.  The structure and the stability of stellar polytropes in
$d$ dimensions has been studied in \cite{lang}. For $d=2$, we can
easily adapt these results to describe polytropic vortices.  Note
that the functional (\ref{poly1}) and the corresponding distribution
(\ref{poly2}) are similar to the so-called Tsallis entropies and
Tsallis distributions introduced in non-standard thermodynamics
\cite{tsallis}. However, in the present context, they have a
completely different interpretation
[63,66-69]. They correspond to
dynamical, not thermodynamical, equilibrium states. In particular, the
maximization of $S_{q}$ at fixed circulation and energy is a condition
of nonlinear dynamical stability for the 2D Euler-Poisson system, not
a condition of generalized thermodynamical stability.  The
distributions (\ref{poly2}) form just a particular family of
stationary solutions of the 2D Euler equation that can sometimes give
a good fit of the quasi-stationary state (QSS) resulting from an {\it
incomplete} violent relaxation \cite{jfm2,brands,fermiHMF}.  This
dynamical interpretation of the functional (\ref{poly1}) is different
from the interpretation given by Boghosian \cite{boghosian} in terms
of Tsallis generalized thermodynamics.  For $q\rightarrow 1$, we
recover the isothermal vortex (\ref{g2}) as a special case. For $q=2$,
we have
\begin{eqnarray}
S_{2}=-\int \omega^{2}d{\bf r},
\label{poly3}
\end{eqnarray}
up to an additive constant. Therefore, the neg-enstrophy
$S_{2}=-\Gamma_{2}$ can be interpreted as a particular $H$-function
(or Casimir integral).  In that case, the relationship (\ref{poly2})
is linear
\begin{eqnarray}
\omega=\mu-{1\over 2}\beta\psi'.
\label{poly4}
\end{eqnarray}
The structure and the stability of these {\it linear vortices} has
been studied in detail in \cite{jfm1,jfm2}. In these studies, the
enstrophy was interpreted as an approximation of the
Miller-Robert-Sommeria mixing entropy [47-50] in a limit
of strong mixing (or low energy). Therefore, the results of
\cite{jfm1,jfm2} were presented as results of thermodynamical
stability for the process of violent relaxation (in a particular limit
of the theory). Alternatively, if we view the enstrophy as an
$H$-function (see Appendix \ref{sec_eul}), the results of
\cite{jfm1,jfm2} can also be interpreted as results of nonlinear
dynamical stability with respect to the 2D Euler equation.

For illustration, let us consider the case of a disk of radius $R$ and
take $\Omega_{L}=0$. We also assume that the vortex fills the whole
domain and that $\omega(R)=\psi(R)=0$. This implies that $\mu=0$ in
Eq. (\ref{poly4}). We then obtain the Helmholtz equation
\begin{eqnarray}
\Delta\psi+k^{2}\psi=0,
\label{poly5}
\end{eqnarray}
where $k^{2}=-\beta/2$ (we have assumed $\beta<0$). The solution is
$\psi=A J_{0}(\alpha r/R)$ where $\alpha=kR=2.40482...$ is the first
zero of the Bessel function $J_{0}(x)$. Setting
$\omega_{0}=\alpha^{2}A/R^{2}$, the vorticity profile is
\begin{eqnarray}
\omega=\omega_{0} J_{0}(\alpha r/R).
\label{poly6}
\end{eqnarray}
The Fokker-Planck equation describing the relaxation of a test vortex in a bath specified by the distribution (\ref{poly6}) is
\begin{eqnarray}
{\partial P\over\partial t}={1\over r}{\partial\over\partial r}\left\lbrack r D(r)\left ({\partial P\over\partial r}+{\alpha \Gamma_{0}\over \gamma R}P{J_{1}(\alpha r/R)\over J_{0}(\alpha r/R)}\right )\right\rbrack.\qquad 
\label{poly9}
\end{eqnarray}
Using $(xJ_{1})'=xJ_{0}$, the profile of angular velocity corresponding to the Bessel distribution (\ref{poly6}) is given by
\begin{eqnarray}
\Omega={R\omega_{0}\over\alpha r} J_{1}(\alpha r/R).
\label{poly7}
\end{eqnarray}
Therefore, according to Eq. (\ref{f10}), the diffusion coefficient can
be written
\begin{eqnarray}
D(r)={\gamma\over 4}\ln\Lambda {{\alpha r\over R}J_{0}(\alpha r/R)\over 2J_{1}(\alpha r/R)-{\alpha r\over R}J_{0}(\alpha r/R)}.
\label{poly8}
\end{eqnarray}
Considering now the limit $\beta\rightarrow 0$ and $\Omega_{L}\rightarrow +\infty$ in such a way that the product $\lambda=\beta\Omega_{L}/2$ remains finite, we obtain
\begin{eqnarray}
\omega(r)=\left\lbrack 1-{\lambda(q-1)\over q} r^{2}\right\rbrack^{1\over q-1}.
\label{poly10}
\end{eqnarray}
For $q\rightarrow 1$, we recover the gaussian vortex (\ref{g3}). For
$q>1$, the vorticity vanishes at a finite radius $R_{*}$ (we take
$\omega=0$ for $r>R_{*}$). For $q<1$, the vorticity decreases like
$\omega\sim r^{-2/(1-q)}$. This algebraic behavior falls in the class
of solutions studied in Sec. \ref{sec_p}.

\subsection{Fermi-Dirac vortex}
\label{sec_fd}

As a last example, we consider a typical prediction of the statistical
mechanics of violent relaxation of the 2D Euler equation
[47-50]. In the two-levels approximation, the
statistical equilibrium state is obtained by maximizing the mixing
entropy
\begin{eqnarray}
S=-\int \left\lbrace {\overline{\omega}\over\sigma_{0}}\ln
{\overline{\omega}\over\sigma_{0}}+\left
(1-{\overline{\omega}\over\sigma_{0}}\right )\ln \left
(1-{\overline{\omega}\over\sigma_{0}}\right )\right \rbrace d{\bf r},\qquad
\label{fd1}
\end{eqnarray}
at fixed circulation, energy and angular momentum. This leads to a distribution of the form
\begin{eqnarray}
\overline{\omega}={\sigma_{0}\over 1+Ke^{\beta\psi'}},
\label{fd2}
\end{eqnarray}
with $K>0$. This will be called the {\it Fermi-Dirac vortex}. This
distribution of the bath can emerge from a complete violent relaxation
of the 2D Euler-Poisson system in the first regime of the dynamics
(see discussion in the Conclusion). According to Appendix
\ref{sec_eul}, a maximum of the entropy (\ref{fd1}) at fixed energy
and circulation (thermodynamical equilibrium state in the theory of
violent relaxation) is nonlinearly dynamically stable with respect to
the 2D Euler-Poisson system. For $K\rightarrow +\infty$, we recover
the isothermal vortex (\ref{g2}) and for $K\rightarrow 0$, we obtain a
vortex with uniform vorticity surrounded by irrotational flow (this is
the analogous of the Fermi distribution in phase space). Considering
now the limit $\beta\rightarrow 0$ and $\Omega_{L}\rightarrow +\infty$
in such a way that the product $\lambda=\beta\Omega_{L}/2$ remains
finite, we obtain
\begin{eqnarray}
\overline{\omega}={\sigma_{0}\over 1+Ke^{\lambda r^{2}}}.
\label{fd3}
\end{eqnarray}
For $K\rightarrow +\infty$, we recover the gaussian vortex
(\ref{g3}). More generally, the bath distribution behaves like
$\overline{\omega}\sim e^{-\lambda r^{2}}$ for $r\rightarrow +\infty$
so that the results of Sec. \ref{sec_g} apply to the present context.

\section{Conclusion}
\label{sec_conclusion}

In this paper, we have developed the kinetic theory of point vortices
in two-dimensional hydrodynamics. The interest of this kinetic theory
goes beyond the realm of fluid mechanics since the point vortex gas
constitutes a system of particles with long-range interactions (like
self-gravitating systems) which displays peculiar features in regard
to statistical mechanics and kinetic theory \cite{houches}. Starting
from the Liouville equation and using the projection operator
formalism, we have derived a kinetic equation (\ref{block1}) that
describes the evolution of the system as a whole. This kinetic
equation conserves all the constraints of the point vortex dynamics
(circulation $\dot\Gamma=0$, energy $\dot E=0$ and angular momentum
$\dot L=0$) and increases the Boltzmann entropy monotonically ($\dot
S\ge 0$). However, it does {\it not} necessarily relax towards the
Boltzmann distribution of statistical equilibrium because this kinetic
equation admits an infinite number of stationary solutions and the
evolution stops when there is no resonance anymore (this happens when
the profile of angular velocity becomes monotonic). In that case, the
vorticity profile can remain frozen in a QSS with a non-Boltzmannian
distribution until more complex (e.g., three body) correlations come
into play. This has been shown numerically in Sec. \ref{sec_num}. We
have next considered the relaxation of a test vortex in a steady bath
of field vortices. This relaxation is described by a Fokker-Planck
equation (\ref{f4}) with a space dependent diffusion coefficient. We
have studied analytically and numerically some properties of this
Fokker-Planck equation regarding the evolution of the front profile in
the tail of the distribution function and the decay of the temporal
correlation function of the position of the test vortex depending on
the distribution of the bath. For a thermal bath (gaussian vortex), we
have shown that the evolution of the front position is logarithmic,
scaling like $r_{f}(t)\propto (\ln t)^{1/2}$ and that the temporal
correlation function of the position decreases algebraically (up to
logarithmic corrections), scaling like $\langle
\Delta r(0)\Delta r(t)\rangle\propto \ln t/t$.  
This is at variance with the exponential decay of correlations for the
usual Brownian motion (with constant diffusion coefficient and linear
drift) but this is similar with results obtained for the HMF model
\cite{bd,cl}.  We would like to conclude this paper by discussing the
different regimes that take place in the dynamics of point
vortices. This discussion is important to clarify certain points in
order to answer some of the criticisms raised by Dubin
\cite{dubin} and discuss specifically the domains of application of the different kinetic
theories.

Basically, a system of point vortices is described by the $N$-body
Hamiltonian equations of motion (\ref{s1})-(\ref{s2}). From these
equations, we can derive the Klimontovich equation for the exact
vorticity field $\omega_{exact}({\bf
r},t)=\sum_{i}\gamma_{i}\delta({\bf r}-{\bf r}_{i}(t))$ expressed in
terms of $\delta$-functions (see Eq. (15) of \cite{kin}) or the
Liouville equation for the $N$-body distribution $P_{N}({\bf
r}_{1},...,{\bf r}_{N},t)$ (see Eq. (59) of \cite{kin}). These exact
equations are equivalent to the $N$-body dynamics and contain too
much information to be of practical use. The object of the kinetic
theory is to derive an equation for the smooth vorticity field
$\omega({\bf r},t)=\langle \sum_{i}\gamma_{i}\delta({\bf r}-{\bf
r}_{i}(t))\rangle$ or the one-body distribution $P_{1}({\bf r},t)$
(see Sec. \ref{sec_k}). By rescaling the variables appropriately, we
have seen that the ``collision term'' could be evaluated in an
expansion in powers of $1/N$ for $N\rightarrow +\infty$ with fixed
normalized energy $\epsilon= E/(N^{2}\gamma^{2})$ and fixed
normalized temperature $\eta=\beta\gamma\Gamma$. This unusual
thermodynamic limit corresponds to $N\rightarrow +\infty$ with
$\gamma\sim 1/N$ and $V\sim 1$ (implying $E\sim 1$ and $\beta\sim
N$). For the moment, only the term of order $O(1/N)$ in the kinetic
theory has been computed [see Eq. (\ref{k1})]. This amounts to
neglecting three-body (or higher) correlations among the vortices.
These points are made clear in \cite{bbgky}.

To leading order in $N\rightarrow +\infty$, the evolution of the
smooth vorticity field $\omega({\bf r},t)$ representing the spatial
distribution of the point vortices is governed by the 2D Euler-Poisson
system (see Eq. (125) of \cite{kin}) as for inviscid and
incompressible 2D flows with continuous vorticity. This Vlasov (or
mean field) limit amounts to neglecting all the correlations among the
vortices. Physically, this regime is valid for $t\ll Nt_{D}$ and its
domain of validity can be very long when $N$ is large. The 2D
Euler-Poisson system has a very complex dynamics. Starting from an
unstable initial condition $\omega_{0}({\bf r})$, the 2D Euler-Poisson
system develops a complicated mixing process leading ultimately to a
Quasi Stationary State (QSS), also called a metaequilibrium state,
which has the form of a vortex or a jet. This ``collisionless
relaxation'' is similar to the process of violent relaxation in
astrophysics based on the Vlasov-Poisson system
\cite{lb,csr,houches}. It takes place on a timescale of the order of a
few dynamical times $t_{D}$. The resulting QSS is a stable stationary
solution of the 2D Euler equation on the coarse-grained scale
$\overline{\omega}({\bf r})$, i.e. if we locally average over the
filaments. A statistical theory has been developed by Miller
\cite{miller} and Robert \& Sommeria
\cite{rs} in order to predict the most probable equilibrium state
resulting from a violent relaxation. This theory is based on a
hypothesis of ergodicity, assuming that the relaxation is complete, so
that the system is expected to be in the most mixed state.  This is
similar to the statistical theory of Lynden-Bell \cite{lb} for
collisionless stellar systems. A kinetic theory of the process of
violent relaxation has been proposed by Robert \& Sommeria
\cite{rsmep}, and further developed in \cite{csr,rr,csprl}, by using
a phenomenological Maximum Entropy Production Principle (MEPP). They
obtained a drift-diffusion equation for the coarse-grained vorticity
$\overline{\omega}({\bf r},t)$ with a time dependent temperature
$\beta(t)$ that evolves in order to satisfy the conservation of energy
(see Eqs. (11)-(12) of \cite{rsmep}). In order to determine the domain
of validity of this phenomenological approach and determine the value
of the diffusion coefficient (which is not given by the MEPP),
Chavanis \cite{prl} attempted to develop a kinetic theory of violent
relaxation, starting directly from the 2D Euler equation and using a
systematic procedure based on a coarse-graining and a linearization of
the equation for the fluctuations.  This quasilinear theory was
developed in analogy with the quasilinear theory of the Vlasov-Poisson
system in astrophysics \cite{kp,sl}. The approach of Chavanis
\cite{prl} remains, however, unsatisfactory because it yields an
equation [see Eq. (20)] which does not conserve energy. This could be
cured as proposed in \cite{kin} [see Eq. (148)] by taking into account
resonances between the motion of the fluid particles but this
procedure may not be completely justified since there is no real
timescale separation between the decrease of correlations and the
evolution of the coarse-grained vorticity field (unlike during the
slow collisional relaxation of point vortices). As a result, memory
(non-Markovian) effects are expected to be important in the process of
violent relaxation. Note that the more general equation (18) derived 
in \cite{prl} may conserve energy and provide a correct description of the 
coarse-grained dynamics \cite{bbgky}.

If we now consider an axisymmetric distribution of point vortices that
is stable with respect to the 2D Euler equation and if we take into
account terms of order $1/N$ in the kinetic theory, we describe the
``collisional'' relaxation process. The kinetic equation governing the
evolution of the system as a whole is given by Eq.  (\ref{block1})
obtained by Chavanis \cite{kin}, or by the more general kinetic
equation taking into account collective effects obtained by Dubin \&
O'Neil \cite{dn}. Physically, this regime is valid on a timescale
$Nt_{D}$. For single species systems, the evolution of the smooth
vorticity field is due to a condition of resonance
$\Omega(r,t)=\Omega(r',t)$ which can be satisfied only when the
profile of angular velocity is non-monotonic. As a result, the
evolution stops when the profile of angular velocity becomes monotonic
so that there is no resonance anymore.  In that case, the system
remains frozen in a long-lived QSS. This is a stationary solution of
the 2D Euler equation which usually differs from the statistical
equilibrium state (Boltzmann distribution) predicted by Joyce \&
Montgomery \cite{jm} and Lundgren \& Pointin
\cite{lp}. The fact that the system can be frozen in a QSS state 
in the absence of resonance was stressed in
\cite{kin}. This implies either that the system will not reach
statistical equilibrium at all or that the kinetic theory is
incomplete. Indeed, the collision term calculated in the kinetic
theory of Dubin \& O'Neil \cite{dn} and Chavanis \cite{kin}
corresponds to a term of order $1/N$ in a systematic expansion carried
out in a proper thermodynamic limit $N\rightarrow +\infty$.  In
particular, it neglects non-trivial three-body correlations.
Therefore, this theory describes the dynamics of the system on a
timescale of order $N t_{D}$ where $t_{D}$ is the dynamical time. If
the collision term cancels out because of the absence of resonance,
i.e. when the profile of angular velocity becomes monotonic, the
relaxation towards statistical equilibrium (if any) will take place on
a timescale larger than $Nt_{D}$ so that next order terms in the $1/N$
expansion must be taken into account in the kinetic theory.  These
higher order terms have not yet been considered. Therefore, there is
no certitude, up to date, that the point vortex gas relaxes towards
statistical equilibrium for $t\rightarrow +\infty$. This could be
checked in numerical simulations of the point vortex dynamics
\footnote{Some interesting work in that direction has been done
recently in \cite{jap,jap2}.} by solving the Hamilton equations
(\ref{s1})-(\ref{s2}). A challenging problem is to determine the
collisional relaxation time $t_{relax}$ of the point vortex gas
towards statistical equilibrium. According to the present kinetic
theory it should scale like $t_{relax}\sim N^{\delta} t_{D}$ with
$\delta\ge 1$ but the precise value of $\delta$ is not yet known (see
Sec. \ref{sec_ss}). In numerical simulations, we must be careful not
to confound the true statistical equilibrium state resulting from a
{\it slow} collisional relaxation for large times (larger that
$Nt_{D}$) with the QSS resulting from a {\it violent} collisionless
relaxation taking place on a much shorter timescale $\sim
t_{D}$. Indeed, there are many confusions in the literature between
the Joyce-Montgomery
\cite{jm} statistical equilibrium state (collisional relaxation) and
the Miller-Robert-Sommeria \cite{miller,rs} statistical equilibrium
state (collisionless relaxation). Although these statistical
distributions are mathematically similar in certain limits (dilute
limit), their physical origin is completely different since they
apply to different regimes of the point vortex dynamics. This
implies in particular that the two theories are not in conflict
since they refer to completely different regimes.

Finally, if we consider the relaxation of a test particle in a
thermal bath of point vortices at statistical equilibrium, we obtain
a Fokker-Planck equation (\ref{f13}) which has the form of a
drift-diffusion equation with a space dependent diffusion
coefficient inversely proportional to the local shear created by the
bath. This Fokker-Planck equation derived by Chavanis
\cite{preR,kin} is {\it similar} to the relaxation equation proposed phenomenologically  by Robert \& Sommeria \cite{rsmep} by using the MEPP to describe the
evolution of the 2D Euler equation on the coarse-grained scale during
the process of violent relaxation. However, this resemblance is
essentially coincidental since the two theories describe completely
different situations and different timescales.  Therefore, the kinetic
theory of point vortices is rich and different regimes can be
evidenced.  For that reason, it is important to delimit the domains of
application of each kinetic theory in order to avoid
misunderstandings.  Indeed, the mathematical structure of the kinetic
equations can be formally similar while describing completely
different situations. When the conditions of validity of the different
theories are properly delineated, there are no contradiction between
our results and those obtained by Dubin \cite{dubin}. The works are
complementary.

In future works, it will be important to evaluate terms of higher
order in the development in $1/N$ of the kinetic equation so as to
describe the evolution of the system on longer timescales. This
implies the evaluation of two and three-body correlation functions,
which is a formidable task. We must also distinguish the case where
the circulations of the point vortices have the same sign or a
different sign. When the circulations of the point vortices take
positive and negative values, the point vortices can group themselves
in dipoles or tripoles. As indicated in Sec. \ref{sec_pvg}, this
situation cannot be described by the mean field theory. The system
will then have a very different dynamics than the one described here
(in Sec. \ref{sec_block}). For example, Sire \& Chavanis \cite{sc}
show that the merging between vortices in 2D decaying turbulence is
due essentially to collisions between dipoles and monopoles. The
inclusion of dipoles and tripoles in the kinetic theory needs to take
into account two and three-body correlation functions as discussed by
Newton \& Mezic \cite{nm}. In our approach, these pairs would form on
a timescale larger than $Nt_{D}$ so they have not been taken into
account. We note, finally, that our approach implicitly assumes that
there are no correlations initially so that the $1/N$ development is
well-defined (see
\cite{bbgky} for more details). If pairs $(+,+)$, dipoles $(+,-)$, 
tripoles $(+,-,+)$ or more complex ``structures'' exist initially,
then we must take into account high order correlations since the start
and the kinetic theory will be different.

Another extension of the kinetic theory is to consider flows that are
not axisymmetric. In that case, the kinetic equation valid at the
order $1/N$ is given by Eq. (\ref{k1}). It describes the evolution of
the system as a whole on a timescale $Nt_{D}$. Using the timescale
separation between the dynamical time $t_{D}$ and the collisional
relaxation time $t_{relax}$ (of order $Nt_{D}$ or larger), it would be
of interest to derive an ``orbit-averaged'' kinetic equation in terms
of appropriate variables similar to the angle-action variables used in
other contexts
\cite{bt,action}. The resulting kinetic equation would
be more complicated than Eq. (\ref{block1}) but the basic concepts
remain the same. For non-axisymmetric flows, we can have a lot of
resonances (see the expectedly similar case in \cite{action}) and they
can ``push'' the system towards a distribution {\it close} to the
Boltzmann distribution on a timescale $\sim Nt_{D}$ (the exact
Boltzmann distribution should be reached however on longer timescales
due to higher order terms $1/N^{2}$, $1/N^{3}$,... in the kinetic
equation). The same remark applies to the inhomogeneous phase of the
HMF model as discussed in
\cite{action}. While there is no evolution at all on a timescale
$Nt_{D}$ in the homogeneous phase of the HMF model (because there is
no resonance), many resonances appear in the inhomogeneous phase which
can ``push'' the system towards a distribution {\it close} to the
Boltzmann distribution on a timescale $N t_{D}$. These interesting
problems should be given further consideration in future works.

\appendix

\section{Nonlinear dynamical stability of stationary solutions of the 2D Euler equation}
\label{sec_eul}

We consider the 2D Euler-Poisson system
\begin{eqnarray}
{\partial\omega\over\partial t}+{\bf u}\cdot \nabla\omega=0, \qquad {\bf u}=-{\bf z}\times \nabla\psi, \qquad \omega=-\Delta\psi.\nonumber\\
\label{eul1}
\end{eqnarray}
Any relation of the form $\omega=f(\psi)$ determines a stationary
solution of the 2D Euler equation $\partial\omega/\partial t=0$
since $\nabla\omega=f'(\psi)\nabla\psi$ is perpendicular to ${\bf
u}$. This result is valid in an arbitrary domain. In a disk or in an
infinite domain, we can also consider pseudo-stationary solutions of
the form $\omega(r,\theta,t)=\omega(r,\theta-\Omega_{L} t)$ that
describe a structure rotating uniformly with angular velocity
$\Omega_{L}$ (the system is stationary in the rotating frame). If we
write the 2D Euler equation in the form
\begin{eqnarray}
{\partial\omega\over\partial t}+{1\over r}{\partial\psi\over\partial\theta}{\partial \omega\over\partial r}-{1\over r}{\partial\psi\over\partial r}{\partial \omega\over\partial \theta}=0,
\label{eul2}
\end{eqnarray}
and substitute for the preceding relation, we find that
\begin{eqnarray}
{\partial\psi\over\partial\theta}{\partial \omega\over\partial
r}-\left (\Omega_{L} r+{\partial\psi\over\partial r}\right ){\partial
\omega\over\partial \theta}=0.
\label{eul3}
\end{eqnarray}
This equation is satisfied by any relation of the form $\omega=f(\psi')$ where $\psi'=\psi+{1\over 2}\Omega_{L} r^{2}$ is the relative streamfunction.

One important question concerns the dynamical stability of a
stationary solution of the 2D Euler equation. Let us introduce the
functionals
\begin{eqnarray}
S=-\int C(\omega)d{\bf r},
\label{eul4}
\end{eqnarray}
where $C$ is any convex function, i.e. $C''>0$. Such functionals form
a particular class of Casimirs. They are also called generalized
$H$-functions for the reasons explained in \cite{tremaine,physicaA}.
It can be shown that the maximization problem
\begin{eqnarray}
{\rm Max} \lbrace S[\omega]\quad |\quad E[\omega]=E, \Gamma[\omega]=\Gamma, L[\omega]=L \rbrace,
\label{eul5}
\end{eqnarray}
determines a steady solution of the 2D Euler equation with monotonic
$\omega=f(\psi')$ relationship that is nonlinearly dynamically stable
\footnote{Note that this maximization problem is formally similar to a
criterion of (generalized) thermodynamical stability in the
microcanonical ensemble but it has a completely different physical
interpretation \cite{gen,antonov}.}. This condition of nonlinear
dynamical stability has been proven by Ellis {\it et al.}
\cite{ellis}. It refines the Arnold's sufficient criteria of
nonlinear dynamical stability and solves some apparent ``paradoxes''
observed in geophysical fluid dynamics (some flows can be dynamically
stable according to the criterion (\ref{eul5}) although they do not
satisfy the Arnold criteria) \cite{ellis,physicaD}. Note that the
nonlinear dynamical stability criterion (\ref{eul5}) has been proven
in the absence of angular momentum constraint. When the angular
momentum is conserved, the optimization problem (\ref{eul5})
determines a family of solutions that can be deduced from each other
by a mere rotation. In that case, the notion of nonlinear dynamical
stability must be reconsidered to account for this property.

Introducing Lagrange multipliers and writing the first order
variations as
\begin{eqnarray}
\delta S-\beta\delta E-{1\over 2}\beta\Omega_{L} \delta L-\alpha\delta\Gamma=0,
\label{eul6}
\end{eqnarray}
we get
\begin{eqnarray}
C'(\omega)=-\beta\psi'-\alpha, \label{eul7}
\end{eqnarray}
where $\psi'$ is the relative stream function defined above. Since
$C$ is monotonic, this relation can be reversed to give
$\omega=F(\beta\psi'+\alpha)=f(\psi')$ where $F(x)=(C')^{-1}(-x)$.
Therefore $f$ is monotonic. It is increasing at negative
``temperatures'' and decreasing at positive ``temperatures'', since
$\omega'(\psi')=-\beta/C''(\omega)$. When $\beta\rightarrow 0$ and
$\Omega_{L}\rightarrow +\infty$ in such a way that the product
$\lambda\equiv {1\over 2}\beta\Omega_{L}$ remains finite, the
previous solution becomes $\omega=F(\lambda r^{2}+\alpha)$.

In conclusion, the critical points of the optimization problem (\ref{eul5})
determine steady states of the 2D Euler equation (in a rotating
frame). Furthermore,  maxima of $S$ at fixed $E$, $\Gamma$ (and $L$)
are nonlinearly dynamically stable.

\section{The calculation of the function $M$}
\label{sec_m}

The velocity (by unit of circulation) created by point vortex $1$
located in ${\bf r}_{1}$ on point vortex $0$ located in ${\bf r}$ is
\begin{eqnarray}
{\bf V}(1\rightarrow 0)={1\over 2\pi}\hat{\bf z}\times {{\bf r}-{\bf r}_{1}\over |{\bf r}-{\bf r}_{1}|^{2}}={1\over 2\pi}\hat{\bf z}\times \nabla \ln |{\bf r}-{\bf r}_{1}|.\nonumber\\
\label{m1}
\end{eqnarray}
Therefore, the radial component $V_{r}={\bf V}\cdot \hat{\bf r}$ of the velocity in the direction of the test vortex $0$ can be written
\begin{eqnarray}
V_{r}(1\rightarrow 0)=-{1\over 2\pi r}{\partial\over\partial\theta}\ln |{\bf r}-{\bf r}_{1}|.
\label{m2}
\end{eqnarray}
Introducing a polar system of coordinates $(r,\theta)$ to localize a
point vortex and using
\begin{eqnarray}
|{\bf r}-{\bf r}_{1}|=r_{1}^{2}+r^{2}-2r r_{1}\cos(\theta-\theta_{1}),
\label{m3}
\end{eqnarray}
we obtain Eqs. (\ref{k6}) and (\ref{k7}). The function $M$ can
be evaluated from these expressions as done in
\cite{kin,houches}. However, it is more convenient to directly use the
expansion
\begin{eqnarray}
\ln|{\bf r}-{\bf r}_{1}|=\ln r_{>}-\sum_{m\neq 0}{1\over 2|m|}\left ({r_{<}\over r_{>}}\right )^{|m|}e^{im(\theta-\theta_{1})},\nonumber\\
\label{m4}
\end{eqnarray}
so that
\begin{eqnarray}
V_{r}(1\rightarrow 0,t)={1\over 4\pi r}\sum_{m\neq 0}{i m\over |m|}\left ({r_{<}\over r_{>}}\right )^{|m|}e^{im(\theta-\theta_{1})}.\qquad
\label{m5}
\end{eqnarray}
Using the equations of motion (\ref{k4}), we have
\begin{eqnarray}
V_{r}(1\rightarrow 0,t-\tau)={1\over 4\pi r}\sum_{m\neq 0}{i m\over |m|}\left ({r_{<}\over r_{>}}\right )^{|m|}e^{im(\theta-\theta_{1}-\Delta\Omega \tau)}.\nonumber\\
\label{m6}
\end{eqnarray}
Therefore, the function (\ref{k9}) takes the form
\begin{eqnarray}
M=-\int_{0}^{+\infty}d\tau\int_{0}^{2\pi} d\theta_{1} {1\over 16\pi^{2}r^{2}}\qquad\qquad\nonumber\\
\times
\sum_{m\neq 0,n\neq 0}{m n\over |m n|}\left ({r_{<}\over r_{>}}\right )^{|m|+|n|}e^{i m(\theta-\theta_{1})}e^{in(\theta-\theta_{1}-\Delta\Omega \tau)}.\nonumber\\
\label{m7}
\end{eqnarray}
Integrating on the angles, we obtain
\begin{eqnarray}
M={1\over 16\pi r^{2}}
\sum_{m\neq 0}\left ({r_{<}\over r_{>}}\right )^{2|m|}\int_{-\infty}^{+\infty}d\tau e^{i m\Delta\Omega \tau}.
\label{m8}
\end{eqnarray}
Finally, the time integration yields
\begin{eqnarray}
M={1\over 8 r^{2}}
\sum_{m\neq 0}\left ({r_{<}\over r_{>}}\right )^{2|m|} \delta(m\Delta\Omega),
\label{m9}
\end{eqnarray}
or equivalently
\begin{eqnarray}
M={1\over 4 r^{2}}\delta(\Delta\Omega)
\sum_{m=1}^{+\infty}\left ({r_{<}\over r_{>}}\right )^{2m}{1\over m}.
\label{m10}
\end{eqnarray}
This  can be rewritten
\begin{eqnarray}
M=-{1\over 4 r^{2}}\delta(\Delta\Omega)\ln\biggl\lbrack 1-\biggl
({r_{<}\over r_{>}}\biggr )^{2}
\biggr\rbrack.
\label{m11}
\end{eqnarray}

As pointed out in Sec. \ref{sec_k}, this expression has no sense for
$r$ and $r_1$ such that $\Omega(r)=\Omega(r_1)$ and ${\partial
\Omega}/{\partial r}(r_1)=0$, because the Dirac function cannot be
defined in this case even in the sense of distributions. However, a
simple regularization can be introduced to overcome this technical
problem. Indeed, for the derivation of the expression $M$ above, one
could restrict the integration to a finite time $t$ (this is in fact
what the kinetic theory says at the start) and get an approximation
$M(t)$ of $M$ as follows:
\begin{eqnarray}
M(t)={1\over 16\pi r^{2}} \sum_{m\neq 0}\left ({r_{<}\over
r_{>}}\right )^{2|m|} \int_{-t}^{+t}d\tau e^{i m\Delta\Omega
\tau},\label{m12}
\end{eqnarray}
which is equivalent to
\begin{eqnarray}
M(t)={1\over 8\pi r^{2}} \sum_{m\geq 1}\left ({r_{<}\over
r_{>}}\right)^{2m} \frac{e^{i mt\Delta\Omega }- e^{-i m
t\Delta\Omega }}{i m\Delta\Omega}.\nonumber\\ \label{m13}
\end{eqnarray}
Using the complex logarithmic function, we get
\begin{eqnarray}
M(t)={1\over 8\pi r^{2}} \frac{1}{i\Delta \Omega}\biggl\lbrace - \ln
\biggl\lbrack 1-\left({r_{<}\over
    r_{>}}\right)^2 e^{i t \Delta\Omega }\biggr\rbrack\nonumber\\
 +\ln \left\lbrack 1-\left({r_{<}\over
    r_{>}}\right)^2 e^{-i t\Delta\Omega }\right\rbrack\biggr\rbrace, \qquad\qquad\label{m14}
\end{eqnarray}
which finally simplifies into
\begin{eqnarray}
M(t)={1\over 4\pi r^{2}} \frac{1}{\Delta \Omega}\arctan\left[
\frac{\left({r_{<}\over
    r_{>}}\right)^2 \sin( t\Delta\Omega) }{ 1-\left ({r_{<}\over
    r_{>}}\right)^2 \cos(t\Delta\Omega )}\right]. \nonumber\\
\label{m15}
\end{eqnarray}
Note that  this expression tends to  expression (\ref{m11}) of $M$
when $t$ goes to  infinity.

\section{First and second moments of the radial increment}
\label{sec_fs}

In this Appendix, we calculate the first and second moments $\langle
\Delta r\rangle$ and $\langle (\Delta r)^{2}\rangle$ of the radial
increment of the test vortex directly from the
Hamiltonian equations of motion (\ref{s1})-(\ref{s2}). We follow a
procedure similar to that used by Valageas \cite{valageas} in a
different context. Since the calculations are similar, we shall only
give the main steps of the derivation. For simplicity, we assume that
all the vortices have the same circulation $\gamma$. In order to
separate the mean field dynamics from the discrete effects which give
rise to the diffusion and the drift of point vortices, we write the Hamiltonian
(\ref{s2}) as
\begin{eqnarray}
H=\gamma (H_{0}+H_{I}),
\label{fs1}
\end{eqnarray}
where we defined the mean field Hamiltonian $H_{0}$ by
\begin{eqnarray}
H_{0}=\sum_{i=1}^{N}\psi_{0}({\bf r}_{i}),
\label{fs2}
\end{eqnarray}
and the interaction Hamiltonian $H_{I}$ by
\begin{eqnarray}
H_{I}=e^{\omega t}\left\lbrack -\frac{\gamma}{4\pi}\sum_{i\neq j}\ln |{\bf r}_{i}-{\bf r}_{j}|-\sum_{i}\psi_{0}({\bf r}_{i})\right\rbrack.
\label{fs3}
\end{eqnarray}
In Eq. (\ref{fs2}) the mean field stream function is given by
\begin{eqnarray}
\psi_{0}({\bf r})=-\frac{1}{2\pi}\int \omega({\bf r}')\ln |{\bf r}-{\bf r}'| d{\bf r}',
\label{fs4}
\end{eqnarray}
where $\omega({\bf r}')$ is the mean field equilibrium vorticity. The
factor $e^{\omega t}$ has been added for the computation of
perturbative eigenmodes and we shall ultimately let $\omega\rightarrow
0^{+}$ (this $\omega$ should not be confused with the vorticity
$\omega({\bf r})$). Thus $H_{0}$ describes the mean field dynamics
whereas $H_{I}$ describes the discrete effects which vanish in the
limit $N\rightarrow +\infty$. Therefore, we consider $H_{I}$ as a
perturbation of $H_{0}$ and we apply a perturbative analysis in powers
of $1/N$.

Using the identity (\ref{m4}), the  interaction Hamiltonian $H_{I}$ is given by
\begin{eqnarray}
H_{I}=e^{\omega t}\biggl\lbrack \frac{\gamma}{4\pi}\sum_{i\neq j}\sum_{m\neq 0}\frac{1}{2|m|}\left (\frac{r_{<}}{r_{>}}\right )^{|m|}e^{im(\theta-\theta')}\nonumber\\
-\frac{\gamma}{4\pi}\sum_{i\neq j}\ln r_{>}+\sum_{i}\int \omega(r')\ln r_{>}r'dr'\biggr\rbrack.
\label{fs5}
\end{eqnarray}
On the other hand, introducing a polar system of coordinates, the equations of motion read
\begin{eqnarray}
\gamma\frac{dr_{i}}{dt}=\frac{1}{r_{i}}\frac{\partial H}{\partial\theta_{i}}, \qquad \gamma r_{i}\frac{d\theta_{i}}{dt}=-\frac{\partial H}{\partial r_{i}}.
\label{fs6}
\end{eqnarray}
We write the trajectories $\lbrace r(t),\theta(t)\rbrace$ as the perturbative expansions $r=r^{(0)}+r^{(1)}+r^{(2)}+...$ where $r^{(k)}$ is formally of order $k$ over $H_{I}$. At zeroth-order, we simply have
\begin{eqnarray}
\frac{dr_{i}^{(0)}}{dt}=0, \qquad \frac{d\theta_{i}^{(0)}}{dt}=\Omega(r_{i}^{(0)}),
\label{fs7}
\end{eqnarray}
which yields the mean field equilibrium orbits
\begin{eqnarray}
r_{i}^{(0)}={\rm constant}, \qquad \theta_{i}^{(0)}=\Omega(r_{i}^{(0)})t+\theta_{i}^{(0)}(0).
\label{fs8}
\end{eqnarray}
At first order, we obtain
\begin{eqnarray}
\frac{dr_{i}^{(1)}}{dt}=\frac{1}{r_{i}^{(0)}}\frac{\partial H_{I}}{\partial\theta_{i}}, \qquad  \frac{d\theta_{i}^{(1)}}{dt}=\Omega'(r_{i}^{(0)})r_{i}^{(1)}-\frac{1}{r_{i}^{(0)}}\frac{\partial H_{I}}{\partial r_{i}},\nonumber\\
\label{fs9}
\end{eqnarray}
where we can substitute the zeroth-order orbits in the r.h.s. A simple calculation yields
\begin{eqnarray}
r_{i}^{(1)}=\frac{1}{r_{i}^{(0)}}\frac{\partial \chi}{\partial\theta_{i}}, \qquad  \theta_{i}^{(1)}=-\frac{1}{r_{i}^{(0)}}\frac{\partial \chi}{\partial r_{i}},
\label{fs10}
\end{eqnarray}
with
\begin{eqnarray}
\chi=e^{\omega t}\frac{\gamma}{4\pi}\sum_{j\neq j'}\sum_{m\neq 0}\frac{1}{2|m|}\left (\frac{r_{<}}{r_{>}}\right )^{|m|}\frac{e^{im(\theta-\theta')}}{\omega+im(\Omega-\Omega')}.\nonumber\\
\label{fs11}
\end{eqnarray}
At second order, we have
\begin{eqnarray}
\frac{dr_{i}^{(2)}}{dt}=\frac{1}{r_{i}^{(0)}}\sum_{j}\frac{\partial^{2} H_{I}}{\partial\theta_{i}\partial r_{j}}r_{j}^{(1)}\nonumber\\
+\frac{1}{r_{i}^{(0)}}\sum_{j}\frac{\partial^{2} H_{I}}{\partial\theta_{i}\partial \theta_{j}}\theta_{j}^{(1)}-\frac{r_{i}^{(1)}}{(r_{i}^{(0)})^2}\frac{\partial H_{I}}{\partial\theta_{i}}.
\label{fs12}
\end{eqnarray}
Considering a test vortex, and taking the average with the distribution $\omega(r)$ over the orbits of other field vortices yields at order $1/N$
\begin{eqnarray}
\langle \dot r^{(2)}\rangle=\frac{\gamma}{8\pi} e^{2\omega t}\frac{\partial}{\partial r} \int r'dr'\omega(r')\nonumber\\
\times\sum_{m\neq 0}\frac{1}{r^{2}}\left (\frac{r_{<}}{r_{>}}\right )^{2|m|}
\frac{\omega}{\omega^{2}+m^{2}(\Omega-\Omega')^{2}}\nonumber\\
-\frac{\gamma}{8\pi} e^{2\omega t}\int dr'\omega(r')\frac{\partial}{\partial r'}\nonumber\\
\times\sum_{m\neq 0}\frac{1}{r}\left (\frac{r_{<}}{r_{>}}\right )^{2|m|}\frac{\omega}{\omega^{2}+m^{2}(\Omega-\Omega')^{2}}.
\label{fs13}
\end{eqnarray}
Then, using ${\rm lim}_{\omega\rightarrow
0}\omega/(\omega^{2}+x^{2})=\pi\delta(x)$, the limit
$\omega\rightarrow 0^{+}$ gives Eq. (\ref{d15}). On the other hand,
from Eqs. (\ref{fs10})-(\ref{fs11}), the mean square radial
displacement $\langle (\Delta r)^{2}\rangle$ reads at order $1/N$:
\begin{eqnarray}
\langle (\Delta r)^{2}\rangle=\frac{\gamma}{8\pi r^{2}}e^{2\omega t_{1}}\int r' dr' \omega(r')\nonumber\\
\times\sum_{m\neq 0}\left (\frac{r_{<}}{r_{>}}\right )^{2|m|}\frac{1}{\omega^{2}+m^{2}(\Omega-\Omega')^{2}}\nonumber\\
\times \left (1+e^{2\omega\Delta t}-2e^{\omega\Delta t}\cos\lbrack m(\Omega-\Omega')\Delta t\rbrack\right ).
\label{fs14}
\end{eqnarray}
The limit $\omega\rightarrow 0^{+}$ now gives
\begin{eqnarray}
\left \langle \frac{(\Delta r)^{2}}{\Delta t}\right \rangle=\frac{\gamma}{4\pi r^{2}}\int r' dr' \omega(r')\nonumber\\
\times\sum_{m\neq 0}\left (\frac{r_{<}}{r_{>}}\right )^{2|m|}\frac{1-\cos\lbrack m(\Omega-\Omega')\Delta t\rbrack}{m^{2}(\Omega-\Omega')^{2}\Delta t}.
\label{fs15}
\end{eqnarray}
Taking $\Delta t\rightarrow +\infty$ and using ${\rm
lim}_{t\rightarrow +\infty}(1-\cos tx)/tx^{2}=\pi\delta(x)$, we obtain
Eq. (\ref{d14}).


\end{document}